\documentclass[english,11pt,aps,prd,a4paper,preprintnumbers, floatfix,nofootinbib,showpacs,superscriptaddress,  notitlepage]{revtex4-1} 
 \pdfoutput=1
\usepackage[usenames,dvipsnames]{color}  
\usepackage{graphicx}
\usepackage{setspace}
\usepackage{upgreek}
\usepackage{braket}
\usepackage{enumitem}
\usepackage{caption}
\usepackage{stackengine}
\usepackage{subcaption}
\usepackage{amsmath}
\usepackage{amssymb}
\usepackage[colorlinks=true,citecolor=darkred,urlcolor=darkred, pdfborder={0 0 0}]{hyperref}
\usepackage[normalem]{ulem}
\usepackage[table]{xcolor}
\usepackage{graphicx}
\makeatletter
\def\p@subsection{}
\makeatother
\usepackage{float}

\usepackage{ctable}
\usepackage{colortbl}
\definecolor{darktangerine}{rgb}{1.0, 0.66, 0.07}
\definecolor{gainsboro}{rgb}{0.86, 0.86, 0.86}
\definecolor{ceruleanblue}{rgb}{0.16, 0.32, 0.75}
\usepackage{fdsymbol}
\usepackage[capitalize]{cleveref}
\usepackage{soul}

\definecolor{darkred}{rgb}{0.6,0,0}

\definecolor{linkcolor}{rgb}{0,0,0.5}

\definecolor{vdrgreen}{rgb}{0.0, 0.7, 0.0}

\def\gsim{\raise0.3ex\hbox{$\;>$\kern-0.75em\raise-1.1ex\hbox{$\sim\;$}}}
\def\lsim{\raise0.3ex\hbox{$\;<$\kern-0.75em\raise-1.1ex\hbox{$\sim\;$}}}

\def\beqn#1{\begin{equation}\label{#1}}
\def\eeqn{\end{equation}}

\def\beqa#1{\begin{eqnarray}\label{#1}}
\def\eeqa{\end{eqnarray}}

\usepackage{caption}
\usepackage{subcaption}
\usepackage{adjustbox}

\def\Z2{$\mathcal{Z_2}$}

\def\vev#1{\left\langle #1\right\rangle}

\usepackage{bbold}
\usepackage{ragged2e}

  \newcommand{\rd}[1]{{\color{red}  #1}}
 
\newcommand {\ignore}[1]{}

\def\cevns{CE$\nu$NS}

\usepackage{mathrsfs}

\makeatletter
\newcommand\footnoteref[1]{\protected@xdef\@thefnmark{\ref{#1}}\@footnotemark}
\makeatother

\def\321{$\mathrm{SU(3) \otimes SU(2) \otimes U(1)}$ }

\def\rd#1{\textcolor{red}{#1}}

\newcommand{\AddrAHEP}{%
  AHEP Group, Institut de F\'{i}sica Corpuscular --
  CSIC/Universitat de Val\`{e}ncia and Departament de F\'{i}sica Te\`{o}rica, Universitat de Val\`{e}ncia,  Parc Cient\'ific de Paterna.\\
 C/ Catedr\'atico Jos\'e Beltr\'an, 2 E-46980 Paterna (Valencia) - SPAIN}

\newcommand{\AddrMiranda}{%
Departamento de F\'{\i}sica, Centro de Investigaci\'on
  y de Estudios Avanzados del IPN,\\ Apartado Postal 14-740 07000 Ciudad de Mexico, Mexico}

\newcommand{\AddrAthens}{%
Department of Physics, National and Kapodistrian University
of Athens, Zografou Campus GR-15772 Athens, Greece}

\usepackage{appendix}
\begin{document}

\title{\textcolor{BrickRed}{Physics implications of a combined analysis of COHERENT CsI and LAr data}}

\author{V. De Romeri}\email{deromeri@ific.uv.es}
\affiliation{\AddrAHEP}

\author{O. G. Miranda}\email{omar.miranda@cinvestav.mx}
\affiliation{\AddrMiranda}

\author{D. K. Papoulias}\email{dkpapoulias@phys.uoa.gr}
\affiliation{\AddrAthens}

\author{G. Sanchez Garcia}\email{gsanchez@fis.cinvestav.mx}
\affiliation{\AddrAHEP}
\affiliation{\AddrMiranda}

\author{M. T\'ortola}\email{mariam@ific.uv.es}
\affiliation{\AddrAHEP}

\author{J. W. F. Valle}\email{valle@ific.uv.es}
\affiliation{\AddrAHEP}

\begin{abstract}
  
  The observation of coherent elastic neutrino nucleus scattering has opened the window to many physics opportunities.
  This process has been measured by the COHERENT Collaboration using two different targets, first CsI and then argon.
  Recently, the COHERENT Collaboration has updated the CsI data analysis with a higher statistics and an improved understanding of systematics.
  Here we perform a detailed statistical analysis of the full CsI data and combine it with the previous argon result.
  We discuss a vast array of implications, from tests of the Standard Model to new physics probes. 
  In our analyses we take into account experimental uncertainties associated to the efficiency as well as the timing distribution of neutrino fluxes,
  making our results rather robust.
  In particular, we update previous measurements of the weak mixing angle and the neutron root mean square charge radius for CsI and argon.
  We also update the constraints on new physics scenarios including neutrino nonstandard interactions and the most general case of neutrino generalized interactions,
  as well as the possibility of light mediators.
  Finally, constraints on neutrino electromagnetic properties are also examined, including the conversion to sterile neutrino states.
  In many cases, the inclusion of the recent CsI data leads to a dramatic improvement of bounds.

\end{abstract}

\maketitle

\section{Introduction}

The discovery of oscillations has brought neutrino physics to the precision era~\cite{Kajita:2016cak,McDonald:2016ixn}.
The field is currently thriving, experiments are growing in size and number, and new detection concepts are being proposed.  
Novel ways to probe fundamental parameters of the Standard Model (SM)
as well as new neutrino interactions beyond the SM are now under scrutiny.
In particular, the structure of the neutral current may reveal novel aspects of the physics associated to neutrino mass generation~\cite{Schechter:1980gr,Valle:1987gv}.
This suggests that studies using neutral current neutrino interactions can play an important complementary role in neutrino physics.
An example is provided by the new generation of neutrino experiments using coherent elastic neutrino-nucleus scattering (\cevns)~\cite{Abdullah:2022zue}.
Originally proposed by Freedman in the 1970's~\cite{Freedman:1973yd}, \cevns ~was finally detected using Spallation Neutron Source (SNS) neutrinos emerging from pion decay at rest ($\pi$-DAR)~\cite{COHERENT:2017ipa}.

So far the COHERENT Collaboration has observed this process at the SNS  using detectors made of CsI~\cite{COHERENT:2017ipa,COHERENT:2021xmm} and liquid argon (LAr)~\cite{COHERENT:2020iec}.
More recently, a suggestive evidence for CE$\nu$NS from reactor antineutrinos was reported by the Dresden-II Collaboration~\cite{Colaresi:2022obx}.
Moreover, several reactor-based CE$\nu$NS experiments aim at measuring this process, such as: CONNIE~\cite{CONNIE:2021ggh}, CONUS~\cite{CONUS:2021dwh}, $\nu$GEN~\cite{nuGeN:2022bmg},
MINER~\cite{MINER:2016igy}, RICOCHET~\cite{Billard:2016giu}, $\nu$-cleus~\cite{Strauss:2017cuu}, TEXONO~\cite{Wong:2015kgl}, vIOLETA~\cite{Fernandez-Moroni:2020yyl} and
Scintillating Bubble Chamber (SBC)~\cite{SBC:2021yal}.
There are also further experimental efforts underway, aiming to use $\pi$-DAR at the European Spallation Source~\cite{Baxter:2019mcx} and at the LANSCE Lujan Center~\cite{CCM:2021leg}.
Finally, the newly formed $\nu$BDX-DRIFT Collaboration using a directional time projection chamber aims to measure CE$\nu$NS using decay-in-flight neutrinos produced in the Long Baseline Neutrino Facility (LBNF) beamline~\cite{AristizabalSierra:2022jgg}.  

Here we analyze the updated data release from the CsI COHERENT experiment~\cite{COHERENT:2021xmm} and combine this result with the LAr COHERENT data~\cite{COHERENT:2020ybo}. 
In so doing, we perform a detailed analysis that includes all the relevant uncertainties for these experiments such as, for instance, the timing information as well as the neutrino-electron scattering (ES) signal for the CsI detector. 
This combined study of the recent CsI data with the LAr results provides a solid and updated analysis of COHERENT ~data,
from which  we can extract valuable information both on SM parameters, such as the weak mixing angle at low momentum transfer~\cite{Cadeddu:2021ijh,Majumdar:2022nby,AristizabalSierra:2022axl}
as well as nuclear physics features~\cite{Papoulias:2019lfi,Coloma:2020nhf}.
Moreover, these data can also be used to constrain new physics scenarios.
 These include neutrino nonstandard interactions (NSI)~\cite{Ohlsson:2012kf,Miranda:2015dra,Farzan:2017xzy,Denton:2020hop, Giunti:2019xpr}
  and neutrino generalized interactions (NGI)~\cite{Lee:1956qn,Lindner:2016wff,AristizabalSierra:2018eqm,Flores:2021kzl},
  light mediators~\cite{Abdullah:2018ykz,AristizabalSierra:2019ykk,Flores:2020lji,Cadeddu:2020nbr,Amaral:2021rzw,AtzoriCorona:2022moj},  CP-violating effects~\cite{AristizabalSierra:2019ufd},
   neutrino electromagnetic (EM) properties, e.g. transition magnetic moments~\cite{Schechter:1981hw,Pal:1981rm,Kayser:1982br,Nieves:1981zt,Shrock:1982sc,Kosmas:2015sqa,Canas:2015yoa,Miranda:2019wdy,Miranda:2020kwy},
  charge radius~\cite{Hirsch:2002uv,Bernabeu:2002pd}, or millicharge~\cite{Babu:1989tq}.
  We also examine \cevns ~sensitivities on conversions to sterile neutrinos, induced either by oscillations~\cite{Kosmas:2017zbh,Blanco:2019vyp,Miranda:2020syh}
  or by active-to-sterile transition magnetic moments~\cite{Miranda:2021kre,Bolton:2021pey,AristizabalSierra:2021fuc}.

  Some of the physics scenarios we discuss have been already constrained using COHERENT~data,
  see for example ~\cite{Cadeddu:2021ijh,Flores:2021kzl,AtzoriCorona:2022moj,AtzoriCorona:2022qrf} and~\cite{Coloma:2022avw,Khan:2022not,Khan:2022bcl}.
 However, in some of these references the new CsI dataset was not included or it was not combined with the liquid argon data.
 Note that we improve upon previous analyses by paying special attention to the experimental details of the COHERENT data.
    For example, in the CsI analysis we include the efficiency and timing information together with their uncertainties,
  all systematic uncertainties and the ES background which could mimic a \cevns~signal. In the LAr case, we also account for all systematic errors. 
  By means of such a careful statistical analysis, in this paper we have made an effort to address both SM precision tests as well as new physics scenarios in a comprehensive manner,
  presenting updated constraints. We show that the new CsI data dramatically improves the sensitivity, in some cases leading to stringent constraints, competitive to existing bounds
  from other observables. \\

This paper is organized as follows. In Sec.~\ref{sec:xsecs} we present all physics scenarios under scrutiny together with the relevant cross sections.
In Sec.~\ref{sec:dataanalysis} we discuss the statistical analysis procedure, highlighting all details, uncertainties and backgrounds.
We discuss our results in Sec.~\ref{sec:results} for each scenario previously introduced in Sec.~\ref{sec:xsecs}.
Finally, in Sec.~\ref{sec:conclusions} we conclude and present a table summarizing all the derived bounds.

\section{Cross sections for \cevns ~and electron scattering}
\label{sec:xsecs}

In this section, we provide the necessary cross sections for the case of \cevns ~and neutrino-electron scattering, for various physics scenarios within and beyond the SM.

\subsection{Standard physics}
\label{sec:SMxsec}

The differential \cevns ~cross section with respect to the nuclear recoil energy  $E_\mathrm{nr}$ is given by~\cite{Freedman:1973yd}
\begin{equation}
\label{eq:xsec_CEvNS_SM}
\frac{d\sigma_{\nu \mathcal{N}}}{dE_\mathrm{nr}}\Big|_\mathrm{CE\nu NS}^\mathrm{SM}=\frac{G_F^2 m_N}{\pi}F_{W}^2(|\vec{q}|^2)\left({Q_V^\mathrm{SM}}\right)^2\left(1-\frac{m_N E_\mathrm{nr}}{2E_\nu^2}\right) \, ,
\end{equation}
where $G_F$ is the Fermi's constant, $E_\nu$ denotes the incident neutrino energy, while $m_N$ is the nuclear mass.
Notice that Eq. (\ref{eq:xsec_CEvNS_SM}) applies for both neutrinos and antineutrinos.  In addition, within the SM, the CE$\nu$NS cross section is flavor independent at tree level,
with small loop corrections that are flavor dependent but have no significant impact for current experimental sensitivities \cite{Tomalak:2020zfh}.
The SM weak charge, $Q_V^\text{SM}$, takes the form 
\begin{equation}
\label{eq:CEvNS_SM_Qw}
    Q_V^\text{SM} = g_V^p Z + g_V^n N \, ,
\end{equation}
where the proton and neutron couplings are given by $g_V^p = 1/2 (1- 4 \sin^2 \theta_W)$  and $ g_V^n = -1/2$, respectively.
Notice that, although the proton contribution is small, it contains the dependence on the fundamental electroweak mixing angle.
Using RGE extrapolation in the minimal subtraction $ (\overline{\text{MS}})$ renormalization scheme, one finds that the weak mixing angle in the low-energy regime takes the value 
$\sin^2 \theta_W=0.23857(5)$~\cite{ParticleDataGroup:2020ssz}.
Nuclear physics effects in CE$\nu$NS are incorporated through the weak nuclear form factors for protons, $F_{p}$, and neutrons, $F_{n}$.
Here we assume the latter to be equal\footnote{This is a valid approximation since the proton coupling in Eq.~(\ref{eq:CEvNS_SM_Qw}) is tiny compared to the one of the neutron.} i.e. $F_{p} \simeq F_{n} \equiv F_{W}$, and we rely on the  Klein-Nystrand~\cite{Klein:1999qj} parametrization:  
\begin{equation}
  \label{eq:KNFF}
   F_{W}(|\vec{q}|^2)=3\frac{j_1(|\vec{q}|R_A)}{|\vec{q}|R_A} \left(\frac{1}{1+|\vec{q}|^2a_k^2} \right)\ ,
\end{equation}
where $j_1(x)=\sin(x)/x^2-\cos(x)/x$ is a spherical Bessel function of order one,
$|\vec{q}| = \frac{\sqrt{2  m_N  E_\mathrm{nr}}}{197.3}~\mathrm{fm^{-1}}$ is the magnitude of the three-momentum transfer, $a_k = 0.7$~fm and $R_A = 1.23 \, A^{1/3}$ represents the nuclear root mean square (RMS) radius (in units of fm), $A$ being the mass number. 
The value of the nuclear RMS radius $R_A$ can be obtained in the spirit of a nuclear structure model, see e.g.~\cite{Sahu:2020kwh,Hoferichter:2020osn}.
Experimentally, however, it is known only for the case of protons, while \cevns ~measurements are valuable probes of the yet unknown neutron RMS radius. 
Our determination of $R_A$ using  \cevns \, will be discussed in Sec.~\ref{sec:results}.

 It is well known that, for incoming neutrino energies in the range of a few tens of MeV, the CE$\nu$NS cross section is dominant when compared to other neutrino interactions.
  However, in some new physics scenarios, the ES process may be important and we will need to include the corresponding contribution to the total number of events. Within the SM,
the ES cross section on an atomic nucleus $\mathcal{A}$ containing $Z$ protons is given by
$\frac{d\sigma_{\nu_\ell \mathcal{A}}}{dE_\mathrm{er}} = Z_{\text{eff}}^{\mathcal{A}}(E_\mathrm{er})  \frac{d\sigma_{\nu_\ell}}{dE_\mathrm{er}}$.  Here $\frac{d\sigma_{\nu_\ell}}{dE_\mathrm{er}}$
represents the cross section of a neutrino scattering off a single electron and $Z_\text{eff}(E_\mathrm{er})$ is the effective number of protons seen by the neutrino for an energy deposition $E_\mathrm{er}$.
The effective charges for Cs and I isotopes are given in Table~\ref{tab:ZeffCsI}.
In the SM, the flavor-dependent differential ES cross section is given by 
\begin{align}
\label{eq:xsec_vES_SM}
\begin{split}
\frac{d\sigma_{\nu_\ell \mathcal{A}}}{dE_\mathrm{er}}\Big|_\mathrm{ES}^\mathrm{SM}=& Z^\mathcal{A}_\mathrm{eff}(E_\mathrm{er}) \frac{G_F^2m_e}{2\pi}\left[(g_V^{\nu_\ell} + g_A^{\nu_\ell})^2 +(g_V^{\nu_\ell} - g_A^{\nu_\ell})^2\left(1-\frac{E_\mathrm{er}}{E_\nu}\right)^2 -\left( (g_V^{\nu_\ell})^2-(g_A^{\nu_\ell})^2 \right)\frac{m_e E_\mathrm{er}}{E_\nu^2}\right] \, ,
\end{split}
\end{align}
where $\ell=\{e, \mu, \tau\}$ stands for the incoming neutrino flavor.
Here  $g_V^{\nu_\ell}=2 \sin^2 \theta_W + 1/2$ and $g_A^{\nu_\ell}=1/2$ for the case of $\nu_e-e^-$ scattering, which acquires contributions from both charged- and neutral-currents,
while  for the case of $\nu_{\mu, \tau}-e^-$ scattering, where only neutral-currents are relevant, one has $g_V^{\nu_\ell}=2 \sin^2 \theta_W - 1/2$ and $g_A^{\nu_\ell}=-1/2$.
Note that, for antineutrino-electron scattering, the substitution $g_A^{\nu_\ell} \to - g_A^{\nu_\ell}$ is required.

\begin{table}[t]
\centering
{\renewcommand{\arraystretch}{1.4}
\begin{tabular}{|c|c||c|c|}
\hline
\large $\mathbf{Z^\mathrm{Cs}_\mathrm{eff}}$  & \large  $\mathbf{E_{\mathrm{er}}}$  & \large  $\mathbf{Z^\mathrm{I}_\mathrm{eff}}$  & \large   $\mathbf{E_{\mathrm{er}}}$  \\
\hline 
55 & $E_{\mathrm{er}}>$ 35.99 keV & 53 & $E_{\mathrm{er}}>$ 33.17 keV  \\
53 & 35.99 keV\ $\geq\ E_{\mathrm{er}} >$ 5.71 keV &  51 & 33.17 keV\ $\geq\ E_{\mathrm{er}} >$ 5.19 keV  \\
51 & 5.71 keV\ $\geq\ E_{\mathrm{er}} >$ 5.36 keV  & 49 & 5.19 keV\ $\geq\ E_{\mathrm{er}} >$ 4.86 keV   \\
49 & 5.36 keV\ $\geq\ E_{\mathrm{er}} >$ 5.01 keV &
47 & 4.86 keV\ $\geq\ E_{\mathrm{er}} >$ 4.56 keV \\
45 & 5.01 keV\ $\geq\ E_{\mathrm{er}} >$ 1.21 keV  & 43 & 4.56 keV\ $\geq\ E_{\mathrm{er}} >$ 1.07 keV   \\
43 & 1.21 keV\ $\geq\ E_{\mathrm{er}} >$ 1.07 keV  & 41 & 1.07 keV\ $\geq\ E_{\mathrm{er}} >$ 0.93 keV   \\
41 & 1.07 keV\ $\geq\ E_{\mathrm{er}} >$ 1 keV & 39 & 0.93 keV\ $\geq\ E_{\mathrm{er}} >$ 0.88 keV   \\
37 & 1 keV\ $\geq\ E_{\mathrm{er}} >$ 0.74 keV   & 35 & 0.88 keV\ $\geq\ E_{\mathrm{er}} >$ 0.63 keV   \\
33 & 0.74 keV\ $\geq\ E_{\mathrm{er}} >$0.73 keV   & 31 & 0.63 keV\ $\geq\ E_{\mathrm{er}} >$0.62 keV   \\
27 & 0.73 keV\ $\geq\ E_{\mathrm{er}} >$0.23 keV   & 25 & 0.62 keV\ $\geq\ E_{\mathrm{er}} >$0.19 keV   \\
25 & 0.23 keV\ $\geq\ E_{\mathrm{er}} >$0.17 keV  & 23 & 0.19 keV\ $\geq\ E_{\mathrm{er}} >$0.124 keV  \\
23 & 0.17 keV\ $\geq\ E_{\mathrm{er}} >$0.16 keV  & 21 & 0.124 keV\ $\geq\ E_{\mathrm{er}} >$0.123 keV \\
19 & $E_{\mathrm{er}} <$ 0.16 keV  & 17 & $E_{\mathrm{er}} <$ 0.123 keV \\ 
\hline
\end{tabular}}
\caption{\centering{Effective electron charge for Cs and I as a function of the energy deposition $E_\mathrm{er}$}~\cite{Thompson:booklet}.}
\label{tab:ZeffCsI}
\end{table}


\subsection{Neutrino NSI }
\label{sec:NGIxsec}

Neutral current neutrino nonstandard interactions~\cite{Schechter:1980gr,Valle:1987gv} have attracted considerable attention in recent
years~\cite{Ohlsson:2012kf,Miranda:2015dra,Farzan:2017xzy,Giunti:2019xpr}, mainly because they constitute the ``side-show'' of neutrino mass generation schemes.
Indeed, they arise in a wide class of motivated scenarios beyond the SM~\cite{Boucenna:2014zba}.
Some would involve an extra $U(1)$ symmetry, that could lead to the existence of a new vector mediator.
Others might be associated to different types of interactions and mediators, both at a high~\cite{Lee:1956qn,Lindner:2016wff,AristizabalSierra:2018eqm}
as well as at a low-mass scale~\cite{Cerdeno:2016sfi,Bertuzzo:2017tuf,Farzan:2018gtr}.
If they exist, these mediators would contribute to \cevns~and ES processes leading to detectable distortions of the event rates, especially at low-energy recoils.
For completeness, here we will consider both light and heavy mediators.

In the presence of neutrino NSI associated to a heavy intermediate vector boson~\footnote{For a concrete example on neutral gauge bosons from strings-inspired models see~\cite{Miranda:2020zji}.},
  the neutral current Lagrangian contains~\cite{Ohlsson:2012kf,Miranda:2015dra,Farzan:2017xzy}
\begin{equation}
{\cal{L}}^\mathrm{NSI}_\mathrm{NC}=-2\sqrt{2}G_F \sum\limits_{q,\ell,\ell '} \varepsilon_{\ell\ell '}^{q X}(\bar{\nu}_\ell\gamma^\mu P_L\nu_{\ell '})(\bar{f}\gamma_\mu P_X f) \, ,
\end{equation}
where $q$ corresponds to quarks of the first family, $q=\{u,d\}$. The $\ell$ and $\ell '$ 
indices run over neutrino flavors (specifically for COHERENT: $e$ and $\mu$), $P_X$ ($X=L,R$) are the left and right chirality projectors, and $\varepsilon_{\ell\ell^{'}}^{q V}$
are the couplings that describe the relative strength of the NSI in terms of $G_F$. These couplings can be either flavor preserving (also called nonuniversal, $\ell = \ell '$)  or flavor changing $(\ell \neq \ell '$).
Once NSI are introduced, the SM weak charge of Eq.~(\ref{eq:CEvNS_SM_Qw}) becomes flavor dependent and is modified\footnote{As noted in Ref.~\cite{Hoferichter:2020osn}, in
    new physics scenarios the nuclear form factor can be modified. However, the effect is subdominant 
    in COHERENT, and it is effectively accounted for by the CE$\nu$NS normalization uncertainty in our fitting procedure (see Sec.~\ref{sec:dataanalysis}).}
as $Q_V^\mathrm{SM} \to Q_V^\text{NSI}$, with~\cite{Barranco:2005yy}
\begin{eqnarray}
Q_V^\text{NSI} &=& 
  \left [ \left ( g_{V}^{p}+2\varepsilon _{\ell \ell}^{uV}+\varepsilon _{\ell \ell}^{dV}  \right )Z +\left ( g_{V}^{n}+\varepsilon _{\ell \ell}^{uV}+2\varepsilon _{\ell \ell}^{dV}  \right ) N  \right ]  \nonumber \\
  &+&\sum_{\ell, \ell'}\left [ \left ( 2\varepsilon _{\ell \ell '}^{uV}+\varepsilon _{\ell \ell '}^{dV} \right ) Z + \left ( \varepsilon _{\ell \ell '}^{uV}+2\varepsilon _{\ell \ell '}^{dV} \right )N   \right ] .
\label{eq:qvNSI}
\end{eqnarray}

\subsection{Neutrino NGI}

Beyond the typical NSI interactions that could arise in gauge extensions of the SM, a more general framework can be considered to accommodate all Lorentz invariant interactions that
might lead to new physics, i.e. NGI with heavy mediators~\cite{Lee:1956qn,Lindner:2016wff,AristizabalSierra:2018eqm}.
The relevant Lagrangian reads~\cite{AristizabalSierra:2018eqm} 
\begin{equation}
\label{equn:NGI+nuclear_Lagrangian}
\mathscr{L}^\mathrm{NGI}_\mathrm{NC}   \sim
   \sum_{X=S,V,T}\bar\nu\,\Gamma_X\nu\,\bar N\,C_X\,\Gamma_X\,N
   +
   \sum_{\substack{(X,Y)=(P,S),\\\qquad\,\,\,(A,V)}}
   \bar\nu\,\Gamma_X\nu\,\bar N\,iD_X\,\Gamma_Y\,N \, ,
\end{equation}
with $\Gamma^X=\{\mathbb{I},i\gamma^5, \gamma^\mu,\gamma^\mu\gamma^5,\sigma^{\mu\nu}\}$, $\sigma^{\mu\nu}=i[\gamma^\mu,\gamma^\nu]/2$. The $C_X$, $D_X$ denote the corresponding neutrino-nucleus couplings.
Note that, for X=$V,S,T$, the couplings $C_X$ correspond to the weak charges given in Eqs.~(\ref{eq:Q_V}), (\ref{eq:QS}) and (\ref{eq:QT}).

The corresponding differential cross section associated to the Lagrangian density in Eq.~\eqref{equn:NGI+nuclear_Lagrangian} takes the form~\cite{AristizabalSierra:2018eqm}  
\begin{equation}
\begin{aligned}
\left.\frac{d\sigma}{dE_\text{nr}}\right|_\mathrm{CE\nu NS}^\mathrm{NGI}=\frac{G_F^2m_N}{4 \pi}F_{W}^2(|\vec{q}|^2)\Bigl\{&C_S^2 \frac{m_N E_\mathrm{nr}}{2E_\nu^2}+\left(C_V+ 2\, Q_V^\text{SM}\right)^2\left(1-\frac{m_N E_\text{nr}}{2E_\nu^2}-\frac{E_\mathrm{nr}}{E_\nu}\right)\\
&+ 8\, C_T^2 \left(1-\frac{m_N E_\text{nr}}{4E_\nu^2}-\frac{E_\text{nr}}{E_\nu}\right)\pm \mathcal{R}\frac{E_\text{nr}}{E_\nu} \Bigr\}\, .
\end{aligned}
\label{equn:NGI_xSec}
\end{equation}
Notice the interference between the SM and vector NGI, as well as the interference between the scalar and tensor terms given by $\mathcal{R} = 2 C_S C_T$, where the plus (minus)
sign accounts for coherent elastic antineutrino (neutrino) scattering off nuclei (see also Ref.~\cite{Flores:2021kzl}).
The weak charge associated to the new vector boson reads
\begin{equation}
    \label{eq:Q_V}
 C_V = g_{\nu V} \left[\left(2g_{uV}+g_{dV}\right)Z+\left(g_{uV}+ 2g_{dV} \right)N\right] \, , 
\end{equation}
with $g_{\nu V}$ and $g_{q V}$ denoting the new mediator couplings with neutrinos and quarks $q=\{u,d\}$. 
The  scalar and tensor weak charges are given by 
\begin{equation}
\label{eq:QS}
   C_S = g_{\nu S} \left(Z\sum_q g_{qS}\frac{m_p}{m_q}f^p_{q}+N\sum_q g_{qS}\frac{m_n}{m_q}f^n_{q}\right) 
\end{equation}
and 
\begin{equation}
    \label{eq:QT}
   C_T = g_{\nu T} \left(Z\sum_q g_{qT}\delta^p_{q}+N\sum_q g_{qT}\delta^n_{q}\right) \, , 
\end{equation}
respectively. The  hadronic structure parameters for the case of scalar interactions: $f_u^p=0.0208$, $f_u^n=0.0189$, $f_d^p=0.0411$,  $f_d^n=0.0451$ and tensor interactions: $\delta^p_u=\delta^n_d=0.54$, $\delta^p_d=\delta^n_u=-0.23$ are taken from Ref.~\cite{AristizabalSierra:2019zmy}. See also Ref.~\cite{Cirelli:2013ufw}.

\subsection{Light mediators}

It has been noticed that low-energy neutrino experiments are very sensitive to interactions involving light mediators~\cite{Cerdeno:2016sfi,Bertuzzo:2017tuf,Farzan:2018gtr, Denton:2022nol}.
In the simplest scenario with light vector-type (LV) interactions, the relevant \cevns ~cross section can be written as~\cite{Cerdeno:2016sfi} 
\begin{equation}
    \label{eq:xsec_CEvNS_LV}
\begin{aligned}
\frac{d\sigma_{\nu_\ell \mathcal{N}}}{dE_\mathrm{nr}}\Big|_\mathrm{CE\nu NS}^\mathrm{LV}=&
 \left(1+ \kappa \frac{C_V}{\sqrt{2}G_F Q_V^\mathrm{SM}\left(2 m_N E_\mathrm{nr}+m_{V}^2\right)}\right)^2 \frac{d\sigma_{\nu_\ell \mathcal{N}}}{dE_\mathrm{nr}}\Big|_\mathrm{CE\nu NS}^\mathrm{SM} \, .
\,
\end{aligned}
\end{equation}
For the case of ES, the corresponding cross section is given by Eq.~(\ref{eq:xsec_vES_SM}) with the following substitution
\begin{equation}
\label{equn:EveS_BSM_xsec_V,A}
g^{\nu_\ell}_{V}=g_{V}^{\nu_\ell}+\kappa \frac{g_{\nu V}\cdot g_{e V}}{2\sqrt{2}G_F(2m_e E_\mathrm{er} + m_{V}^2)} \,   ,
\end{equation}
where $g_{e V}$ denotes the new mediator coupling with the electrons.
Let us note that, for the case of \cevns, $\kappa=1$ for universal couplings (see e.g.~\cite{Langacker:2008yv,Billard:2018jnl} for reviews)
and $\kappa=-1/3$ in the $\mathrm{B-L}$ model~\cite{Langacker:2008yv,Okada:2018ktp} while, for ES, $\kappa=1$ for both the universal and $\mathrm{B-L}$ models~\footnote{Similar to the case of universal couplings, one may consider theoretically consistent anomaly-free U(1) extensions of the SM, such as those discussed in~\cite{AtzoriCorona:2022moj}.
A specially interesting example is the model proposed in~\cite{Bonilla:2017lsq,Allanach:2019mfl} as it is motivated by the explanation of the anomalies observed in B meson systems.
}. As a simple phenomenological reference, we will show results for the universal-coupling scenario, as any other case can be obtained by adequate coupling rescaling. 

Going one step further, one may also consider more general scalar and tensor interactions allowed by Lorentz invariance, and involving light mediators.
Such neutrino generalized interactions have been previously searched for using \cevns ~in Refs.~\cite{Majumdar:2021vdw,AristizabalSierra:2022axl}
and ES in Ref.~\cite{A:2022acy}. For the case of a new light scalar (LS) mediator, the \cevns ~cross section reads
\begin{align}
\label{eq:xsec_CEvNS_LS}
\begin{split}
\frac{d\sigma_{\nu_\ell \mathcal{N}}}{dE_\mathrm{nr}}\Big|_\mathrm{CE\nu NS}^\mathrm{LS}=& \frac{m_N^2 E_{\mathrm{nr}} \, C_S^2 }{4\pi E_\nu^2\left(2 m_N E_\mathrm{nr}+m_S^2\right)^2}  F_{W}^2(|\vec{q}|^2) 
\, ,
\end{split}
\end{align}
while the corresponding cross section for the case of a new light tensor (LT) mediator takes the form
\begin{align}
\label{eq:xsec_CEvNS_LT}
\begin{split}
\frac{d\sigma_{\nu_\ell \mathcal{N}}}{dE_\mathrm{nr}}\Big|_\mathrm{CE\nu NS}^\mathrm{LT}=& \frac{m_N (4E_\nu^2-m_N E_{\mathrm{nr}}) \, C_T^2}{2\pi E_\nu^2\left(2 m_N E_\mathrm{nr} +m_{T}^2\right)^2}  F_{W}^2(|\vec{q}|^2) 
\, .
\end{split}
\end{align}
 Finally, the cross sections for the case of scalar- and tensor-mediated ES processes read~\cite{Link:2019pbm}
\begin{equation}
\label{eq:xsec_vES_LS}
\frac{d\sigma_{\nu_\ell \mathcal{A}}}{dE_\mathrm{er}}\Big|_\mathrm{ES}^\mathrm{LS}= Z^\mathcal{A}_\mathrm{eff}(E_\mathrm{er}) \left[\frac{g^2_{\nu S}\cdot g^2_{eS}}{4\pi(2m_e E_{\mathrm{er}} + m_{S}^2)^2}\right]\frac{m_e^2 E_{\mathrm{er}}}{E_\nu^2}\, ,
\end{equation}
and
\begin{equation}
\label{eq:xsec_vES_LT}
\frac{d\sigma_{\nu_\ell \mathcal{A}}}{dE_\mathrm{er}}\Big|_\mathrm{ES}^\mathrm{LT}=Z^\mathcal{A}_\mathrm{eff}(E_\mathrm{er}) \frac{m_e\cdot g^2_{\nu T}\cdot g^2_{eT}}{2\pi(2m_e E_{\mathrm{er}} + m_{T}^2)^2}\cdot\left[1+2\left(1-\frac{E_{\mathrm{er}}}{E_\nu}\right)+\left(1-\frac{E_{\mathrm{er}}}{E_\nu}\right)^2-\frac{m_e E_{\mathrm{er}}}{E_\nu^2}\right]\, .
\end{equation}


\subsection{Neutrino electromagnetic properties} 
\label{sec:EMxsec}

Turning our attention to neutrino electromagnetic properties~\footnote{For a review see Ref.~\cite{Giunti:2014ixa}.}, our aim is to explore the associated phenomenological parameters using \cevns ~experiments.
These include the neutrino effective magnetic moment (MM), the neutrino electric charge (EC) and the neutrino charge radius (CR).
It should be noted that the former is given in terms of the fundamental transition magnetic moments (TMMs)~\cite{Schechter:1981hw}
and the corresponding effective parameter combinations relevant to each experimental setup are given in Refs.~\cite{Grimus:2000tq, Miranda:2019wdy,AristizabalSierra:2021fuc}.

\subsubsection{Effective neutrino magnetic moment}

Neutrino magnetic moment interactions flip chirality and do not interfere with the SM terms. Therefore, in the presence of a nonzero effective neutrino MM,
the differential cross section is incoherently added to the SM one and can be cast in the form~\cite{Vogel:1989iv} 
\begin{equation}
   \frac{d\sigma_{\nu_\ell \mathcal{N}}}{dE_\mathrm{nr}}\Big|_\mathrm{CE\nu NS}^\mathrm{MM}
=
\dfrac{ \pi \alpha^2_\mathrm{EM} }{ m_{e}^2 }
\left( \dfrac{1}{E_\mathrm{nr}} - \dfrac{1}{E_\nu} \right)
Z^2 F_{W}^2(|\vec{q}|^2)
\left| \dfrac{\mu_{\nu_{\ell}}}{\mu_{\text{B}}} \right|^2 \, ,
\label{eq:xsec_CEvNS_magmom}
\end{equation}
with $\alpha_\mathrm{EM}$ being the fine structure constant and $\mu_B$ the Bohr magneton. For the case of ES, the corresponding cross section is given in a similar form to Eq.~(\ref{eq:xsec_CEvNS_magmom}), as 
\begin{equation}
\frac{d\sigma_{\nu_\ell \mathcal{N}}}{dE_\mathrm{er}}\Big|_\mathrm{ES}^\mathrm{MM}
=
\dfrac{ \pi \alpha^2_\mathrm{EM} }{ m_{e}^2 }
\left( \dfrac{1}{E_\mathrm{er}} - \dfrac{1}{E_\nu} \right)
Z_\text{eff}^\mathcal{A}(E_\text{er})
\left| \dfrac{\mu_{\nu_{\ell}}}{\mu_{\text{B}}} \right|^2 \, . 
\label{eq:xsec_ES_magmom}
\end{equation}
At this point, we should stress that, while the neutrino magnetic moment is given as an effective parameter in Eqs.~(\ref{eq:xsec_CEvNS_magmom}) and (\ref{eq:xsec_ES_magmom}),
it should actually be expressed in terms of the fundamental TMM matrix~\footnote{For Majorana (Dirac) neutrinos $\lambda_{ij}$ is an antisymmetric (general complex) matrix~\cite{Canas:2015yoa}.}, $\lambda_{ij}$, as
\begin{equation}
 \mu^\text{eff}_{\nu_\ell} =\sum_k \left | \sum_j U^*_{\ell k} \lambda_{jk} \right |^2,
\end{equation}
where $U$ denotes the lepton mixing matrix. Relevant expressions for SNS-produced neutrinos are given in Refs.~\cite{Miranda:2019wdy, AristizabalSierra:2021fuc}.

\subsubsection{Neutrino charge radius}

The neutrino charge radius is the only EM neutrino parameter that is different from zero within the SM framework. 
Indeed, flavor-diagonal CR are generated via radiative corrections from the $\gamma-Z$ boson mixing and box diagrams involving $W$ and $Z$ bosons, leading to~\cite{Bernabeu:2002pd}
\begin{equation}
  \langle r^2_{\nu_{\ell \ell}} \rangle_\text{SM} = -\frac{G_F}{2 \sqrt{2} \pi^2} \left[ 3 - 2 \ln{\frac{m_\ell^2 }{M_W^2}}\right] \, , 
\end{equation}
where $m_\ell$ denotes the mass of the corresponding charged lepton and $M_W$ is the mass of $W$ boson.
Numerically, the SM values are $\left( \vev{ r^2_{\nu_{ee}} }, \vev{r^2_{\nu_{\mu \mu}}}, \vev{ r^2_{\nu_{\tau \tau}} } \right) = \left(-0.83,-0.48,-0.30\right) \times 10^{-32}~\mathrm{cm^2}$, close 
to the sensitivity reach of \cevns ~experiments. 
The contribution to the \cevns ~cross section due to flavor-diagonal charge radii is obtained through the substitution $g_V^p \to g_V^p - Q_{\ell \ell}^\text{CR}$ in Eq.~(\ref{eq:CEvNS_SM_Qw}) with 
\begin{equation}
  Q_{\ell \ell}^\text{CR} = \frac{\sqrt{2} \pi \alpha_\text{EM}}{3\, G_F} \langle r^2_{\nu_{\ell \ell}} \rangle\, .
\end{equation}
For the case of ES, the corresponding cross section is obtained via the substitution $g_V^{\nu_\ell} \to g_V^{\nu_\ell} + Q_{\ell \ell}^\text{CR} $.
Before closing this discussion, let us finally note that transition charge radii with $\ell \neq \ell^\prime$ can be generated via neutrino 
mixing~\cite{Kouzakov:2017hbc} and/or physics beyond the SM, as explained in Ref.~\cite{Cadeddu:2018dux}.

\subsubsection{Neutrino millicharge}

Another interesting EM neutrino property that can be probed at low-energy neutrino scattering experiments is the existence of a tiny neutrino EC (also referred to as the neutrino millicharge).
Such neutrino millicharges could arise in gauge models that include right-handed neutrinos~\cite{Babu:1989tq}. According to Ref.~\cite{AtzoriCorona:2022qrf}, 
the contribution of a nonvanishing EC to the differential \cevns ~cross section is given by Eq.~(\ref{eq:xsec_CEvNS_SM}) with the substitution $g_V^p \to g_V^p - Q_{\ell \ell}^\text{EC} $,
where the quantity $ Q_{\ell \ell}^\text{EC} $ is given as
\begin{equation}
 Q_{\ell \ell}^\text{EC}  = \frac{2 \sqrt{2} \pi \alpha_\text{EM}}{G_F q^2} q_{\nu_{\ell \ell} }\, ,
\end{equation}
with $q^2 = -2 m_N E_\text{nr}$ denoting the four-momentum transfer and $q_{\nu_{\ell \ell}}$ being the neutrino millicharge. 
Similarly, the EC contribution to ES is taken by substituting $g_V^{\nu_\ell} \to g_V^{\nu_\ell} + Q_{\ell \ell}^\text{EC} $ in Eq.~(\ref{eq:xsec_vES_SM}) where in this case the four-momentum transfer is $q^2= - 2 m_e E_\text{er}$
(see also Ref.~\cite{Kouzakov:2017hbc}). Interactions with flavor-nondiagonal EC are also possible and have been explored in Ref.~\cite{AtzoriCorona:2022qrf}.


\subsection{Conversion to sterile neutrinos} 
\label{sec:vsterile}

Given the distances between source and detector used in current \cevns~ experiments, active-to-sterile oscillations can develop.
Moreover, active-sterile neutrino TMMs may also lead to conversions to light sterile neutrinos.
We now examine conversion to sterile neutrinos, induced either by oscillations or transition magnetic moments.

\subsubsection{Active-to-sterile neutrino oscillations}
\label{sec:active-ster-neutr}

Because of its sensitivity to the total active neutrino flux, the COHERENT experiment can be exploited in order to search for sterile neutrinos.
  Here, we assume the simplest (3+1) scheme with three active neutrinos and one light sterile neutrino. The survival probabilities read  
 \begin{equation}
   P_{ee}(E_\nu)  \simeq 1 - \sin^2 2\theta_{1 4} \sin^2 \left( \frac{\Delta m^2_{41} L}{4E_\nu}\right)\, ,
   \label{eq:prob_nuest}
 \end{equation}
for electron neutrinos, and 
 \begin{equation}
 P_{\mu \mu}(E_\nu)  \simeq  1 - \sin^2 2\theta_{2 4} \sin^2 \left( \frac{\Delta m^2_{42} L}{4E_\nu}\right)\, , 
 \label{eq:prob_numust}
 \end{equation}
 for muon (anti-)neutrinos. In the previous expressions, $\theta_{14}$ and $\theta_{24}$ are the active-sterile mixing angles, $\Delta m^2_{41} \approx \Delta m^2_{42}$ are
 the active-sterile mass splittings, where we have assumed $\Delta m^2_{41} \gg |\Delta m^2_{31}|,   \Delta m^2_{21}$. 
 The presence of a light sterile neutral fermion is taken into account when computing the \cevns ~number of events by changing $Q_V^\mathrm{SM} \to Q_V^\text{SM}\times P_{\alpha \alpha}(E_\nu) $ in Eq.~\eqref{eq:xsec_CEvNS_SM}.

 \subsubsection{Active-to-sterile EM interactions}
 \label{sec:active-sterile-em}

 The presence of active-sterile neutrino TMMs leads to up-scattering processes of the form $\nu_L + \mathcal{N} \to \nu_4 + \mathcal{N}$ with production of a neutral massive outgoing fermion (sterile neutrino) with mass $m_4$.
 This process will contribute to the CE$\nu$NS cross section \cite{McKeen:2010rx} through  
\begin{equation}
  \label{eq:xsec_dipole_portal}
\begin{aligned}
 \frac{d\sigma_{\nu_\ell \mathcal{N}}}{dE_\mathrm{nr}}\Big|_\mathrm{CE\nu NS}^\mathrm{DP} = &
  \dfrac{ \pi \alpha^2_\mathrm{EM} }{ m_{e}^2 }\, Z^2 F_{W}^2(|\vec{q}|^2)
\left| \dfrac{\mu_{\nu_{\ell}}}{\mu_{\text{B}}} \right|^2 \\
 & \times \left[\frac{1}{E_\text{nr}} - \frac{1}{E_\nu} 
    - \frac{m_4^2}{2E_\nu E_\text{nr} m_N}
    \left(1- \frac{E_\text{nr}}{2E_\nu} + \frac{m_N}{2E_\nu}\right)
    + \frac{m_4^4(E_\text{nr}-m_N)}{8E_\nu^2 E_\text{nr}^2 m_N^2}
  \right]  \,  ,
  \end{aligned}
\end{equation}
where the mass of the produced sterile neutrino final state  is subject to the kinematic constraint 
\begin{equation}
  \label{eq:kinematic_constraint}
  m_4^2\lesssim 2m_N E_\text{nr}\left(\sqrt{\frac{2}{m_N E_\text{nr}}}E_\nu -1\right)\ .
\end{equation}
For ES scattering, the corresponding cross section is given by Eqs.~(\ref{eq:xsec_dipole_portal}) and (\ref{eq:kinematic_constraint}) and the following substitutions:
$E_\text{nr}\to E_\text{er}$, $m_N \to m_e$, $Z^2 F_W^2(|\vec{q}|^2) \to Z_\text{eff}^\mathcal{A}$.
Finally, the effective magnetic moment for active-sterile transitions is expressed in terms of the fundamental TMMs as given in Ref.~\cite{Miranda:2021kre}.

\section{Data analysis details} 
\label{sec:dataanalysis}


In this section we provide the necessary details of our analysis. 
Regarding the COHERENT-LAr data, we update the results of Refs.~\cite{Miranda:2020tif,Miranda:2021kre} by including all systematic errors on COHERENT-LAr data,
following the method used in Ref.~\cite{AtzoriCorona:2022moj} and the guidance provided in~\cite{COHERENT:2020ybo}. 
For the analysis of the recent COHERENT-CsI data, we update previous results~\cite{Papoulias:2017qdn,Miranda:2019wdy,Papoulias:2019txv,AristizabalSierra:2018eqm} in two ways.
First, in addition to \cevns, we also include possible contributions from ES events, as in~\cite{Coloma:2022avw, AtzoriCorona:2022qrf}.
 This contribution can get significantly enhanced in some new physics scenarios, e.g. light mediators or millicharges, and dangerously mimic the \cevns~signal.
 Note also that we add the ES contribution only in the CsI analysis since in the LAr case the measurement of the ratio of the integrated photomultiplier amplitude and the total amplitude
 in the first 90~ns (called $F_{90}$) allows a clear separation between \cevns-induced nuclear recoils and ES-induced electron recoils.
Moreover, following closely the analysis method of the COHERENT Collaboration~\cite{COHERENT:2021xmm}, we perform a new comprehensive analysis including further nuisance parameters which account for signal shape uncertainties.
Finally, we also perform a combined analysis of the full CsI+LAr data. Whenever available, we will compare our results to existing constraints in the
literature~\cite{Cadeddu:2021ijh,Flores:2021kzl,Coloma:2022avw,Khan:2022not,AtzoriCorona:2022moj,Khan:2022bcl,AtzoriCorona:2022qrf}.

\subsection{CsI number of events}
\label{sec:csi-number-events}

For the case of the COHERENT-CsI detector, the predicted number of  \cevns ~events per neutrino flavor,  $n$, in each nuclear recoil energy bin $i$, is 
\begin{align}\label{eq:Nevents_CEvNS}
N_{i, n}^{\mathrm{CE}\nu\mathrm{NS}, \mathcal{N} }
&= \nonumber
N_\mathrm{target}
\int_{E_{\mathrm{nr}}^{i}}^{E_{\mathrm{nr}}^{i+1}}
\hspace{-0.3cm}
d E_{\mathrm{nr}}\,
\epsilon_E(E_{\mathrm{nr}})
\int_{0}^{E^{\prime\text{max}}_{\text{nr}}}
\hspace{-0.3cm}
dE'_{\text{nr}}
\,
P(E_{\text{nr}},E'_{\text{nr}}) \times \\
& \int_{E_\nu^{\text{min}}(E'_{\text{nr}})}^{E_\nu^{\text{max}}}
\hspace{-0.3cm}
d E_\nu
\frac{d N_{n}}{d E_\nu}(E_\nu)
\frac{d\sigma_{\nu_\ell \mathcal{N}}}{dE'_\mathrm{nr}}\Big|_\mathrm{CE\nu NS}(E_\nu, E'_{\mathrm{nr}})
,
\end{align}
where $\mathcal{N}=$ Cs or I,  and $N_\mathrm{target} = N_{\mathrm{A}} m_{\mathrm{det}} / M_{\mathrm{\mathrm{target}}}$ is the number of target atoms in the detector. 
Here,  $m_{\rm det} = 14.6$ kg denotes the CsI detector mass, while $N_{\mathrm{A}}$ is the Avogadro number and $M_{\mathrm{\mathrm{target}}}$ is the molar mass.
Moreover, $E_{\text{nr}}$ is the reconstructed nuclear recoil energy, $E'_{\text{nr}}$ is the true nuclear recoil energy,
$P(E_{\text{nr}},E'_{\text{nr}})$ is the energy resolution function,
$E^{\prime\text{max}}_{\text{nr}} \simeq \frac{2  (E_\nu^{\text{max}})^2}{m_N}$, 
$E_\nu^{\text{max}} = m_\mu/2 \sim 52.8$~MeV and
$E_\nu^{\text{min}}(E'_{\text{nr}}) \simeq \sqrt{\frac{m_NE'_\text{nr}}{2}}$.\footnote{Note that, for the case of TMM-induced conversion to a sterile neutrino the integration limits depend also on $m_4$, according to Eq.~(\ref{eq:kinematic_constraint}).}
The SNS neutrino flux consists of a prompt and a delayed beam. The three components to the total differential neutrino flux, $n$,  produced by pions decaying at rest are
\begin{equation}
\begin{aligned} 
\frac{d N_{\nu_\mu}}{d E_\nu}(E_\nu) & = \eta \, \delta\left(E_\nu-\frac{m_{\pi}^{2}-m_{\mu}^{2}}{2 m_{\pi}}\right) \quad &(\text{prompt})\, , \\ 
\frac{d N_{\bar{\nu}_\mu}}{d E_\nu}(E_\nu) & = \eta \frac{64 E^{2}_\nu}{m_{\mu}^{3}}\left(\frac{3}{4}-\frac{E_\nu}{m_{\mu}}\right) \quad &(\text{delayed})\, ,\\ 
\frac{d N_{\nu_e}}{d E_\nu}(E_\nu) & = \eta \frac{192 E^{2}_\nu}{m_{\mu}^{3}}\left(\frac{1}{2}-\frac{E_\nu}{m_{\mu}}\right) \quad &(\text{delayed}) \, ,
\end{aligned}
\label{labor-nu}
\end{equation}
which are normalized to $\eta = r N_{\mathrm{POT}}/4 \pi L^2$, where $L=19.3$~m is the detector distance from the SNS source, and $r=0.0848$ denotes the number of neutrinos per flavor produced for each proton on target (POT),
where $N_{\mathrm{POT}}=3.198 \times 10^{23}$. The energy-dependent detector efficiency is given by 
\begin{equation}
\label{eq:CsI_E_efficiency}
\epsilon_E(x) = \frac{a}{1+e^{-b(x-c)}}+d \, , 
\end{equation}
where $x = \mathrm{PE} + \alpha_7$ 
and $a = 1.32045$, $b = 0.285979$, $c = 10.8646$, $d = -0.333322$~\cite{COHERENT:2021xmm}.
We account for the $1\sigma$ uncertainty on the efficiency curve through the parameter $\alpha_7$, which can vary between $[-1,+1]\times$PE (PE being the number of photoelectrons), see Appendix~\ref{sec:appendix}.
Quenched recoils are given through the light yield $\text{LY} = 13.35~\mathrm{PE/keV_{ee}}$, with PE$= \mathrm{LY}\times E_\mathrm{er}$, where the electron-equivalent energy is given in terms of $E_{\mathrm{nr}}$ as
$E_\mathrm{er}= x_1 E_{\mathrm{nr}}^{\prime} + x_2 E_{\mathrm{nr}}^{\prime 2} +x_3 E_{\mathrm{nr}}^{\prime 3} + x_4 E_{\mathrm{nr}}^{\prime 4}$ ($x_1 = 0.0554628,~x_2 = 4.30681,~x_3 = -111.707,~x_4 = 840.384$)~\cite{COHERENT:2021xmm}. 
Finally, smearing is applied through the Gamma function 
\begin{equation}
P(E_{\text{nr}},E'_{\text{nr}}) =  \frac{(\mathfrak{a} (1 + \mathfrak{b}))^{1 + \mathfrak{b}}}{\Gamma(1 + \mathfrak{b})} \cdot x^\mathfrak{b} \cdot e^{-\mathfrak{a}(1 +\mathfrak{b})x},
\end{equation}
where $x$ is the reconstructed recoil energy in PE, PE$(E_{\mathrm{nr}})$, while $\mathfrak{a}$ and $\mathfrak{b}$ instead depend on the true quenched energy deposition:
 $\mathfrak{a} = 0.0749/E_\mathrm{er}(E'_{\text{nr}})$, $\mathfrak{b} = 9.56 \times E_\mathrm{er}(E'_{\text{nr}})$~\cite{COHERENT:2021xmm}.
Moreover, we include the timing information in our analysis, by distributing the predicted $N_{i}^{\mathrm{CE}\nu\mathrm{NS}, \mathcal{N}}$ in each time bin $j$.  
We take  the time distributions $f^n_T(t_{\mathrm{rec}})$ of $n = \nu_{e}, \nu_{\mu}, \bar{\nu}_{\mu}$ from~\cite{Picciau:2022xzi,COHERENT:2021xmm} and we normalize them to 6 $\mathrm{\mu s}$.
Finally, the predicted \cevns ~event number, per observed nuclear recoil energy and time bins $i, j$ is  
\begin{equation}
\label{eq:Nevents_CEvNS_ij}
 N_{ij}^{\mathrm{CE}\nu\mathrm{NS}, \mathcal{N} } = \sum_{n=\nu_{e}, \nu_{\mu}, \bar{\nu}_{\mu}}\int_{t_{\mathrm{rec}}^{j}}^{t_{\mathrm{rec}}^{j+1}} dt_{\mathrm{rec}} f^n_T(t_{\mathrm{rec}}, \alpha_6)\epsilon_T(t_{\mathrm{rec}}) N_{i,  n}^{\mathrm{CE}\nu\mathrm{NS}, \mathcal{N} }.
\end{equation}
We also include an additional nuisance parameter on beam timing, $\alpha_6$, thus allowing the \cevns ~timing distribution to vary in a $\pm 250$ ns
range~\footnote{Private communication with D. Pershey and~\cite{COHERENT:2021xmm}.}. Finally, 
\begin{equation}
\label{eq:CsI_T_efficiency}
  \epsilon_T(t_{\mathrm{rec}}) = \begin{cases}
    1, & \text{if}\ t_{\mathrm{rec}}< t^\prime \\
    e^{-k(t_{\mathrm{rec}}-t^\prime)}, & \text{if}~ t_{\mathrm{rec}}\geq t^\prime 
    \end{cases}
\end{equation}
is a time-dependent efficiency with $t^\prime = 0.52~\mathrm{\mu s}$ and $k = 0.0494\mathrm{/\mu s}$. The total number of predicted \cevns ~events is in the end given by $N_{ij}^{\mathrm{CE}\nu\mathrm{NS}} = N_{ij}^{\mathrm{CE}\nu\mathrm{NS}, \mathrm{Cs}}+N_{ij}^{\mathrm{CE}\nu\mathrm{NS}, \mathrm{I}}$, where the indices $i$ and $j$ run over the 9 PE-bins and 11 time-bins, as reported in~\cite{COHERENT:2021xmm}.

The predicted ES event number, $N^\mathrm{ES}_{i,n}$,  per neutrino flavor, $n$,  in each observed electron recoil energy bin $i$ is
\begin{align}\label{eq:Nevents_ES}
N_{i,n}^{\mathrm{ES}, \mathcal{A}}
&= \nonumber
N_\mathrm{target}
\int_{E_{\mathrm{er}}^{i}}^{E_{\mathrm{er}}^{i+1}}
\hspace{-0.3cm}
d E_{\mathrm{er}}\,
\epsilon_E(E_{\mathrm{er}})
\int_{0}^{E^{\prime\text{max}}_{\text{er}}}
\hspace{-0.3cm}
dE'_{\text{er}}
\,
P(E_{\text{er}},E'_{\text{er}}) \times \\
& \int_{E_\nu^{\text{min}}(E'_{\text{er}})}^{E_\nu^{\text{max}}}
\hspace{-0.3cm}
d E_\nu
\frac{d N_{n}}{d E_\nu}(E_\nu)
\frac{d\sigma_{\nu_\ell \mathcal{A}}}{dE'_\mathrm{er}}\Big|_\mathrm{ES}(E_\nu, E'_{\mathrm{er}})
,
\end{align}
where $\mathcal{A}$ refers to the atom (Cs or I), $E_\nu^{\text{min}}(E'_{\text{er}}) = (E'_\text{er}+\sqrt{E_\text{er}^{'2} + 2m_e E'_\text{er}})/2$, and $E^{\prime\text{max}}_{\text{er}} = 2 (E_\nu^{\text{max}})^2 / (2E_\nu^{\text{max}}+m_e)$.
We also include the timing information in our analysis, by distributing the predicted $N_{i}^{\mathrm{ES}, \mathcal{A}}$ in  $j$ time bins.
Similarly to the \cevns ~case, the predicted ES event number per observed nuclear recoil energy and time bins $i, j$, is 
\begin{equation}
\label{eq:Nevents_ES_ij}
 N_{ij}^{\mathrm{ES}, \mathcal{A} } = \sum_{n=\nu_{e}, \nu_{\mu}, \bar{\nu}_{\mu}}\int_{t_{\mathrm{rec}}^{j}}^{t_{\mathrm{rec}}^{j+1}} dt_{\mathrm{rec}} f^n_T(t_{\mathrm{rec}}, \alpha_6)\epsilon_T(t_{\mathrm{rec}}) N_{i,n}^{\mathrm{ES}, \mathcal{A} }.
\end{equation}
The total number of predicted ES events is finally given by $N_{ij}^\mathrm{ES} = N_{ij}^{\mathrm{ES}, \mathrm{Cs}}+N_{ij}^{\mathrm{ES}, \mathrm{I}}$.   
 
\subsection{LAr number of events}
\label{sec:lar-number-events}

For the analysis of COHERENT-LAr data, the detector mass is $m_\text{det}=24$~kg and it is located at a distance $L=27.5$~m from the SNS source. 
In this case, the \cevns ~event rate is given by Eq.~(\ref{eq:Nevents_CEvNS}), with $r=0.09$ and $N_\mathrm{POT}=1.38 \times10^{23}$.
The corresponding efficiency function is taken from the data release~\cite{COHERENT:2020ybo}, while the conversion between nuclear and electron recoil energy is given through the quenching factor
$\mathrm{QF}(E_\mathrm{nr}^\prime) = 0.246 + 7.8 \times 10^{-4} E_\mathrm{nr}^\prime (\mathrm{keV_{nr}})$~\cite{COHERENT:2020iec}.
The reconstructed event rate is obtained by smearing the true event spectrum with a normalized Gaussian function $P(E_\mathrm{er},E_\mathrm{er}^\prime)$,
whose energy resolution is given by $\frac{\sigma_{E_\mathrm{er}}}{ E_\mathrm{er}}= \frac{0.58}{\sqrt{E_\mathrm{er}(\mathrm{keV_{ee}})}}$~\cite{COHERENT:2020ybo}.
For the COHERENT-LAr detector there is no reported time efficiency, hence in Eq.~(\ref{eq:Nevents_CEvNS_ij}) we consider $\epsilon_T(t_\text{rec})=1$, while the time
distributions~\footnote{We do not introduce any nuisance parameter in this case.}
$f_T^n(t_\mathrm{rec})$ are taken from~\cite{COHERENT:2020ybo}.
In our analysis, the index $i$ runs over the 12 $E_\mathrm{er}$-bins while the index $j$ runs over the 10 time-bins~\cite{COHERENT:2020iec}. 
Finally, as noted in Ref.~\cite{AtzoriCorona:2022qrf}, we do not include ES events in the analysis of COHERENT-LAr data, since the $F_{90}$ data provided by the COHERENT Collaboration already ensure a successful
discrimination of CE$\nu$NS- versus ES-induced signals in the detector.

\subsection{Statistical analysis}
\label{sec:statistical-analysis}

In order to perform the analysis of the COHERENT-CsI data we use the following Poissonian least-squares function 
\begin{equation}
	\chi^2_{\mathrm{CsI}}\Big|_\mathrm{CE \nu NS (+ ES)}
	 =
	2
	\sum_{i=1}^{9}
	\sum_{j=1}^{11}
	\left[ N_\mathrm{th}^\mathrm{CsI}  -  N_{ij}^{\text{exp}} 
	 +  N_{ij}^{\text{exp}} \ln\left(\frac{N_{ij}^{\text{exp}}}{ N_\mathrm{th}^\mathrm{CsI}} \right)\right]\\
	+ \sum_{k=0}^{4 (5)}
	\left(
	\dfrac{ \alpha_{k} }{ \sigma_{k} }
	\right)^2  
	.
	\label{chi2CsI}
\end{equation}
In the analyses where ES events are relevant, the predicted number of events is defined as 
\begin{align}\nonumber
N_\mathrm{th}^\mathrm{CsI, CE\nu NS + ES} &= (1 + \alpha_{0} +\alpha_{5})  N_{ij}^{\mathrm{CE\nu NS}} (\alpha_{4}, \alpha_{6}, \alpha_{7}) + (1 + \alpha_{0} )  N_{ij}^{\mathrm{ES}} (\alpha_{6}, \alpha_{7})\\
 &+ (1 + \alpha_{1}) N_{ij}^\mathrm{BRN}(\alpha_{6}) + (1 + \alpha_{2}) N_{ij}^\mathrm{NIN}(\alpha_{6})
 + (1 + \alpha_{3}) N_{ij}^\mathrm{SSB}  \,.
	\label{eq:Nth_CsI_chi2}
\end{align}
In the previous expressions, $\sigma_{0} = 11\%$  encodes efficiency and flux uncertainties, $\sigma_{1} = 25\%$ is associated to Beam Related Neutrons (BRN),
$\sigma_{2} = 35\%$ corresponds to Neutrino Induced Neutrons (NIN), $\sigma_{3} = 2.1\%$ to Steady State Background (SSB) and
$\sigma_{5} = 3.8\%$
is the systematic uncertainty associated to the quenching factor (QF). 
Moreover, $N_{ij}^{\mathrm{CE\nu NS}}$  and $N_{ij}^{\mathrm{ES}}$ include three further nuisance parameters: $\alpha_{4}$, which enters the nuclear form factor and thus affects only the \cevns ~number of events.
Note that the expression of $R_A$ that appears in Eq.~(\ref{eq:KNFF}) should be modified to $R_A = 1.23 \, A^{1/3} (1+\alpha_4)$, with  $\sigma_{4} = 5\%$;
$\alpha_{6}$ accommodates the uncertainty in beam timing with no prior assigned, see Eq.~(\ref{eq:Nevents_CEvNS_ij}),
while $\alpha_{7}$ allows for deviations of the uncertainty in the \cevns ~efficiency, see Eq.~(\ref{eq:CsI_E_efficiency}).
We also note that, according to Ref.~\cite{COHERENT:2021xmm}, the efficiency function is already applied in the BRN and NIN backgrounds provided in the data release, hence in our calculations
the latter are not weighted by the nuisance parameter $\alpha_{7}$.  

In scenarios where ES is not playing a relevant role, we instead rely on a simplified least-squares function, where the predicted number of events is defined as
\begin{align}\nonumber
N_\mathrm{th}^\mathrm{CsI, CE\nu NS} &= (1 + \alpha_{0} ) N_{ij}^{\mathrm{CE\nu NS}} (\alpha_{4}, \alpha_{6}, \alpha_{7})
 + (1 + \alpha_{1}) N_{ij}^\mathrm{BRN}(\alpha_{6}) + (1 + \alpha_{2}) N_{ij}^\mathrm{NIN}(\alpha_{6}) \\ 
 & + (1 + \alpha_{3}) N_{ij}^\mathrm{SSB} \,. 
	\label{eq:Nth_CsI_chi2_noES}
\end{align}
In this case, the relevant systematic uncertainties are $\sigma_{0} = 11.45\%$ (which encodes efficiency, flux and QF uncertainties), $\sigma_{1} = 25\%$ (BRN), $\sigma_{2} = 35\%$ (NIN),
$\sigma_{3} = 2.1\%$ (SSB), $\sigma_{4} = 5\%$ ($R_A$). 

In contrast, for the analysis of the COHERENT-LAr dataset we consider the following $\chi^2$ function~\cite{AtzoriCorona:2022moj}
\begin{equation}
    \begin{aligned}
	\chi^2_{\mathrm{LAr}} = 
	\sum_{i=1}^{12}
	\sum_{j=1}^{10}
	\frac{1}{\sigma_{ij}^2}\Bigl[ & (1 + \beta_0 +  \beta_1 \Delta_\mathrm{CE\nu NS}^{F_{90+}} + \beta_1 \Delta_\mathrm{CE\nu NS}^{F_{90-}} + \beta_2 \Delta_\mathrm{CE\nu NS}^\mathrm{t_{trig}}) N_{ij}^{\mathrm{CE\nu NS}}  \\
    & + (1 + \beta_3) N_{ij}^\mathrm{SSB}  \\
    & + (1 + \beta_4 + \beta_5 \Delta_\mathrm{pBRN}^{E_+} + \beta_5 \Delta_\mathrm{pBRN}^{E_-}
    + \beta_6 \Delta_\mathrm{pBRN}^{t_\text{trig}^+} + \beta_6 \Delta_\mathrm{pBRN}^{t_\text{trig}^-} + \beta_7 \Delta_\mathrm{pBRN}^{t_\text{trig}^\text{w}}) N_{ij}^\mathrm{pBRN}\\
    & + (1 + \beta_8) N_{ij}^\mathrm{dBRN}  -  N_{ij}^{\text{exp}} \Bigr]^2  \\ 
     &+ \sum_{k=0,3,4,8}
	\left( 	\dfrac{ \beta_{k} }{ \sigma_{k} } 	\right)^2  +  \sum_{k=1,2,5,6,7} \left(\beta_{k}  	\right)^2 \,  ,
\end{aligned}
\label{chi2LAr}
\end{equation}
where $\sigma_{ij}^2 = N_{ij}^\text{exp} +  N_{ij}^\mathrm{SSB}/5$.  
The nuisance parameters $\beta_0,~\beta_3,~\beta_4$ and $\beta_8$ are introduced to account for the normalization of CE$\nu$NS~\footnote{\label{foot:uncert}This component includes the uncertainties on the flux (10\%), efficiency (3.6\%), energy calibration (0.8\%), the calibration of the pulse-shape discrimination parameter $F_{90}$  (7.8\%), QF (1\%), and nuclear form factor (2\%)~\cite{COHERENT:2020iec}.}, SS, prompt BRN and delayed BRN, respectively, with the corresponding uncertainties being $\{\sigma_{0},~\sigma_{3},~\sigma_{4},~\sigma_{8}\}$=$\{0.13,~0.0079,~0.32,~1.0\}$~\cite{COHERENT:2020iec}. 
The shape uncertainties are taken into account by introducing the nuisance parameters $\beta_1, \beta_2, \beta_5, \beta_6$ and $\beta_7$.
The first two parameters modify the shape of the CE$\nu$NS prediction, while the last three affect the shape of the prompt BRN background. 
In particular, for the case of CE$\nu$NS, the relevant sources of systematic uncertainty
 are the $\pm 1\sigma$ energy distributions of the $F_{90}$ parameter, given by $\Delta_\mathrm{CE\nu NS}^{F_{90+}}$ and $\Delta_\mathrm{CE\nu NS}^{F_{90-}}$, and
 the mean time to trigger distribution, $\Delta_\mathrm{CE\nu NS}^\mathrm{t_{trig}}$.
Similarly, for the shape of the prompt BRN background, the relevant distributions are the $\pm 1\sigma$ energy distributions ($\Delta_\mathrm{pBRN}^{E_+}$ and  $\Delta_\mathrm{pBRN}^{E_-}$), the $\pm 1 \sigma$ mean time to trigger  distributions ($\Delta_\mathrm{pBRN}^{t_\text{trig}^+}$ and $\Delta_\mathrm{pBRN}^{t_\text{trig}^-}$) and the 
trigger width distribution ($\Delta_\mathrm{pBRN}^{t_\text{trig}^\text{w}}$). 
These distributions, introduced in Eq.~(\ref{chi2LAr}), are defined as
\begin{equation}
    \Delta_{\lambda}^{\xi_\lambda} = \frac{N_{ij}^{\lambda,\xi_\lambda} - N_{ij}^{\lambda,\mathrm{CV}}}{N_{ij}^{\lambda,\mathrm{CV}}} \, ,
\end{equation}
where $\lambda=$ \{CE$\nu$NS, pBRN\}, $\xi_\lambda$ is any of the sources relevant to the given $\lambda$ as described above, while CV denotes the central values of the \cevns ~or
prompt BRN distributions, all taken from the COHERENT-LAr data release~\cite{COHERENT:2020ybo}. 

\section{Results}
\label{sec:results}

In this section, we discuss the results that we have obtained from the combined analysis of the COHERENT CsI~\cite{COHERENT:2021xmm} and LAr~\cite{COHERENT:2020iec} data. 
We will start with  standard electroweak and nuclear physics, presenting the determination of the weak mixing angle and neutron RMS radius, 
then proceed to scenarios beyond the SM, including neutrino NSI as well as NGI with both heavy and light mediators.
Next, we discuss neutrino EM properties, including the effective neutrino magnetic moments, charge radius and neutrino millicharge. 
Finally, we discuss conversions to sterile neutrinos, both active-to-sterile neutrino oscillations as well as active-to-sterile EM interactions. 

In what follows, we include the contribution from ES events whenever it becomes relevant, i.e. for the neutrino millicharge, neutrino magnetic moment,
TMM to sterile neutrinos, and the light mediator analyses. 
We will see that the most important effect of the presence of ES events is found for the neutrino millicharge and the light vector and tensor mediator analyses. 
This is due to the fact that their corresponding cross sections are proportional to $\sim 1/E_\text{er}^2$, which enhances the event rates significantly. 
For the neutrino magnetic moment conversions and the light scalar mediator analyses, the cross sections are proportional to $~\sim 1/E_\text{er}$, yielding to a moderate enhancement of the event rates. 
Notice that for the cases of the weak mixing angle, neutrino charge radius and heavy NSI and NGI analyses, the ES event rate is too small and hence neglected.

\subsection{Standard physics} 
\label{sec:electr-nucl-phys}

\begin{figure}[t]
\includegraphics[width= 0.4 \textwidth]{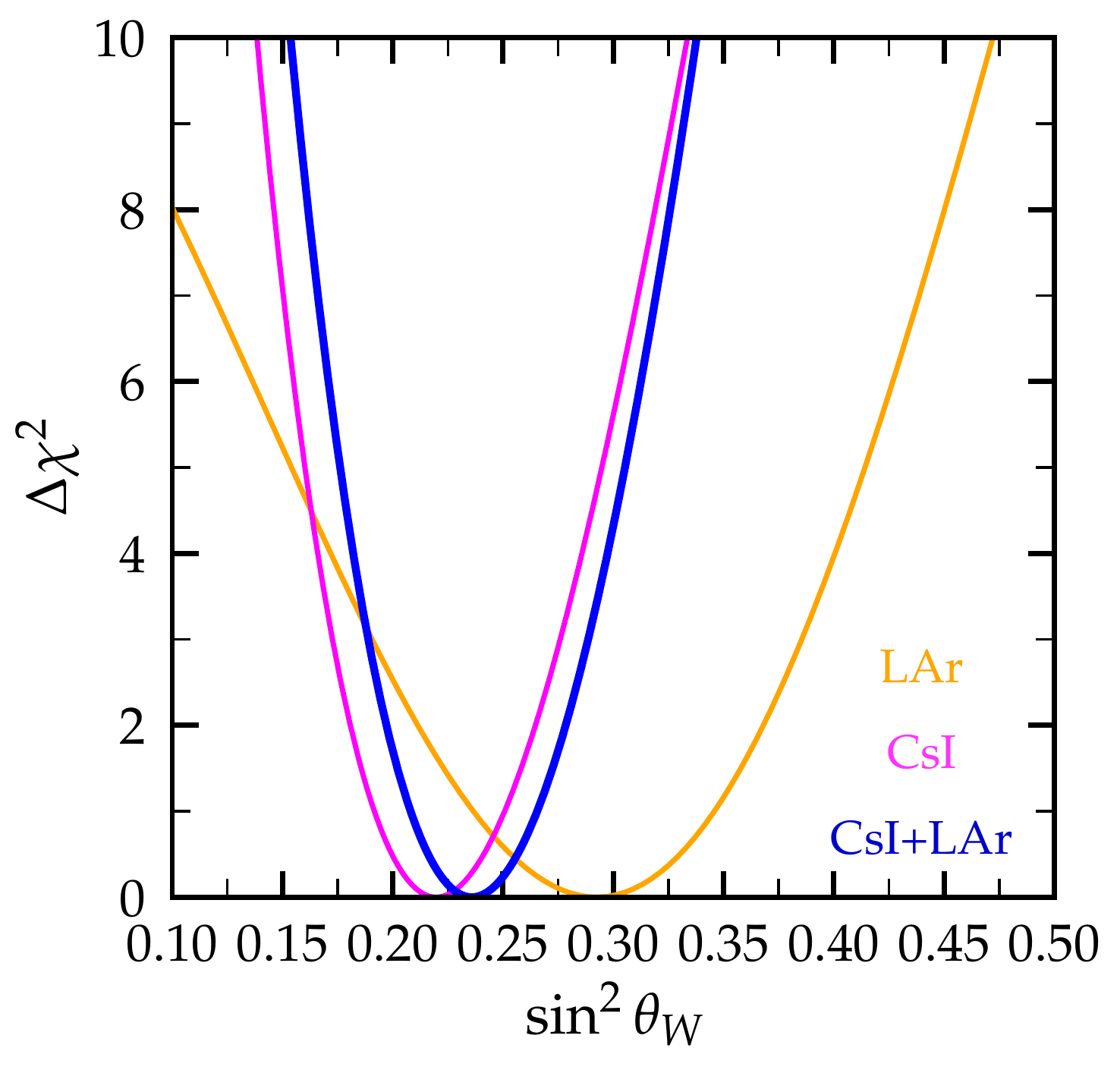}
\includegraphics[width= 0.5 \textwidth]{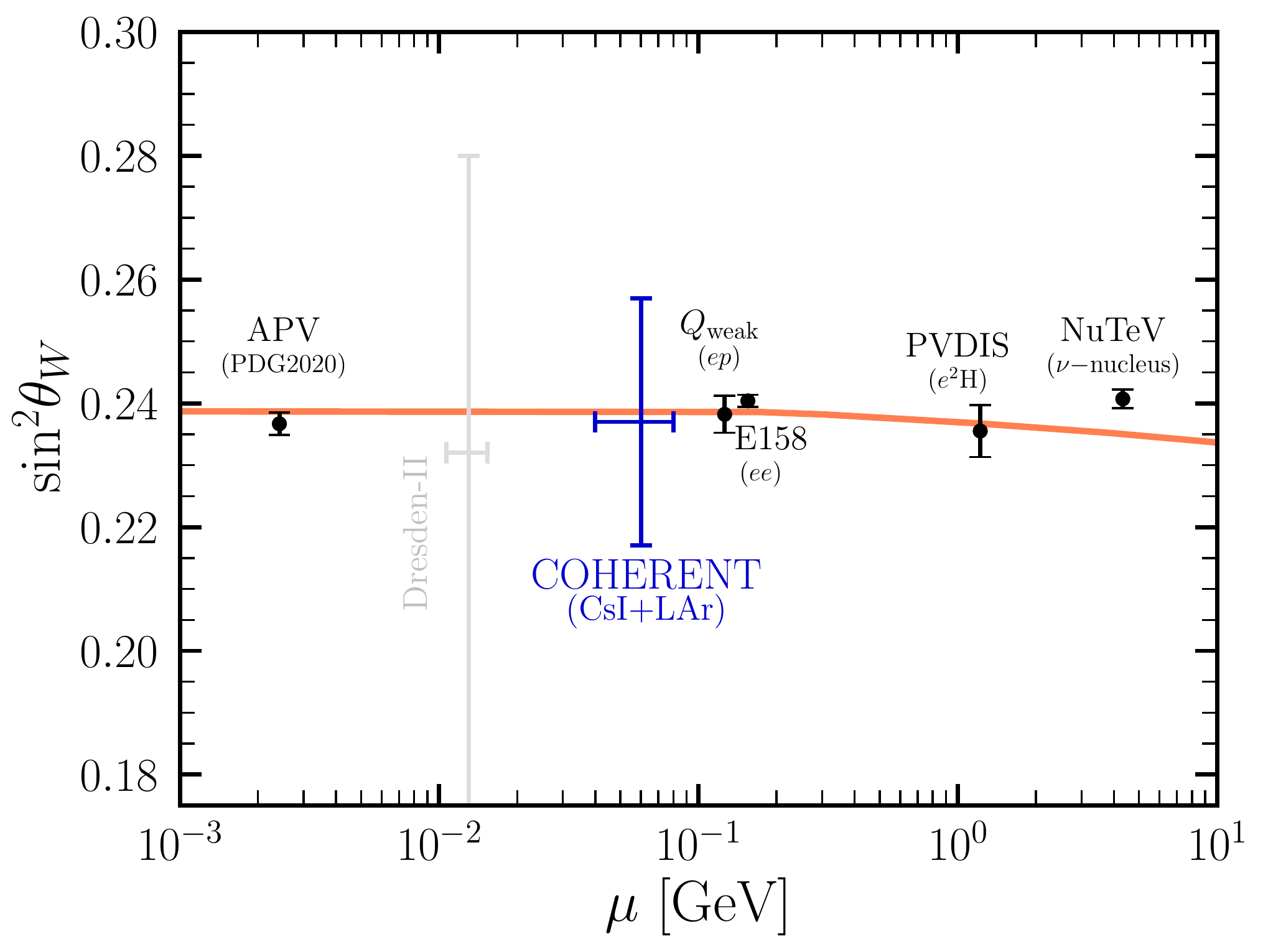}
\caption{ Left panel: $\Delta\chi^2$ profile
for
$\sin^2 \theta_{\text{W}}$
obtained from the analysis of CsI (magenta), LAr (orange) and CsI + LAr (blue) data. Right panel: $\sin^2 \theta_{\text{W}}$ RGE running in the SM for the  $\overline{\text{MS}}$ renormalization scheme (light red line), compared with our CsI + LAr determination  (blue) and other measurements at different scales.
  }
\label{fig:thweak-CsI}
\end{figure}

In the left panel of Fig.~\ref{fig:thweak-CsI} we show the $\Delta\chi^2$ profile 
obtained from the analysis of CsI (magenta), LAr (orange) and CsI + LAr (blue) data as a function of  $\sin^2 \theta_{\text{W}}$.
The combined fit of CsI + LAr data leads to the following best fit value for this parameter at $1\sigma$ 
\begin{equation}
    \sin^2 \theta_{\text{W}} = 0.237 \pm 0.029 \, .
\end{equation}
As can be seen from the plot, clearly the result is mainly driven by the recent CsI data. 
We note that for this analysis the ES events on CsI can be safely neglected.
The right panel of Fig.~\ref{fig:thweak-CsI} shows the $\sin^2 \theta_{\text{W}}$ RGE evolution in the SM (light red line), calculated in the 
$\overline{\text{MS}}$ renormalization scheme as obtained in Ref.~\cite{Erler:2004in}, together with our combined CsI + LAr determination (blue) and other measurements at different
scales~\cite{Wood:1997zq,Derevianko:2010kz,Androic:2018kni,Anthony:2005pm,Wang:2014bba,Zeller:2001hh,AristizabalSierra:2022axl,Majumdar:2022nby}. 
It is interesting to note that the full COHERENT data provide a determination of the weak mixing angle at low-energies, in a region where other data-driven constraints are absent.
Moreover, one may notice the complementarity with the results obtained in Refs.~\cite{AristizabalSierra:2022axl,Majumdar:2022nby} using the Dresden-II \cevns ~reactor data. 

\begin{figure}[h]
\includegraphics[width= 0.4 \textwidth]{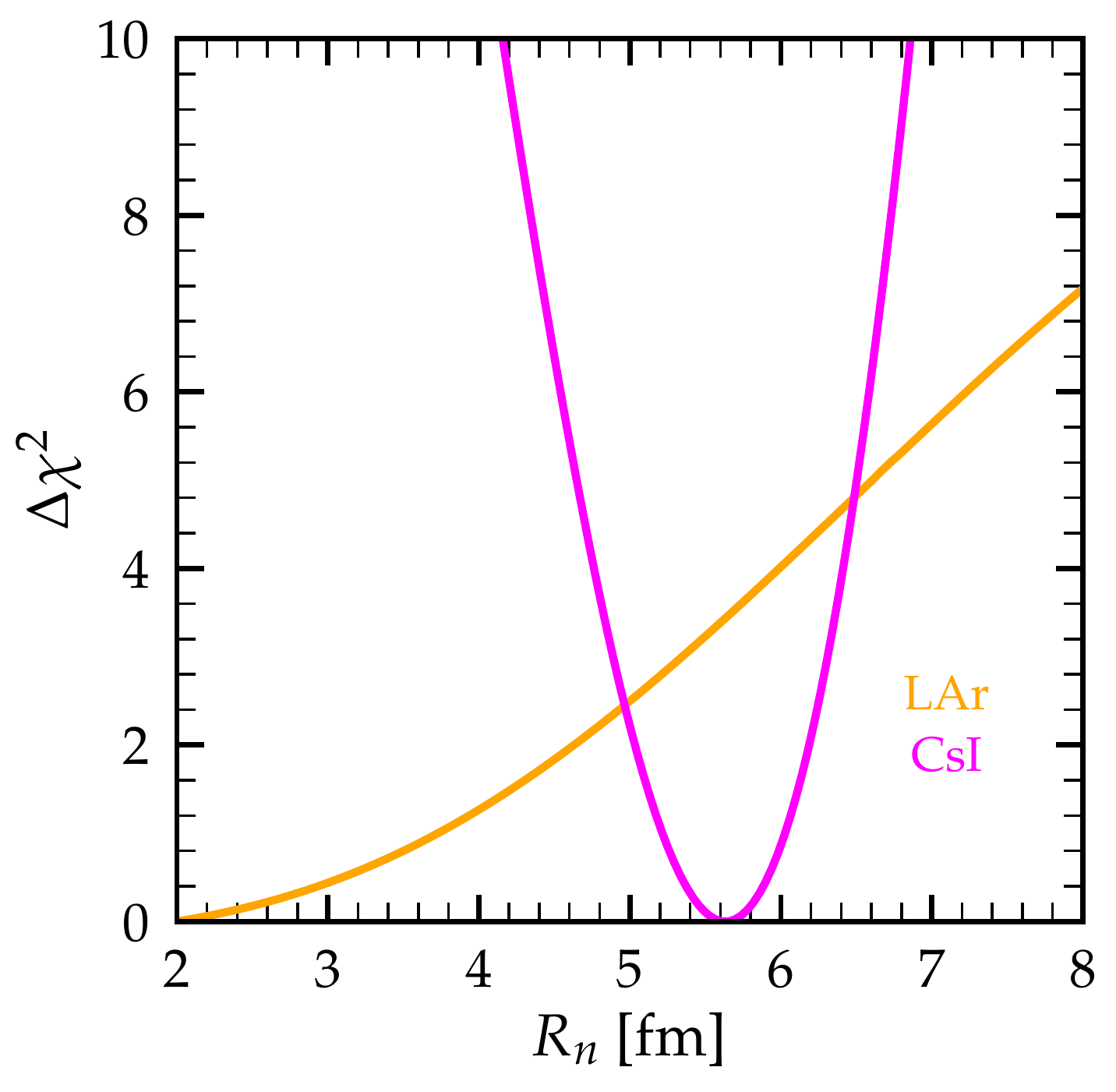}
\caption{\centering{Sensitivity on the neutron RMS radii of argon and CsI.}}
\label{fig:Rn}
\end{figure}

At this point, we will  explore the implications of COHERENT data on nuclear physics, with the goal of probing the neutron density distributions in CsI and Ar. 
The relevant phenomenological parameter is the RMS value of the neutron radius $R_n$ inside the nucleus.  
Let us note that, for this particular analysis, different RMS radii have been considered for protons and neutrons, i.e. we take $F_p(|\vec{q}|) \neq F_n(|\vec{q}|)$. 
Specifically, we fix the proton RMS radius to $R_p(\text{Ar})=3.448~\text{fm}$,  $R_p(\text{I})=4.766~\text{fm}$, $R_p(\text{Cs})=4.824~\text{fm}$~\cite{Cadeddu:2020lky}, while $R_n$ is left to vary freely.
We furthermore deactivate the $\alpha_4$ nuisance parameter for the case of CsI, while for the case of LAr the effect of neglecting the nuclear physics uncertainty in $\beta_0$ is tiny, see footnote~\ref{foot:uncert}. 

In Fig.~\ref{fig:Rn} we display the sensitivity on the neutron RMS radius, $R_n$, of argon and CsI obtained from the analysis of COHERENT data.  
For the case of LAr, here we have updated the result on $R_n(\text{Ar})$ obtained in Ref.~\cite{Miranda:2020tif} using not only the energy data but also the timing information and the shape uncertainties described in Eq.~(\ref{chi2LAr}).
As for the CsI detector, ours is the first result obtained using the actual COHERENT-CsI \rd{(2021)} data~\cite{COHERENT:2021xmm} and the comprehensive $\chi^2$ function given in Eq.~(\ref{eq:Nth_CsI_chi2}), including timing and efficiency shape uncertainties
as well as the small NIN background~\footnote{Note that Ref.~\cite{Cadeddu:2021ijh} analyzed the preliminary CsI data reported in~\cite{Pershey:M7s} neglecting the shape uncertainties.}. 
The 1$\sigma$ regions on the neutron RMS radii of argon and CsI from our analysis are
\begin{equation}
\begin{aligned}
 R_n(\mathrm{Ar}) \in & \, [0.00, 3.72]  \, \text{fm}\, ,\\
   R_n(\mathrm{CsI}) \in &  \, [5.22, 6.03]  \, \text{fm}\, .
    \end{aligned}
\end{equation}

Before closing this discussion, we would like to comment on the level of improvement of the 
current determination of $R_n$  in comparison with the results obtained with the first COHERENT-CsI data.
To this end we compare our results with those extracted in Ref.~\cite{Coloma:2020nhf} which analyzed the energy and timing data of the 2017 COHERENT-CsI release~\cite{COHERENT:2017ipa}.
 Our analysis gives $R_n = 5.62^{+0.41}_{-0.40}$ fm (at $1 \sigma$) compared to $R_n = 5.80^{+0.89}_{-0.93}$ fm reported in ~\cite{Coloma:2020nhf}. While comparing the best fit points may not be straightforward because they belong to different data sets, it is however interesting to notice that in our case the $1 \sigma$ uncertainty is reduced by a factor of 2.

\subsection{Neutrino NSI }

\begin{figure}[t]
\includegraphics[width= 0.4 \textwidth]{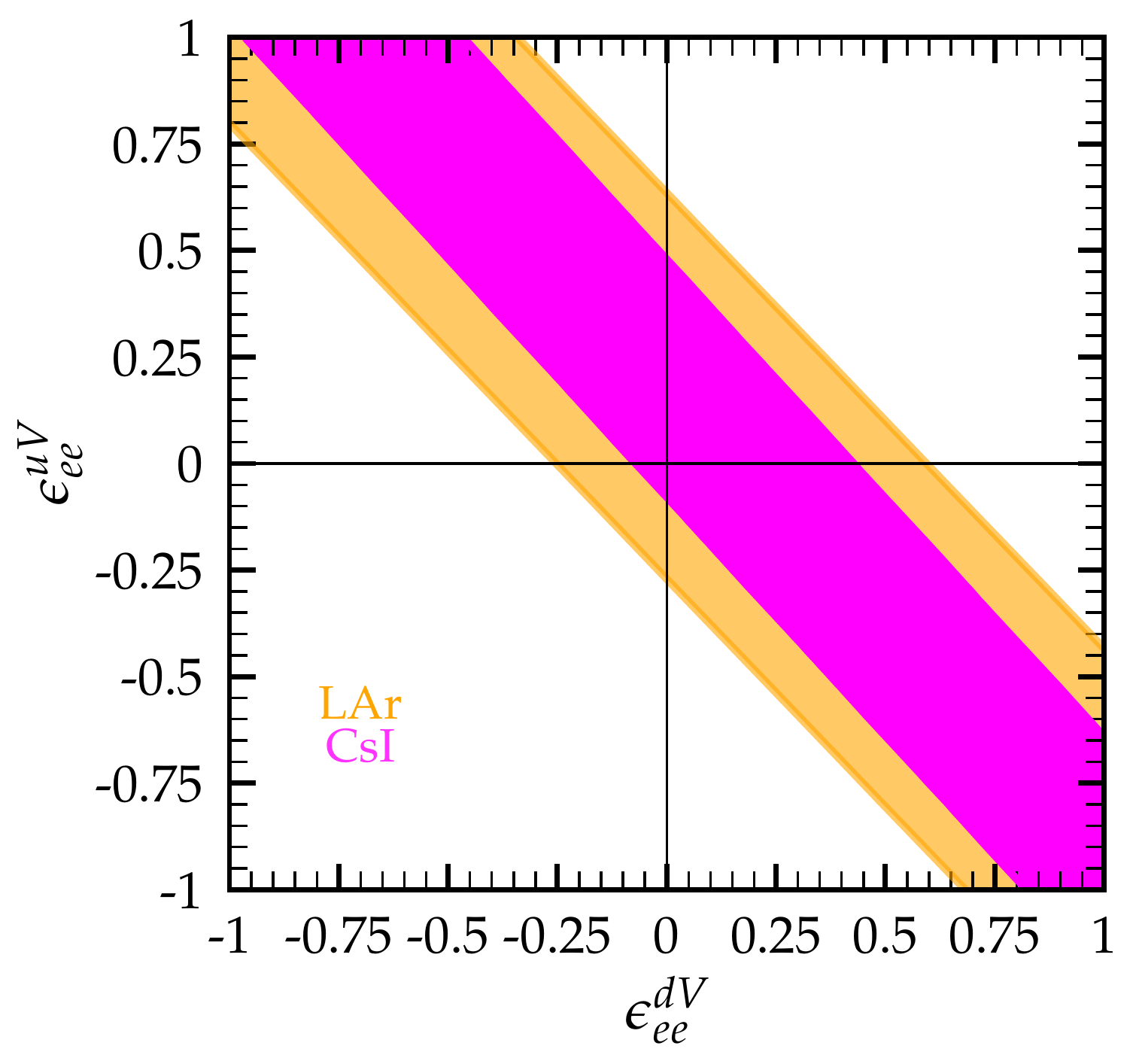}
\includegraphics[width= 0.4 \textwidth]{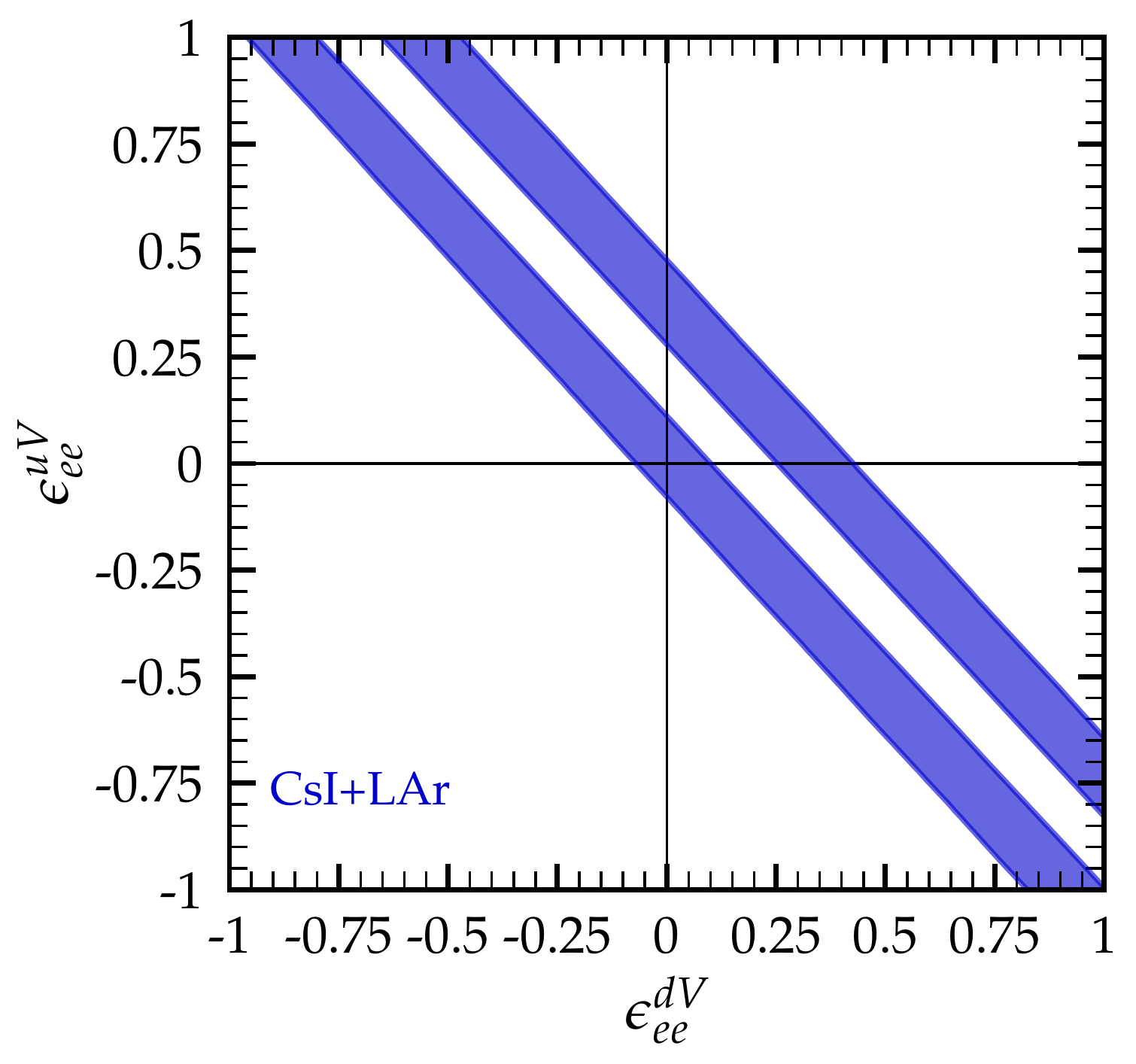}
\includegraphics[width= 0.4 \textwidth]{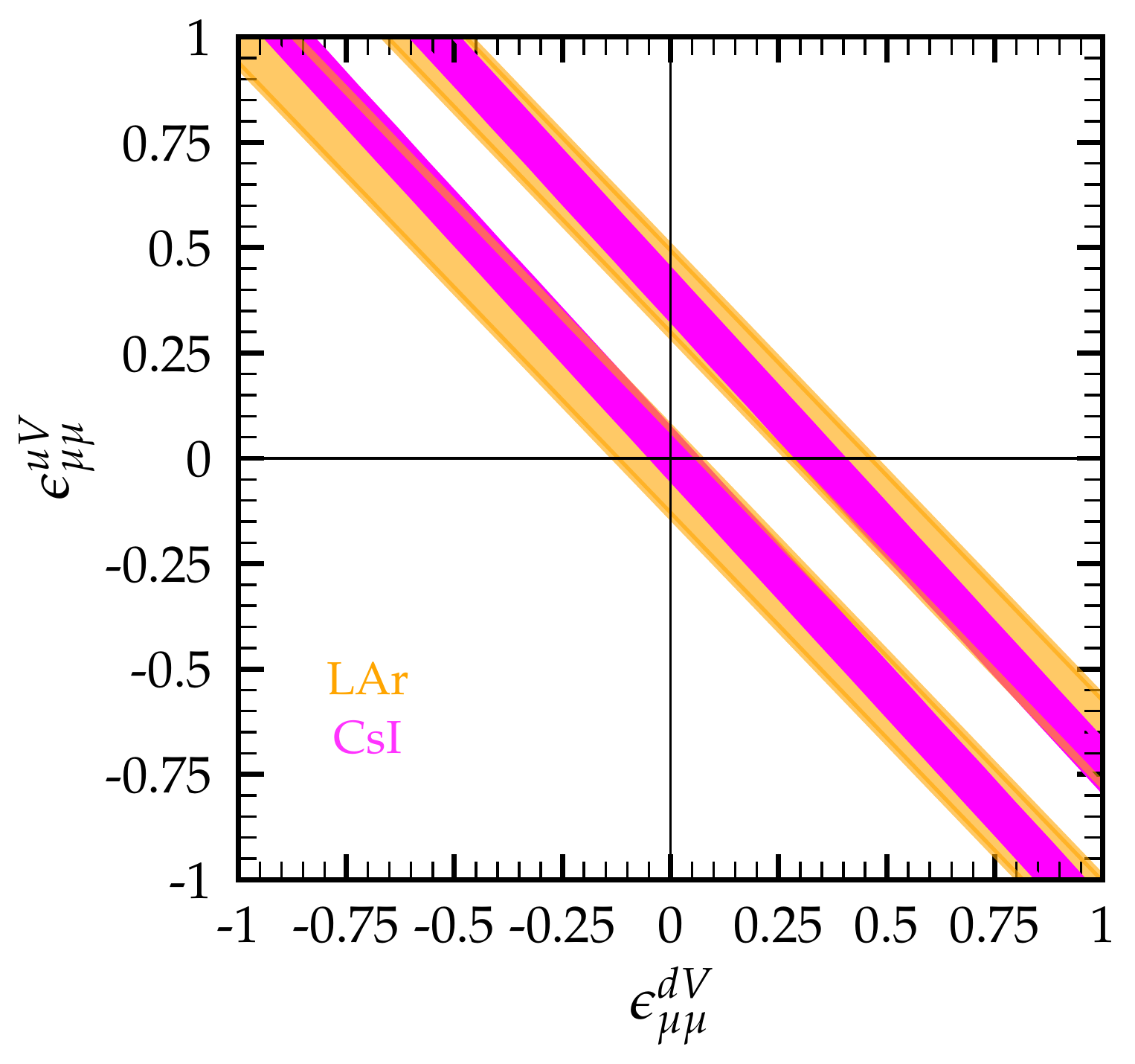}
\includegraphics[width= 0.4 \textwidth]{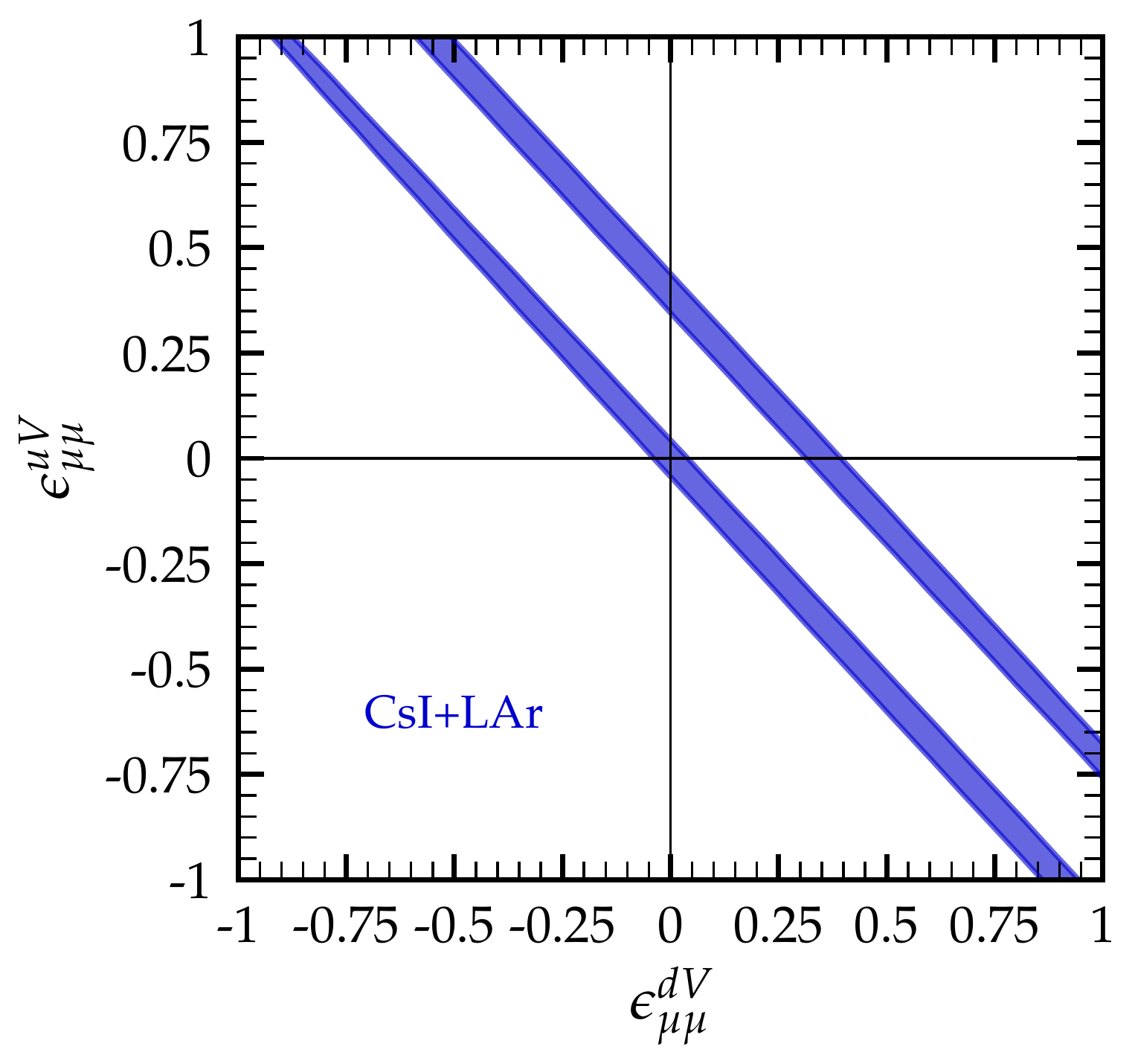}
\caption{ $90 \%$ C.L. (2 d.o.f.) allowed regions on flavor-preserving neutrino NSI from the CsI (magenta), LAr (orange) and CsI + LAr (blue) analyses, allowing two NSI parameters at a time. The analysis of CsI data includes only CE$\nu$NS interactions.} 
\label{fig:NSI-CsI_ee}
\end{figure}

Here we explore NSI involving both the flavor-preserving and flavor-changing terms in Eq.~(\ref{eq:qvNSI}). 
In our analysis we consider two NSI parameters at a time, with the rest set to zero, and our results are presented as $90 \%$ C.L. (2 d.o.f.) preferred regions. 
First, in Fig.~\ref{fig:NSI-CsI_ee} we consider nonuniversal NSI  involving electron or muon neutrinos.
In the upper-left panel we explore the $(\epsilon_{ee}^{dV}, \epsilon_{ee}^{uV})$ plane, noting that a single allowed band is present when analysing CsI or LAr data separately. 
The reason is that, even when the $\nu_e$-induced \cevns ~rate is suppressed, the $\nu_\mu+\bar{\nu}_\mu$ contributions can still fit the data reasonably well.  
Nevertheless, the combination of CsI and LAr data can break the degeneracy between different NSI parameter combinations, resulting in two bands (see upper-right panel).
One band contains the SM solution $(\varepsilon_{ee}^{uV} = \varepsilon_{ee}^{dV} = 0)$, while the other  corresponds to the region where $ Q_V^{\text{NSI}}\approx-Q_V^{\text{SM}}$,
thus mimicking the SM signal.  The width of the two bands is reduced by $\sim 20\%$ when compared to the result presented in \cite{Sinev:2020bux},
  where a combined CsI (2017) + LAr analysis was performed.  
In contrast, having the NSI confined to the muon neutrino sector leads to a different situation (see lower-left panel).
Here, we find two bands in the $(\varepsilon_{\mu \mu}^{dV}, \varepsilon_{\mu \mu}^{uV})$ plane even before combining CsI and LAr data.  
This happens because the contribution to the \cevns ~event rate from $\nu_\mu+\bar{\nu}_\mu$ is larger compared to $\nu_e$. 
Of course, the combined CsI+LAr analysis leads to significantly narrower allowed bands, as seen in the lower-right panel.

\begin{figure}[!htb]
\includegraphics[width= 0.4 \textwidth]{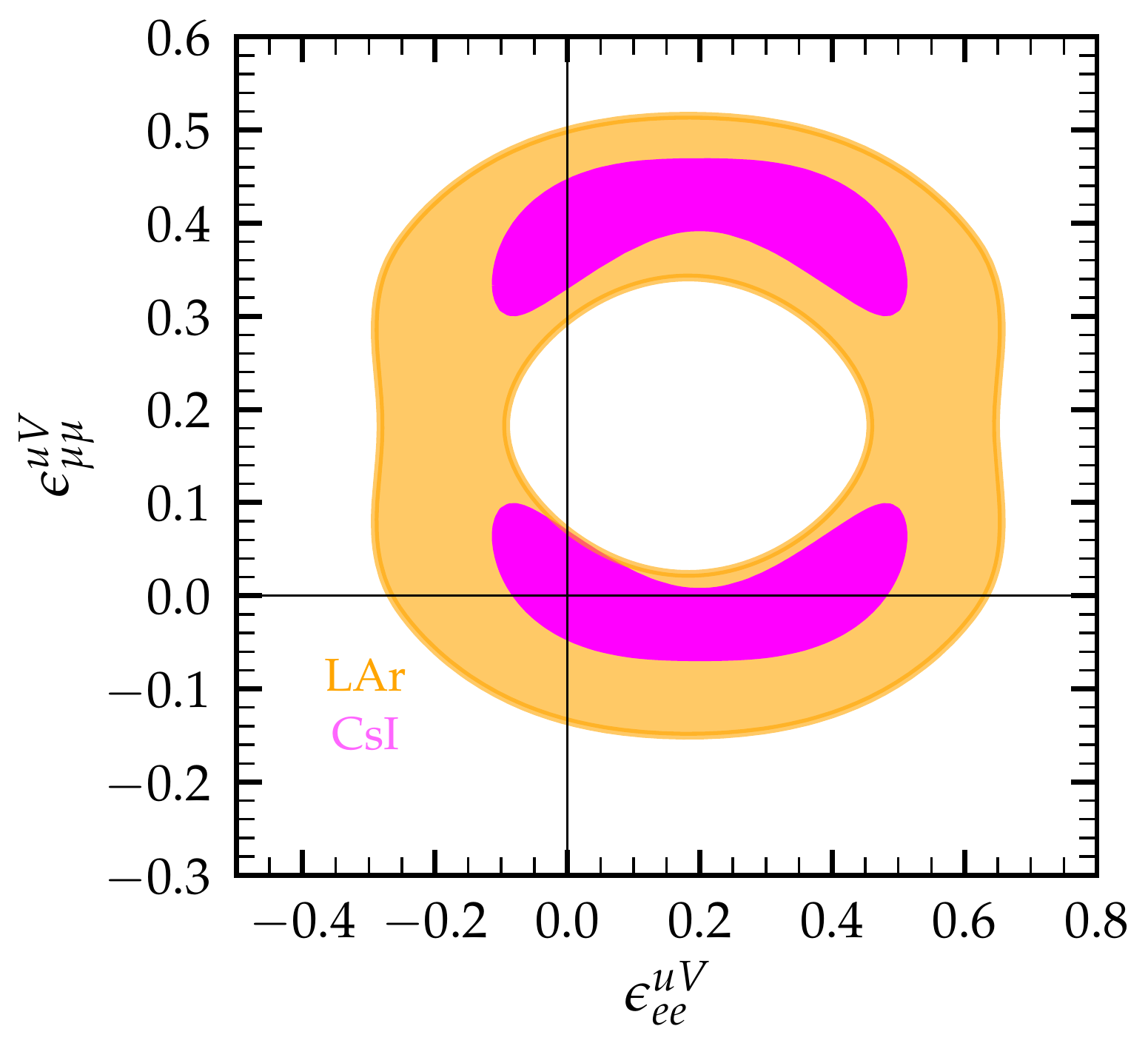}
\includegraphics[width= 0.4 \textwidth]{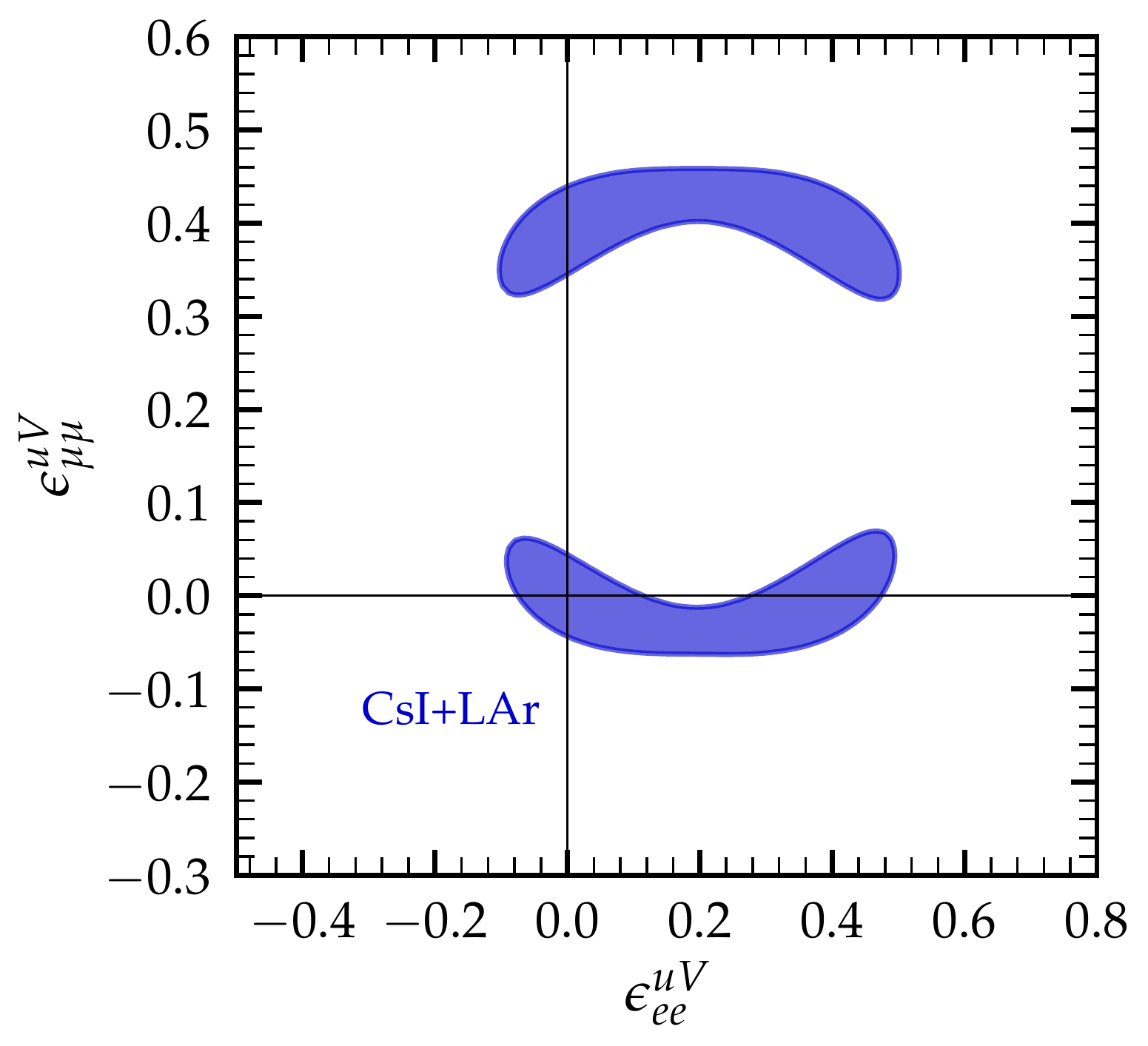}
\includegraphics[width= 0.4 \textwidth]{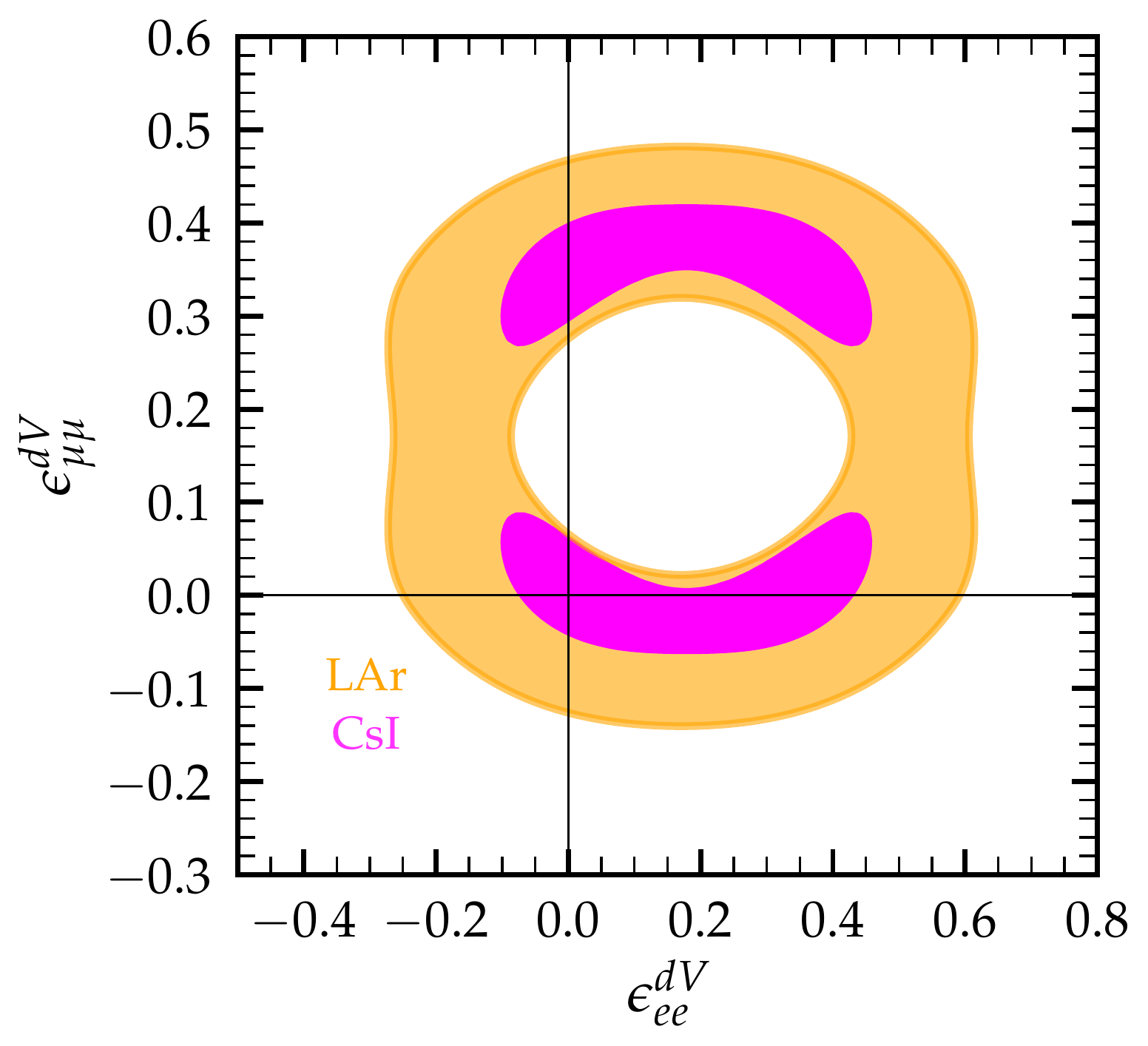}
\includegraphics[width= 0.4 \textwidth]{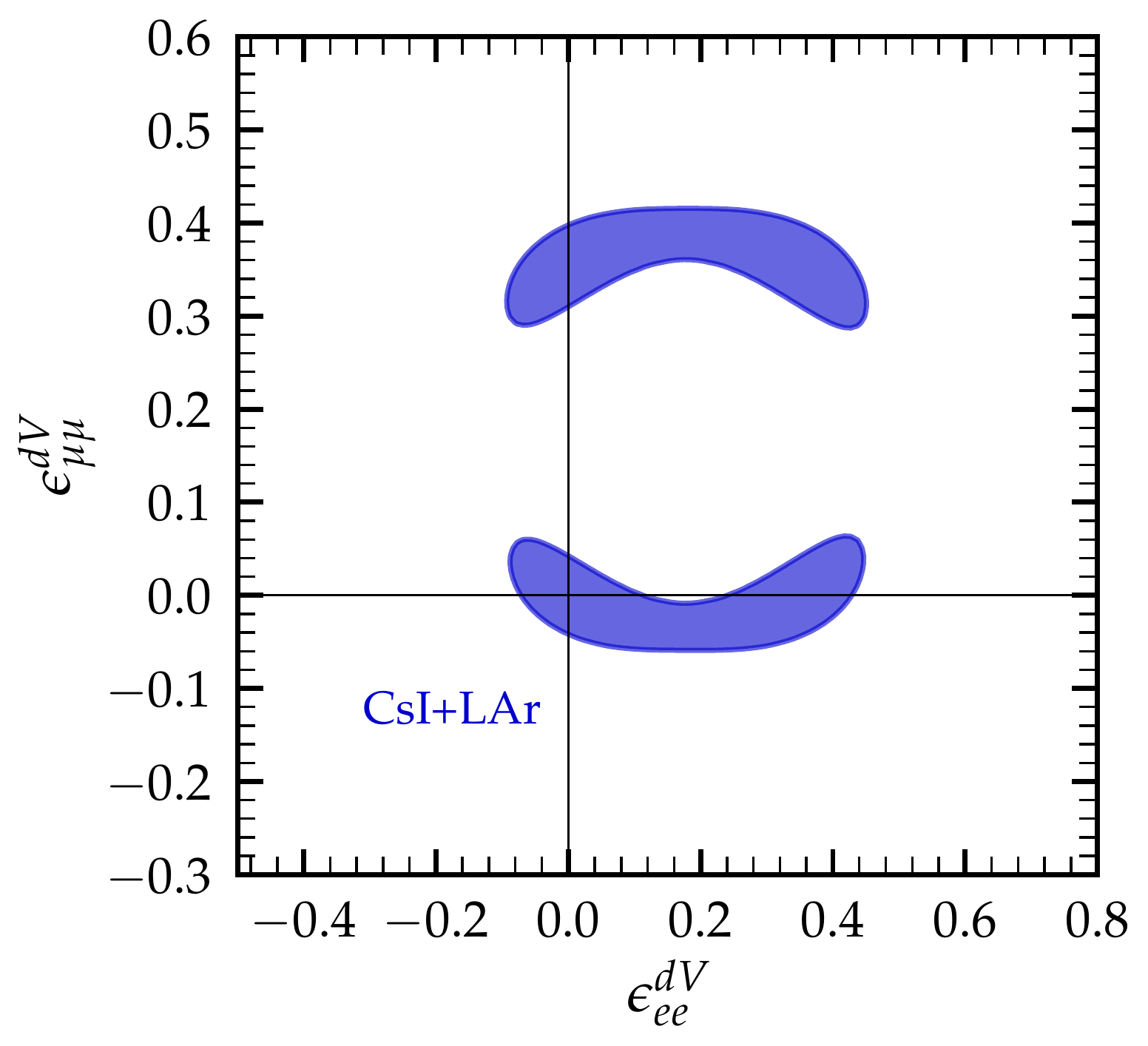}
\caption{\centering Same as Fig.~\ref{fig:NSI-CsI_ee} but in the $(\epsilon_{ee}^{qV}, \epsilon_{\mu \mu}^{qV})$ planes.}
\label{fig:NSI-CsI_ee_mumu}
\end{figure}
\begin{figure}[!htb]
\includegraphics[width= 0.4 \textwidth]{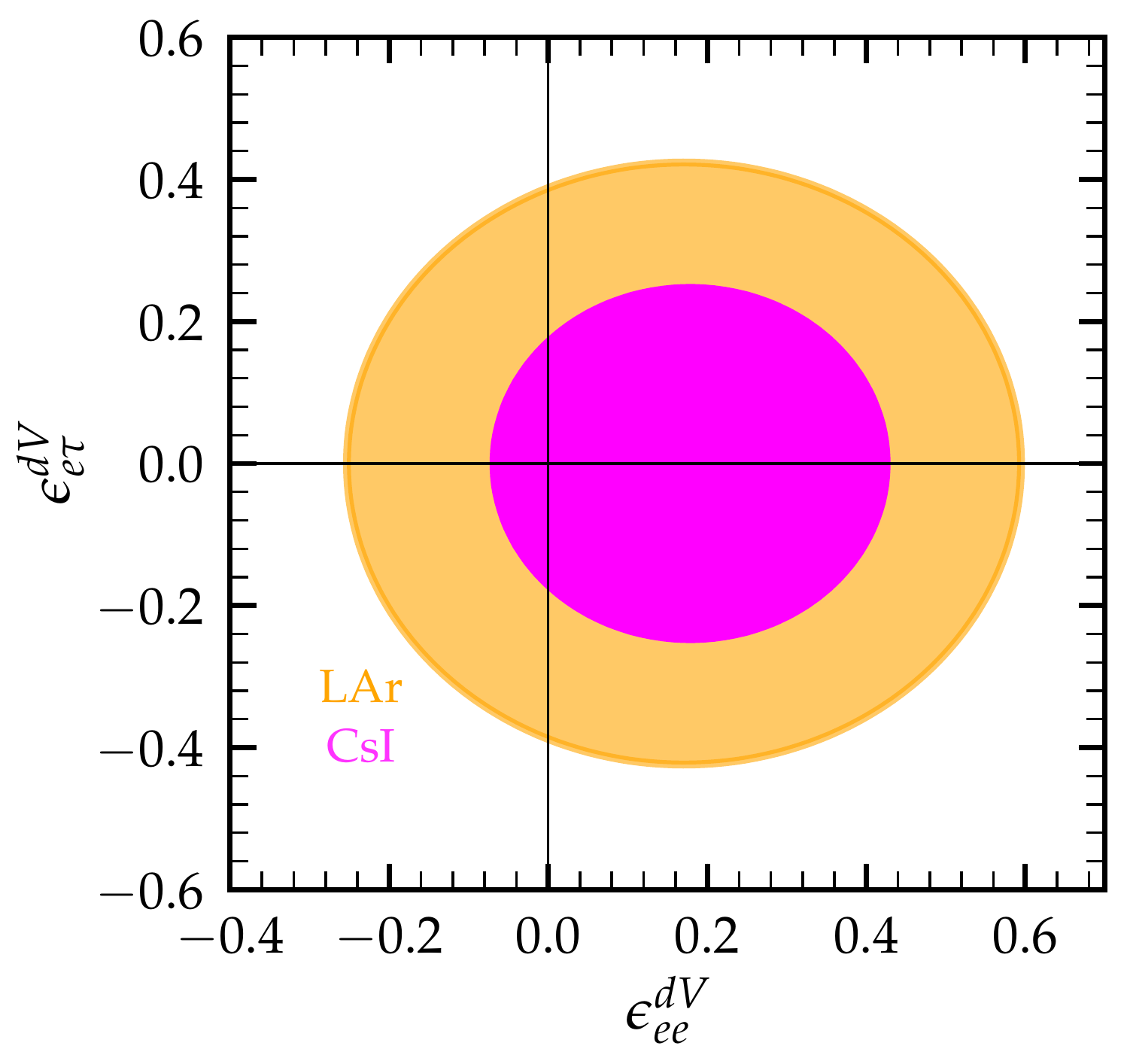}
\includegraphics[width= 0.4 \textwidth]{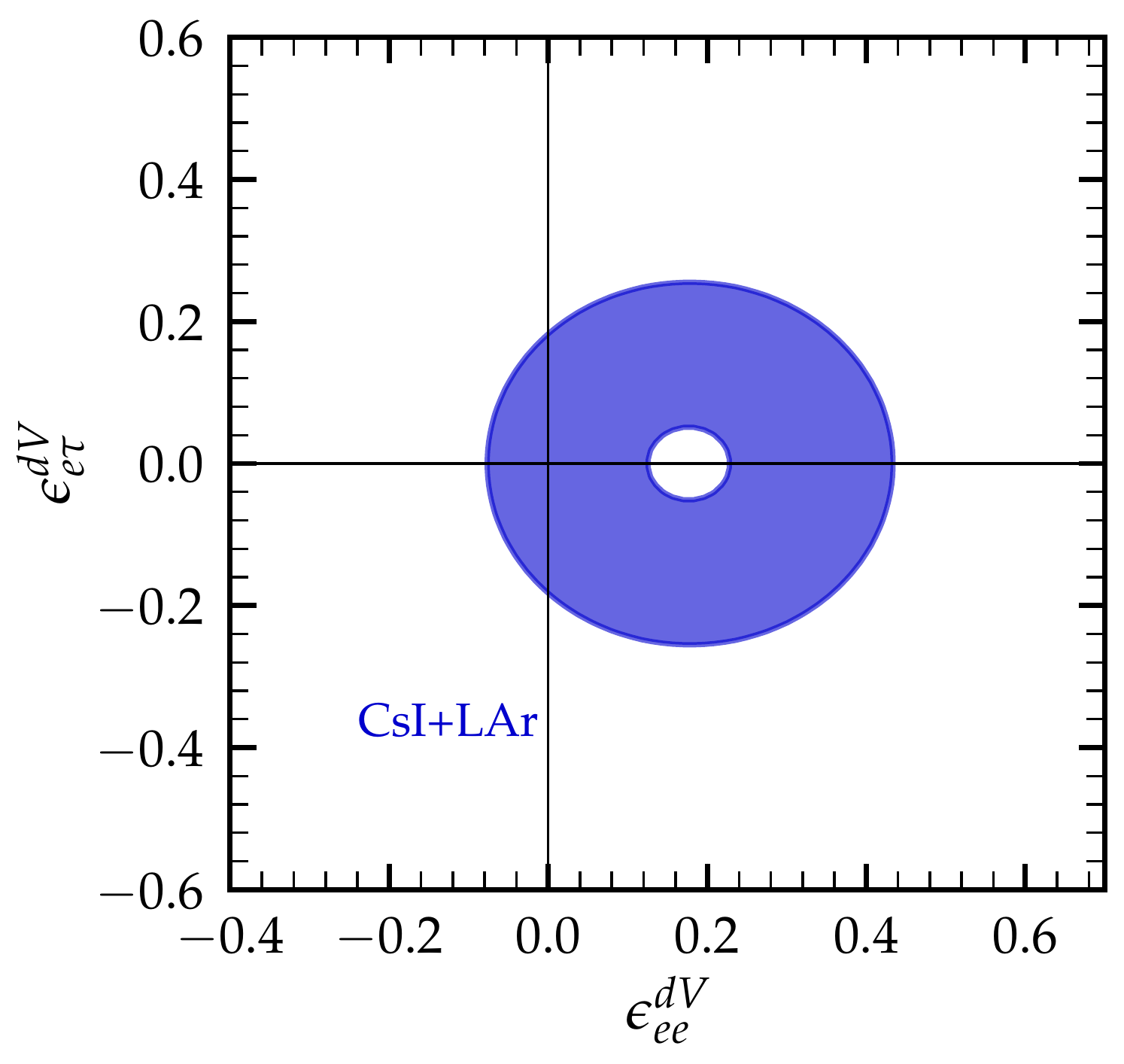}
\includegraphics[width= 0.4 \textwidth]{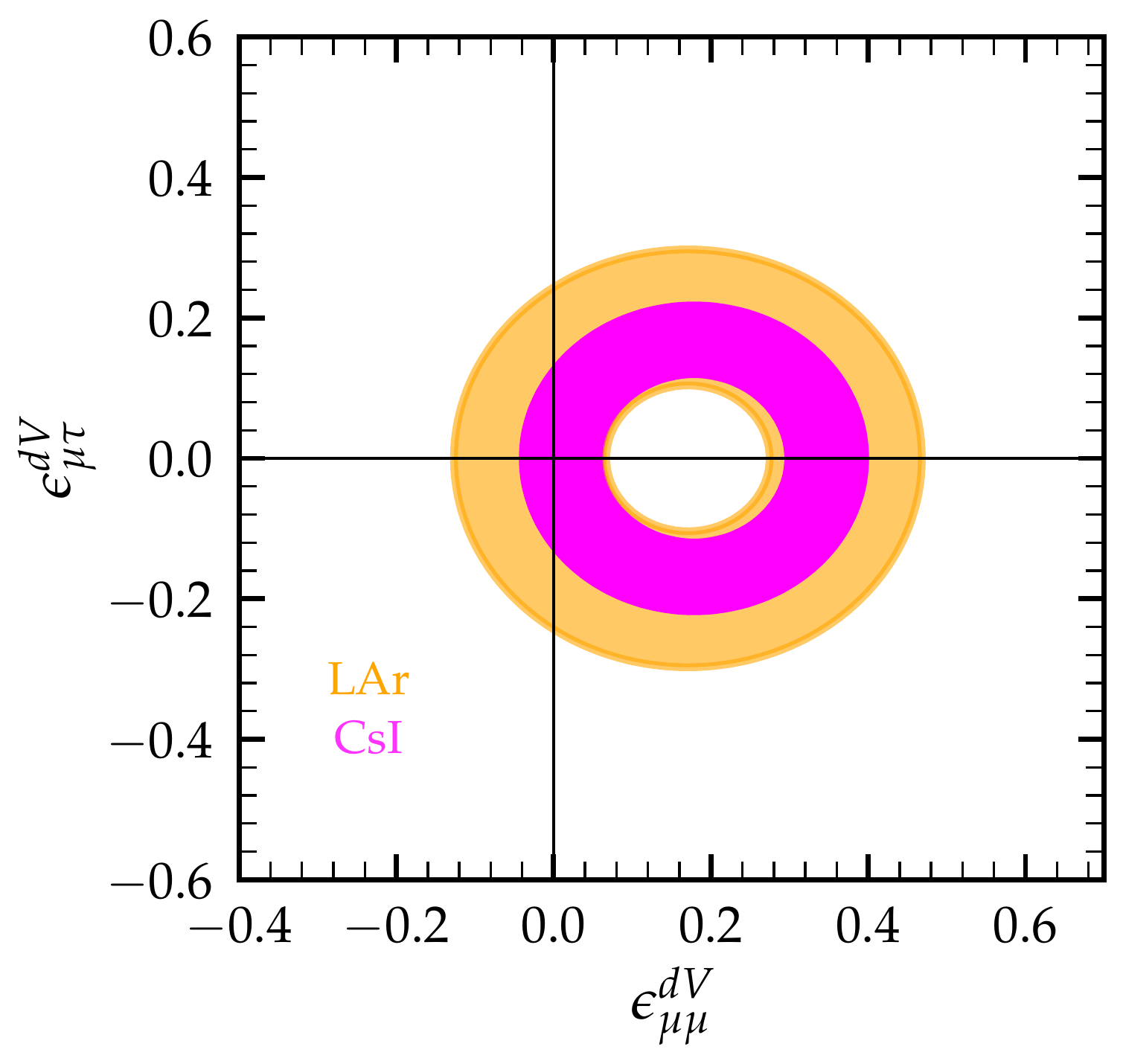}
\includegraphics[width= 0.4 \textwidth]{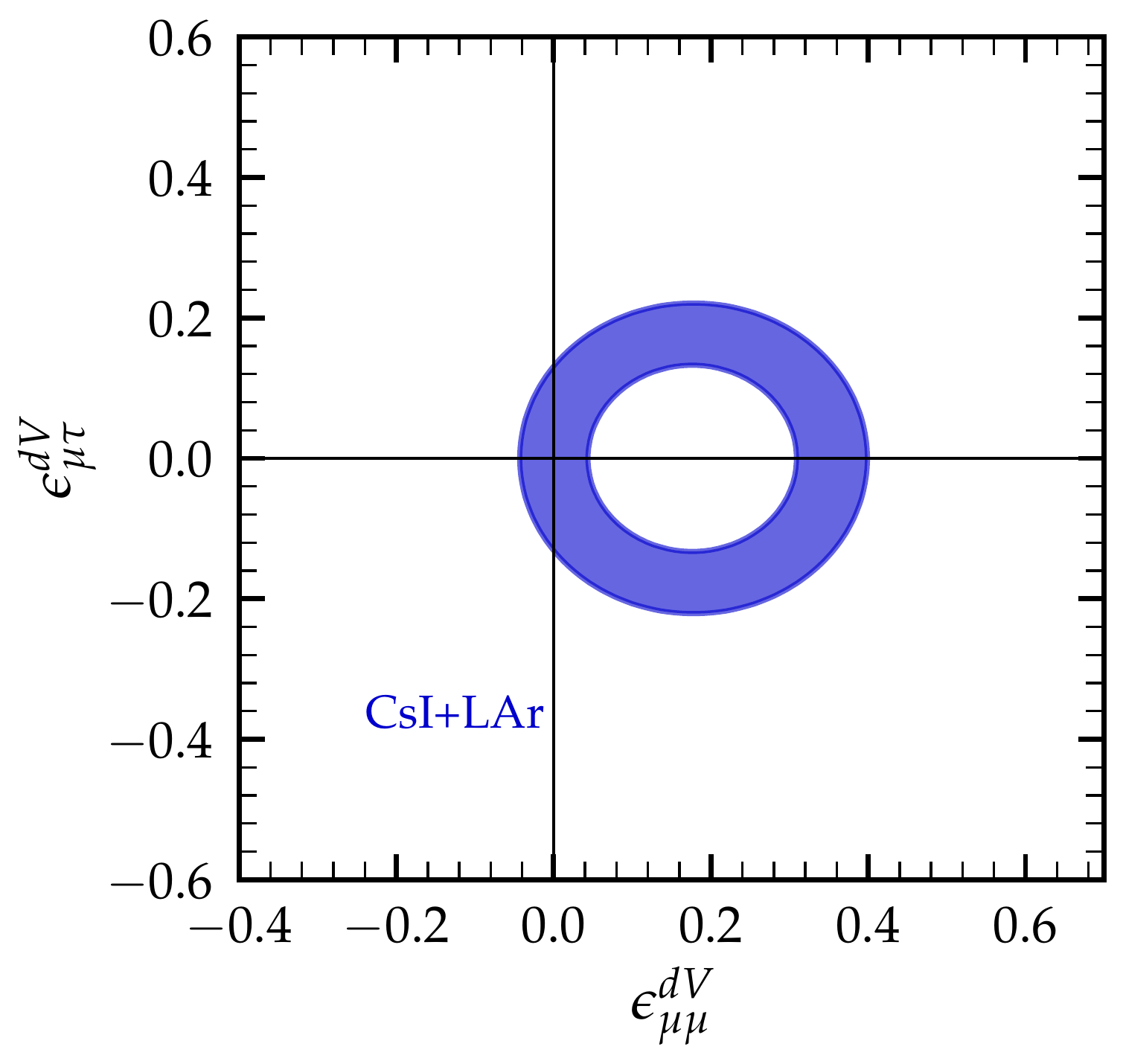}
\caption{$90 \%$ C.L. (2 d.o.f.) allowed regions assuming one nonuniversal and one flavor-changing neutrino NSI, extracted from the analysis of CsI (magenta), LAr (orange) and the combined CsI + LAr (blue) data. The upper panel shows the result in the $(\epsilon_{ee}^{dV}, \epsilon_{e \tau}^{dV})$ plane while the lower panel shows the corresponding result in $(\epsilon_{\mu \mu}^{dV}, \epsilon_{\mu \tau}^{dV})$. The analysis of CsI data includes only CE$\nu$NS interactions.}  
\label{fig:NSI-CsI_FC-a}
\end{figure}
\begin{figure}[!htb]
\includegraphics[width= 0.4 \textwidth]{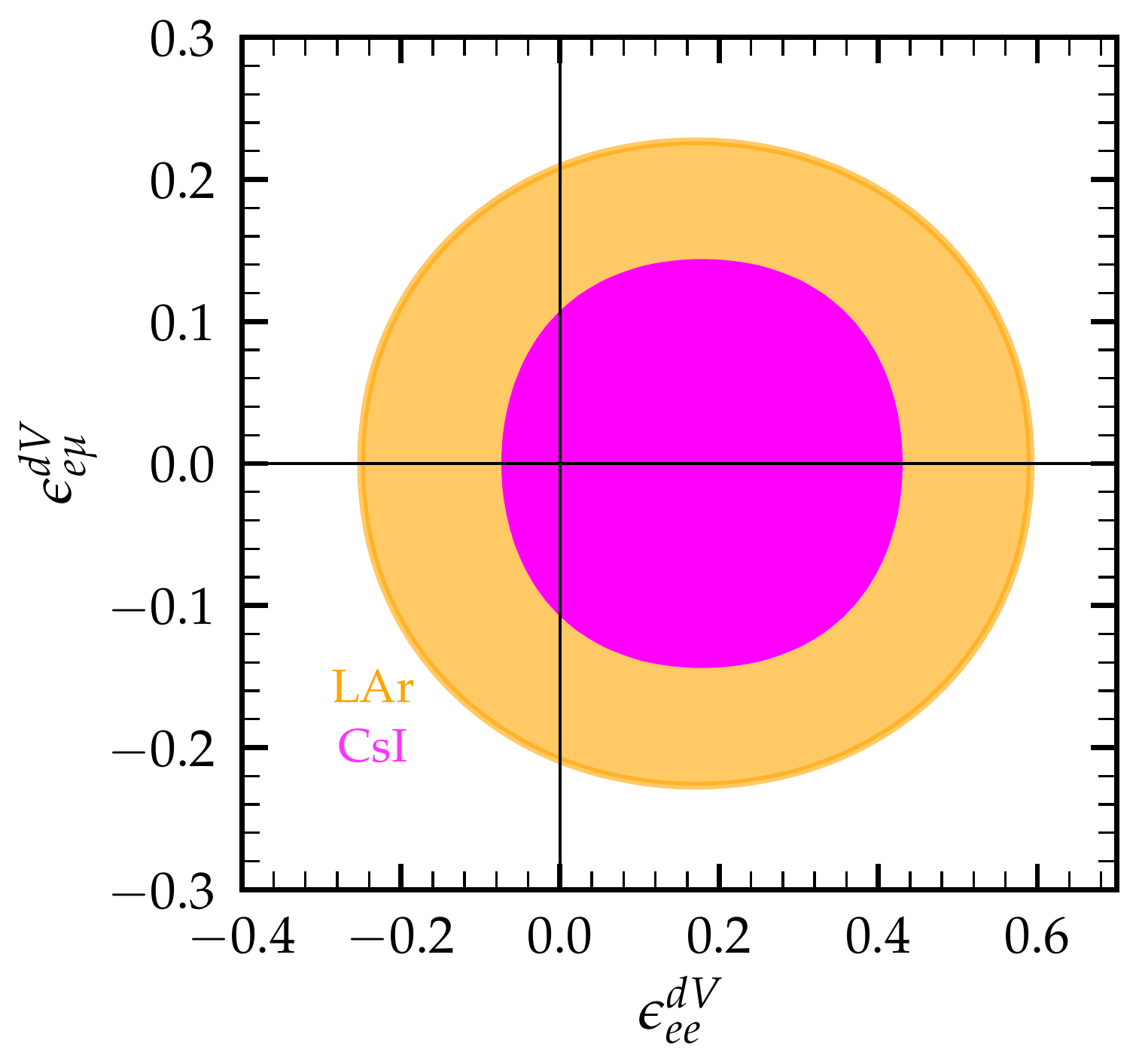}
\includegraphics[width= 0.4 \textwidth]{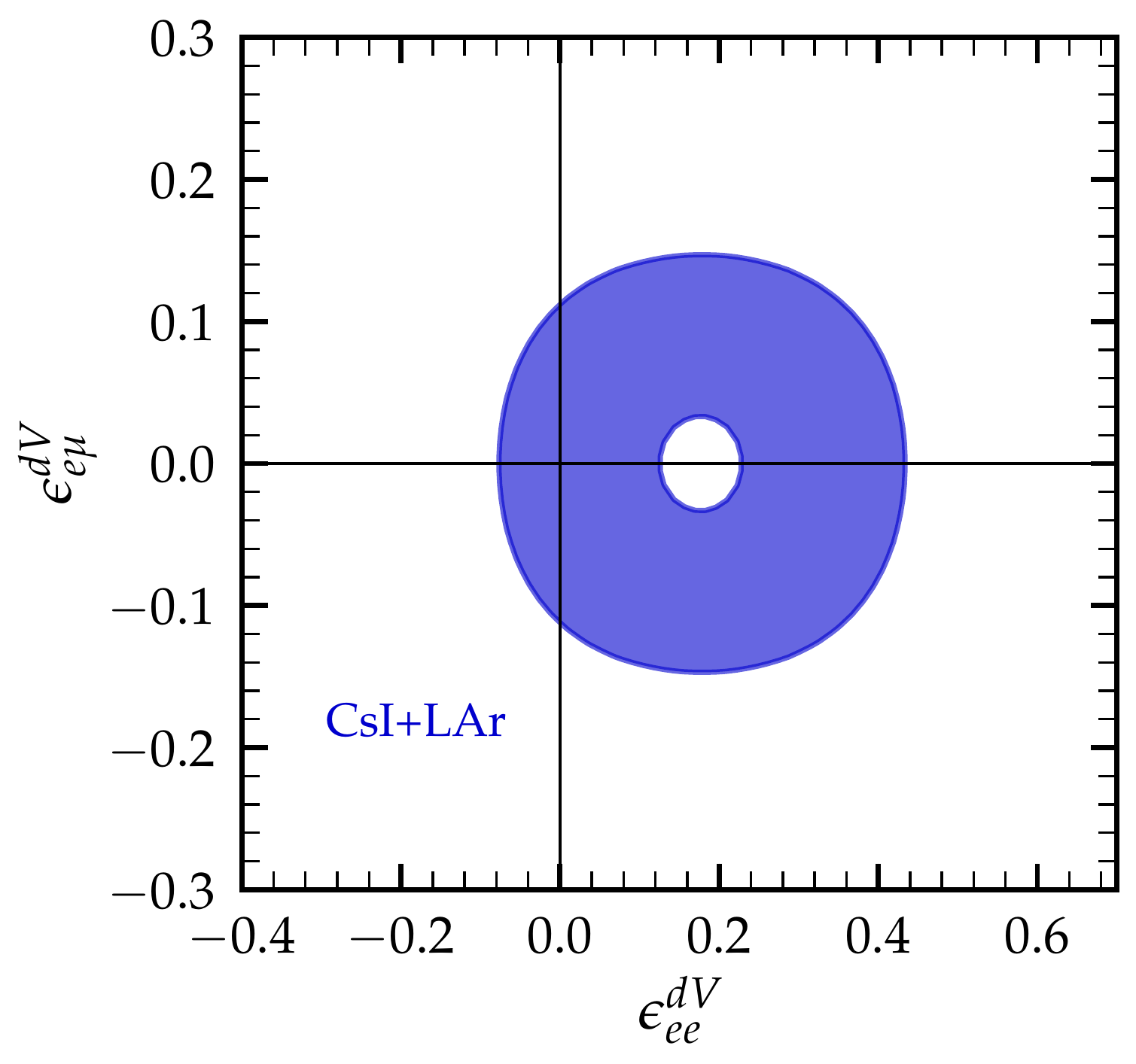}
\includegraphics[width= 0.4 \textwidth]{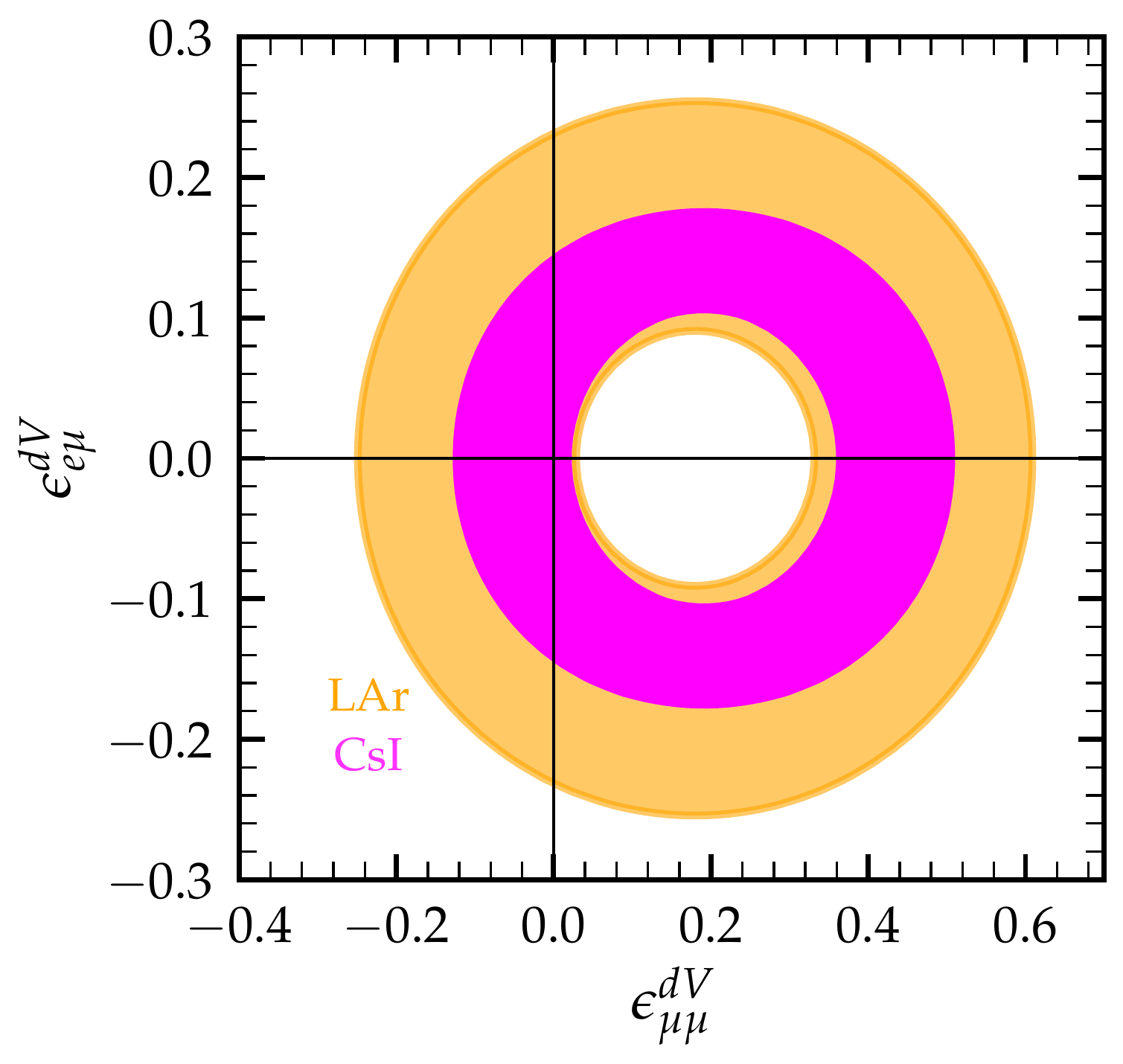}
\includegraphics[width= 0.4 \textwidth]{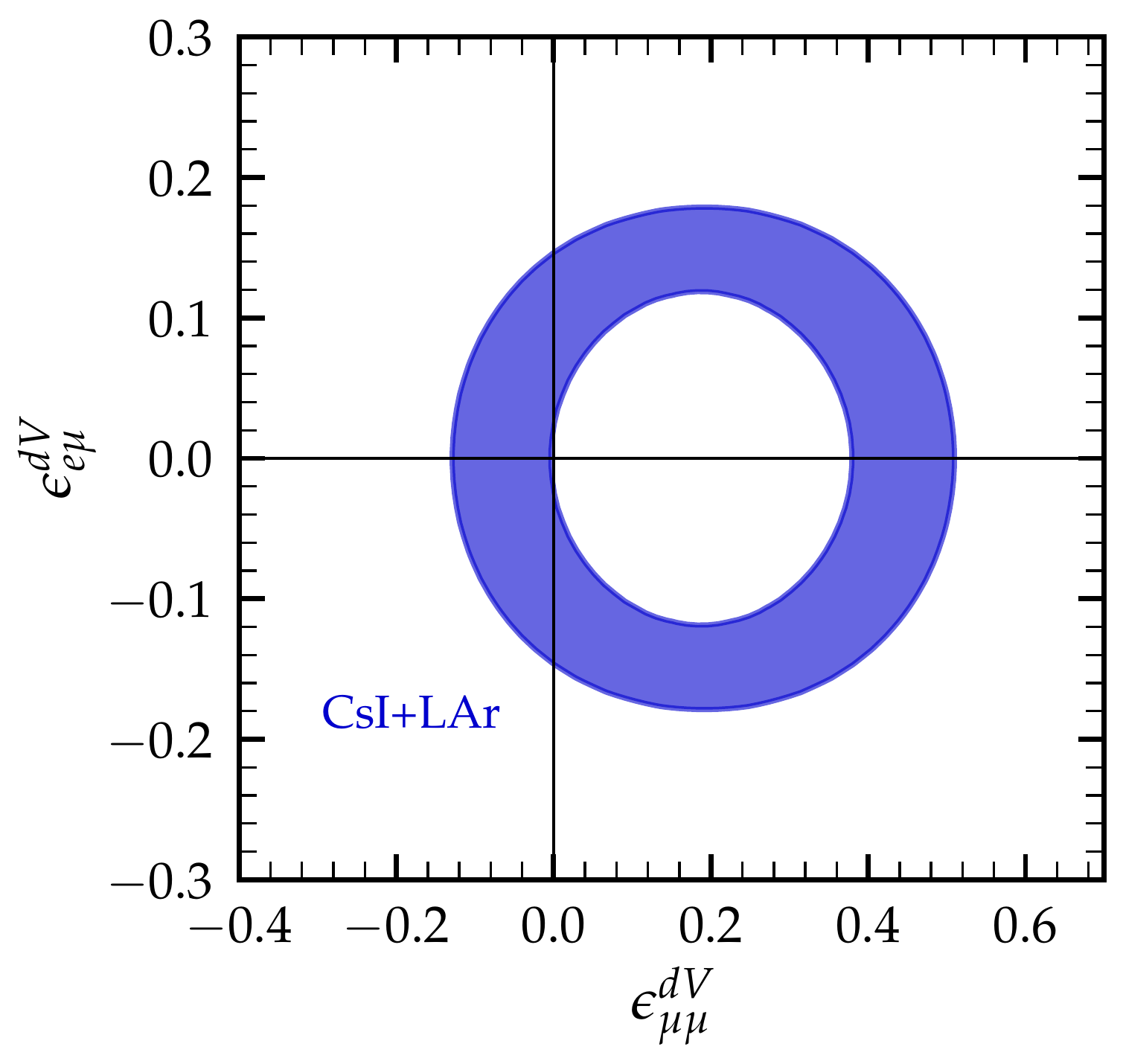}
\caption{ Same as Fig.~\ref{fig:NSI-CsI_FC-a}  for the planes $(\epsilon_{ee}^{dV}, \epsilon_{e \mu}^{dV})$ and  $(\epsilon_{\mu \mu}^{dV}, \epsilon_{e \mu}^{dV})$, in the upper and lower panel, respectively.}  
\label{fig:NSI-CsI_FC-b}
\end{figure}

In Fig.~\ref{fig:NSI-CsI_ee_mumu} we consider nonuniversal NSI for both electron and muon neutrinos.
We present the allowed regions in the $(\epsilon_{ee}^{uV}, \epsilon_{\mu \mu}^{uV})$ plane (upper panel) and in the $(\epsilon_{ee}^{dV}, \epsilon_{\mu \mu}^{dV})$ plane (lower panel). 
One sees that, due to the significantly reduced uncertainty of the recent COHERENT-CsI measurement, there are two separated allowed regions for CsI, but just a single region is obtained from the LAr data analysis (left panel).
By comparing the NSI constraints for $u$ or $d$ quarks, one sees that the latter case is slightly more constrained. 
Moreover, the combined CsI+LAr analysis is mostly driven by the CsI data, leading to a mild improvement compared to the CsI alone analysis (see right panels).
At this point, we stress that our results in the $(\epsilon_{ee}^{uV}, \epsilon_{\mu \mu}^{uV})$ plane regarding the CsI data alone are in excellent agreement with the original result reported by COHERENT Collaboration~\cite{COHERENT:2021xmm,Abdullah:2022zue}, a result which is further strengthened by our simulation of event spectra shown in Appendix~\ref{sec:appendix}. 
This serves also as a calibration~\footnote{Although not shown here, our result in the $(\epsilon_{ee}^{dV}, \epsilon_{ee}^{uV})$ is consistent with Refs.~\cite{COHERENT:2020iec, COHERENT:2021xmm} when assuming 1 d.o.f. as taken by the COHERENT Collaboration. In contrast, here we have considered 2 d.o.f., which leads to a single band.} for the robustness and accuracy of our calculational procedure.  In addition, our result is consistent with the analysis of CsI (2017) data presented in \cite{Coloma:2022avw}.

Finally, in Figs.~\ref{fig:NSI-CsI_FC-a} and \ref{fig:NSI-CsI_FC-b} we display the allowed regions for one flavor-preserving NSI versus one flavor-changing NSI parameter for  various combinations, ($\epsilon_{ee}^{dV}, \epsilon_{e\tau}^{dV})$,
($\epsilon_{\mu\mu}^{dV}, \epsilon_{\mu\tau}^{dV})$, and ($\epsilon_{ee}^{dV}, \epsilon_{e\mu}^{dV})$  ($\epsilon_{\mu\mu}^{dV}, \epsilon_{e\mu}^{dV})$, respectively. 
As previously, the results show the constraints from the CsI or LAr measurements (left panels), as well as those resulting from the combined CsI+LAr analysis (right panels),
leading to the same general conclusions as discussed above.  The implications of the enlarged CsI (2021) dataset can be seen, for instance, in the ($\epsilon_{ee}^{dV}, \epsilon_{e\tau}^{dV})$ plane,
 for which the allowed region is reduced by $\sim 40\%$ in comparison with the results obtained using the former CsI (2017) data~\cite{Miranda:2020tif}.
To summarise, the $1 \sigma$ limits obtained from the combined CsI+LAr analysis on the single nonuniversal NSI parameters (setting all the others to zero) read
\begin{equation}
\begin{aligned}
\epsilon_{ee}^{dV} =& [-0.027, 0.048] \cup
[0.30, 0.39] \, ,\\
\epsilon_{ee}^{uV} =& [-0.024, 0.045] \cup
[0.34, 0.43]\, ,\\
\epsilon_{\mu \mu}^{dV} =& [-0.012, 0.016] \cup
[0.33, 0.37] \, ,\\
\epsilon_{\mu \mu}^{uV} =& [-0.002, 0.001] \cup
[0.37, 0.41] \, .
\end{aligned}
\end{equation}
Similarly, the $1 \sigma$ limits on flavor-changing NSI parameters are
\begin{equation}
\begin{aligned}
\epsilon_{e\mu}^{dV} =& [-0.071, 0.071] \, , \quad
\epsilon_{e\mu}^{uV} = [-0.081, 0.081]\, , \\
\epsilon_{e\tau}^{dV} =& [-0.12, 0.12] \, , \quad \quad
\epsilon_{e\tau}^{uV} = [-0.13, 0.13]\, , \\
\epsilon_{\mu\tau}^{dV} =& [-0.087, 0.087] \, , \quad
\epsilon_{\mu\tau}^{uV} = [-0.098, 0.098] .
\end{aligned}
\end{equation}

\subsection{Neutrino NGI} \label{sec:NGI} 

We now turn our attention on a more general description of exotic new physics, namely NGI that go beyond the typical vector NSI that arise in gauge extensions of the SM.
In this context, we aim to explore additional types of Lorentz-invariant interactions involving also scalar and tensor terms. 
In full generality, all types of NGI could in principle exist at the same time.
However, for simplicity, in our analysis we only allow up to two new interactions to be present at the same time (while setting the third one to zero),
assuming universal couplings (i.e. $C_X^u=C_X^d \equiv C_X^q$) and neglecting the ES data in the analysis.

\begin{figure}[t]
\includegraphics[width= 0.95\textwidth]{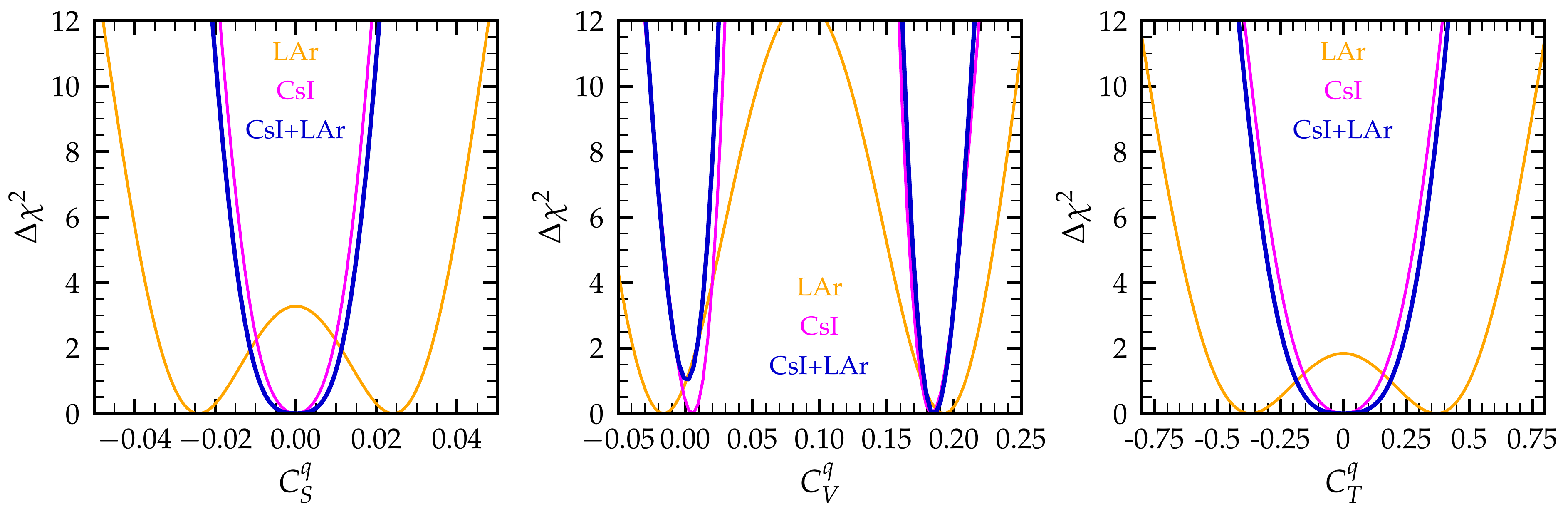}\\
\includegraphics[width= \textwidth]{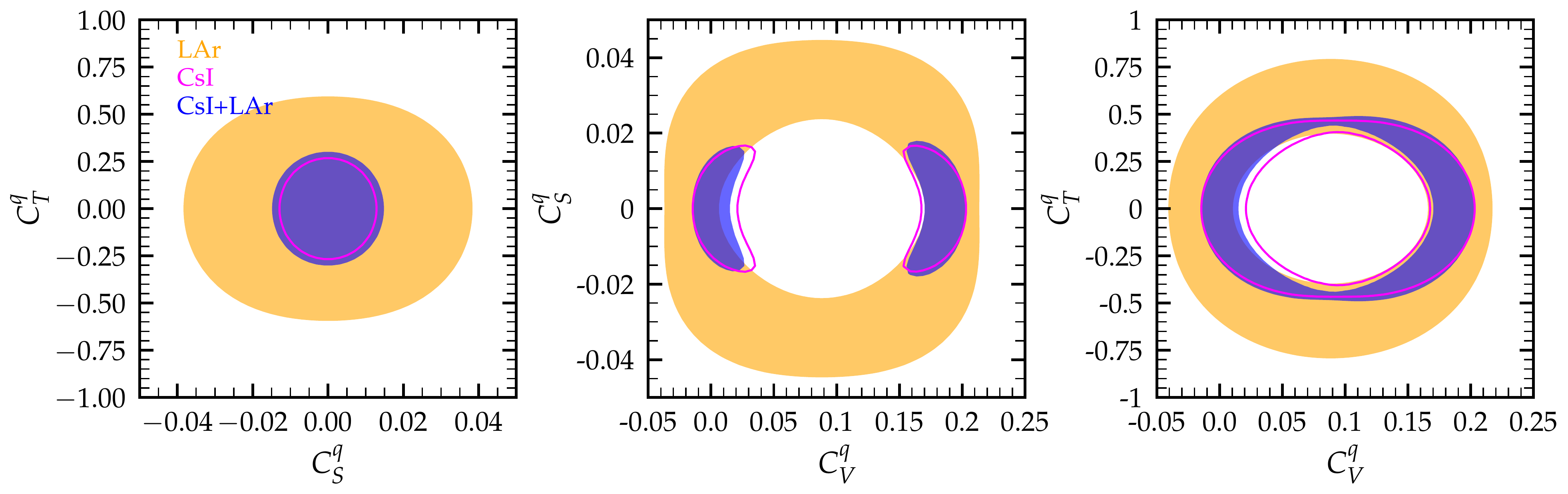}
\caption{ Upper panels: $\Delta\chi^2$ profiles for single NGI from the CsI (magenta), LAr (orange) and CsI + LAr (blue) analyses.
  The analysis of CsI data includes only CE$\nu$NS interactions. Lower panels: $90 \%$ C.L. (2 d.o.f.) allowed regions on neutrino NGI allowing two NGI couplings at a time. }
\label{fig:NGI-CsI}
\end{figure}

Figure~\ref{fig:NGI-CsI} shows the $\Delta\chi^2$ profiles for a single NGI (upper row) given in terms of the $C_X^q$, and the $90 \%$ C.L. (2 d.o.f.)
  allowed regions for neutrino NGI in the $(C_X^q,C_Y^q)$ plane (lower row), with $C_X^q = g_{\nu X} \cdot g_{qX}$ and $X = S, V, T$, see Eqs.~(\ref{eq:Q_V}), (\ref{eq:QS}) and (\ref{eq:QT}). 
  The results show that, although a new scalar or tensor interaction actually improves the fit of LAr data alone, such a new NGI does not fit the CsI data set,
  so that the combined analysis prefers a SM explanation. For the scenario with a new vector interaction instead, both LAr and CsI lead to two $\chi^2$ minima,
    one of them at $C_V^q = 0$ and a second one at $C_V^q \sim 0.18$. In the region between the two minima, the extra vector contribution cancels the SM one, worsening the fit.
If two interactions are simultaneously present, see lower panels, one can also expect interferences between the NGI and the SM interactions, as well as between different NGI.
For the scalar versus tensor (S-T) case, the allowed region, indicated in the left panel, is centered around the SM solution, with $C_T^q=C_S^q = 0$,
with no other solutions allowed at the same confidence level.
 In the other two cases, however, degenerate solutions with similar goodness of fit as the SM solution appear.
 In particular, in the scalar versus vector (S-V) case, shown in the middle panel, the combination of the two data sets reduces the ring-shaped contour obtained with LAr data to two smaller disjoint
 regions, one centered around the SM and a second one around $C_V^q \sim 0.18$, as already seen for the one-dimensional profiles.
Similarly, for the vector versus tensor (V-T) case, the combined analysis of LAr and CsI data results in a ring-shaped allowed region including the SM solution $C_V^q=C_T^q = 0$.
In all cases, the addition of the CsI data substantially improves the sensitivity to these new interactions and is competitive to existing results~\cite{Majumdar:2022nby}, extracted recently from the analysis of Dresden-II data~\cite{Colaresi:2022obx}. 
Furthermore, one sees a tiny part of the parameter space where the combined contour extends outside the CsI allowed region. 
This can be understood from the LAr and CsI $\chi^2$ profiles in the upper panels of Fig.~\ref{fig:NGI-CsI},
which show how in some cases the best fit values for the couplings in the LAr analysis lie outside the region preferred by the CsI data.

The $1 \sigma$ limits obtained from the combined CsI+LAr analysis on vector, scalar and tensor NGI parameters (setting all the others to zero) read
\begin{equation}
\begin{aligned}
C_V^{q} =& [-0.003, 0.003] \cup[0.18, 0.19] \, ,\\
C_S^{q} =& [-0.009, 0.009] \, ,\\
C_T^{q} =& [-0.19, 0.19]\, .
\end{aligned}
\end{equation}

\subsection{Light mediators}  
\label{sec:lightmed}

The NGI discussed above may appear mediated by a light particle. In such a case, the relevant parameters entering the scattering cross sections given
in Sec.~\ref{sec:NGIxsec} are the mass of the mediator, $m_X$, and a coupling $g_X$, which, for the sake of simplicity, we define as $g_X = \sqrt{g_{\nu X} g_{f X}}$, with
$f = \{ u, d\}$ for CE$\nu$NS and $f = e$ for ES. 
Notice that we are assuming universal couplings to quarks, i.e. $g_{uX}=g_{dX} \equiv g_{qX}$. 

\begin{figure}[t]
\includegraphics[width= 0.48 \textwidth]{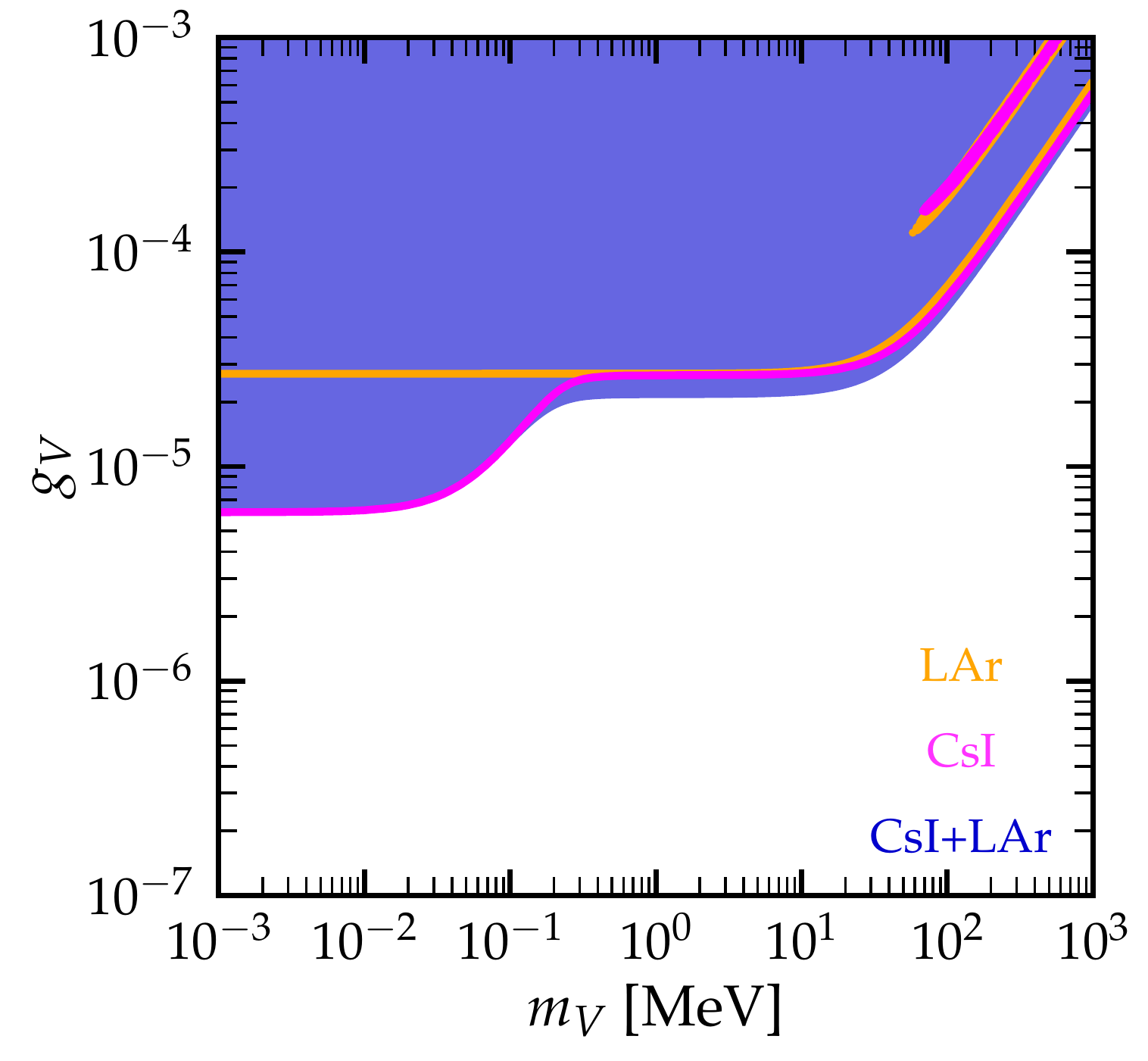}
\includegraphics[width= 0.48 \textwidth]{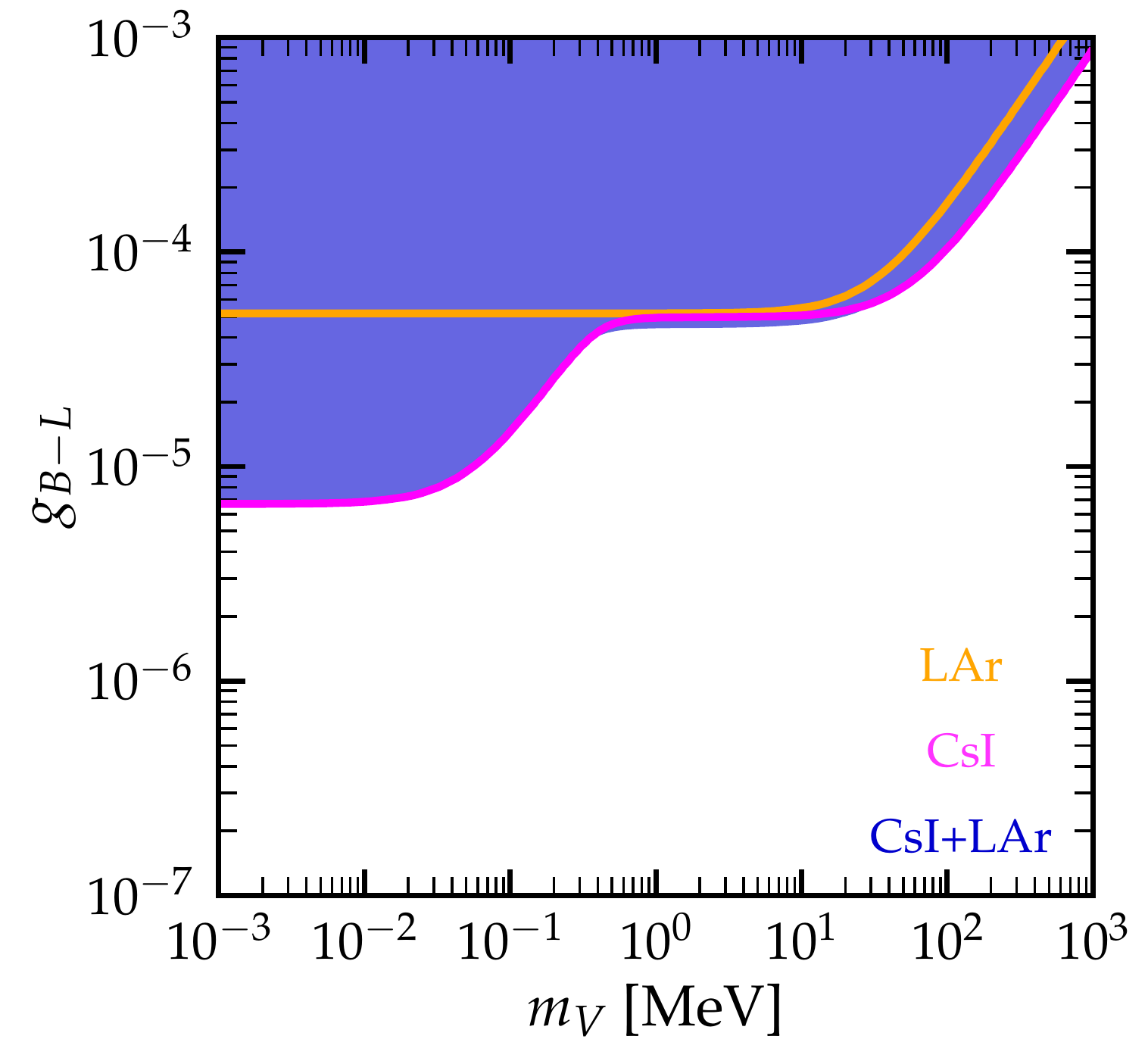}
\caption{ $90 \%$ C.L. (2 d.o.f.) exclusion regions from the analysis of CsI (magenta), LAr (orange) and CsI + LAr (blue) data,
  for the universal light vector model (left) and the B-L scenario (right). The analysis of CsI data includes CE$\nu$NS + ES interactions. }
\label{fig:lightV}
\end{figure}

\begin{figure}[h]
\includegraphics[width= 0.48 \textwidth]{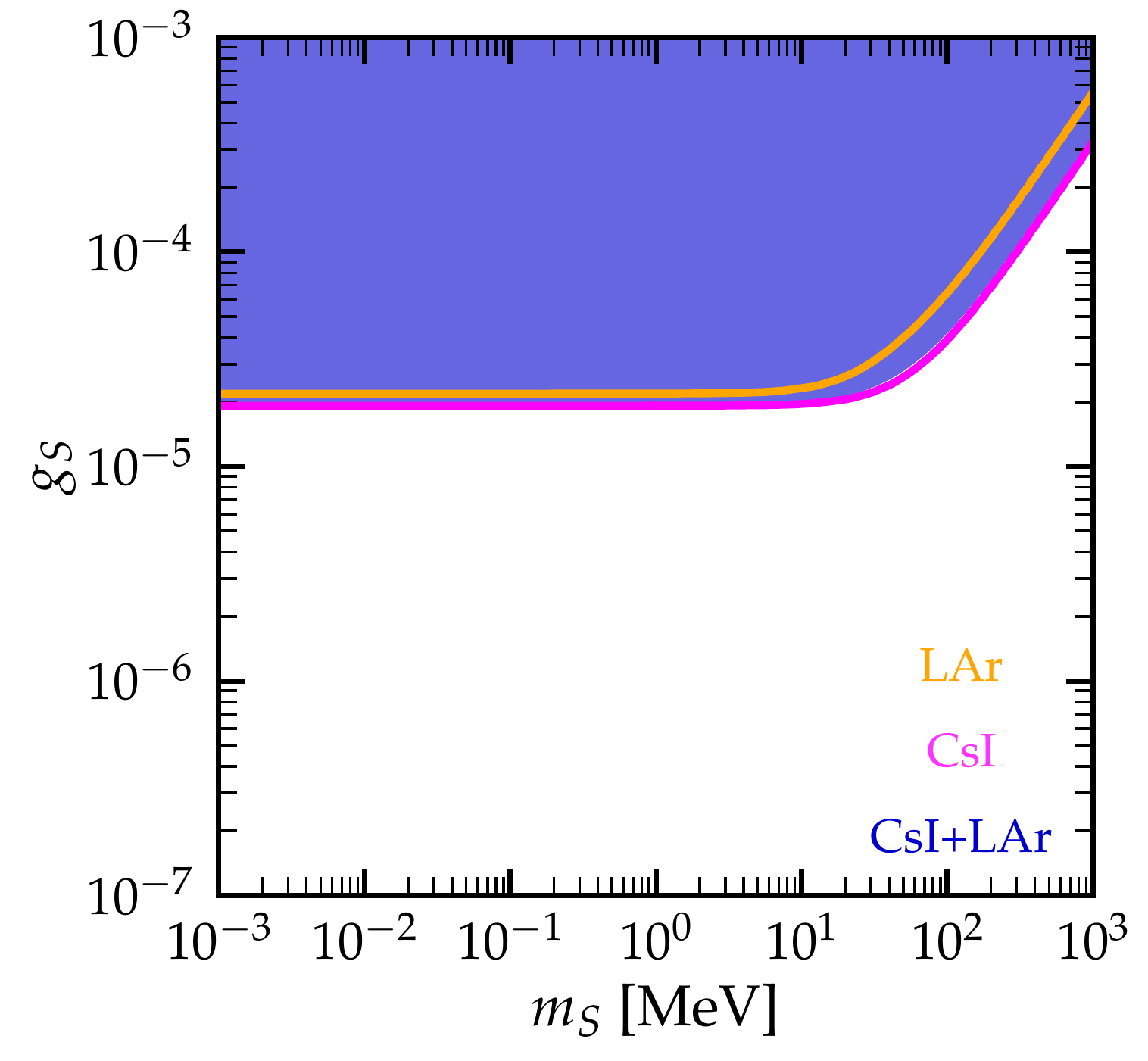}
\includegraphics[width= 0.48 \textwidth]{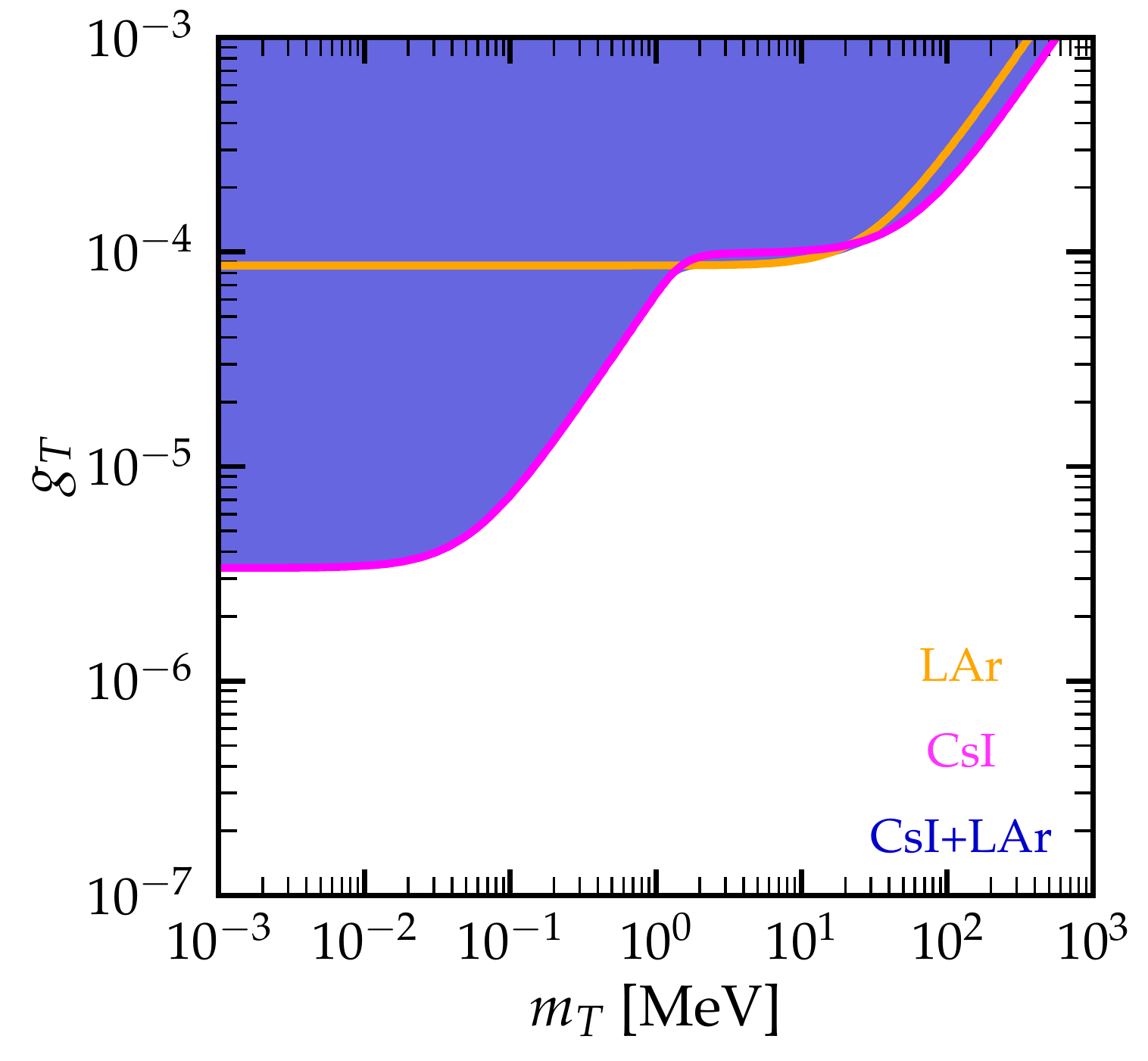}
\caption{ $90 \%$ C.L. (2 d.o.f.) exclusion regions from the analysis of CsI (magenta), LAr (orange) and CsI + LAr (blue) data,
  for the universal light scalar model (left) and the universal light tensor model (right). The analysis of CsI data includes CE$\nu$NS + ES interactions.   }
\label{fig:lightST}
\end{figure}

 We first focus on light vector mediators and we present in Fig.~\ref{fig:lightV} the $90 \%$ C.L. (2 d.o.f.) exclusion regions for the universal light vector model (left) and the B-L scenario (right). 
 In the universal model, a cancellation with the SM is possible thus leading to a tiny unconstrained band at $m_V \sim 0.05-1$ GeV and $g_V \gtrsim 10^{-4}$. 
In the B-L scenario, by contrast, the charge assignment does not allow for this destructive interference. 
We can also notice that the addition of CsI ES events significantly improves the bounds by a factor $\sim 8$, at $m_V \lesssim 30$ MeV. 
The corresponding results for the light scalar and tensor interactions are given in Fig.~\ref{fig:lightST}. The conclusions are similar to the vector case, although without the ES-driven improvement in the scalar scenario. This behavior is understood since, in the scalar case, the cross section is proportional to  $\sim 1/E_\text{er}$, unlike the vector and tensor cases where the ES enhancement is more significant as the cross section of this process is  proportional to $\sim 1/E^2_\text{er}$.
Let us finally comment that, keeping in mind the differences in the data sets considered and in the statistical analyses performed, our current bounds are in general agreement with those given
  previously, e.g. in Refs.~\cite{Cadeddu:2020nbr,AtzoriCorona:2022moj,Coloma:2022avw}.

\begin{figure}[!htb]
\includegraphics[width=  0.48 \textwidth]{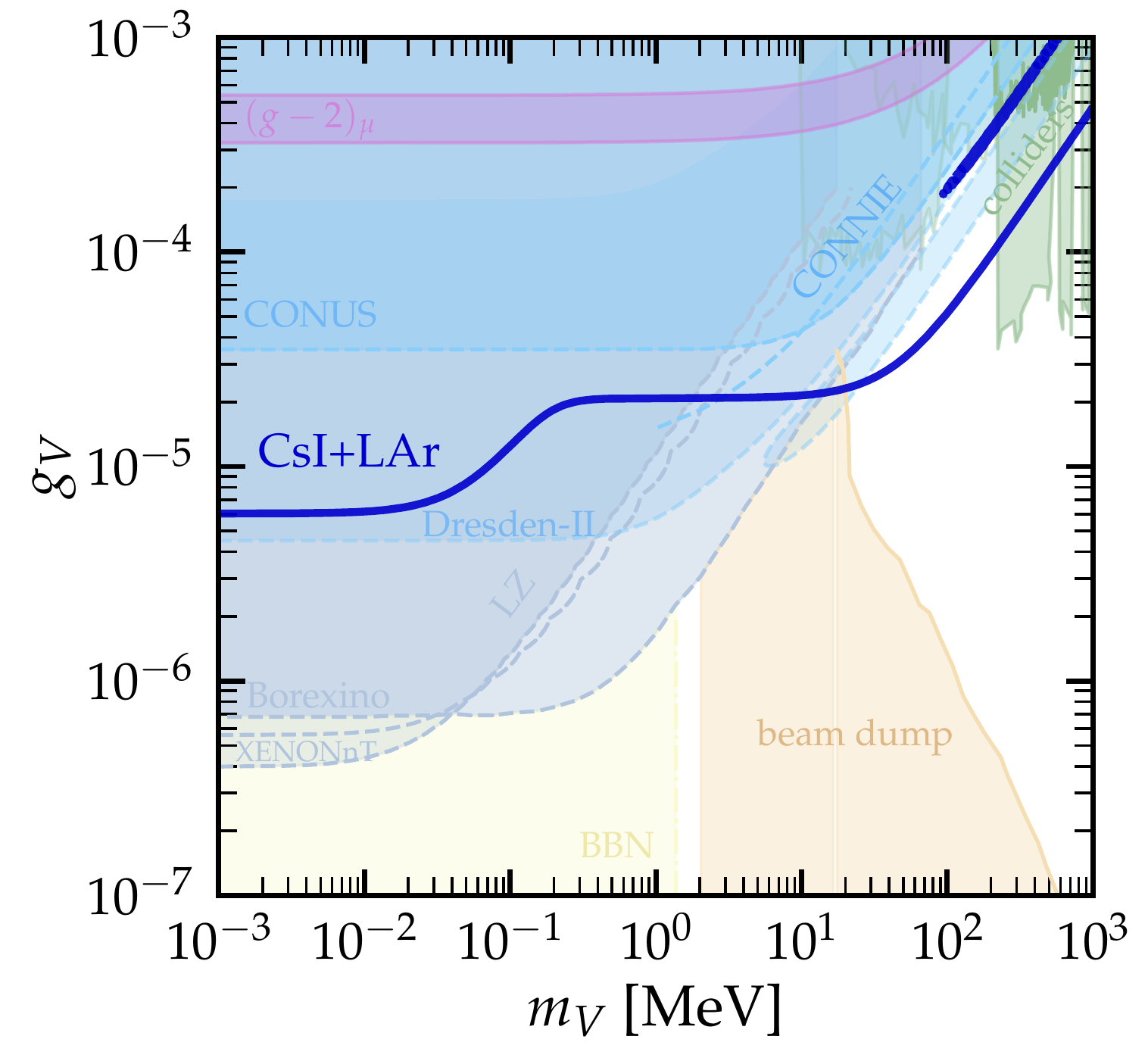}
\includegraphics[width=  0.48 \textwidth]{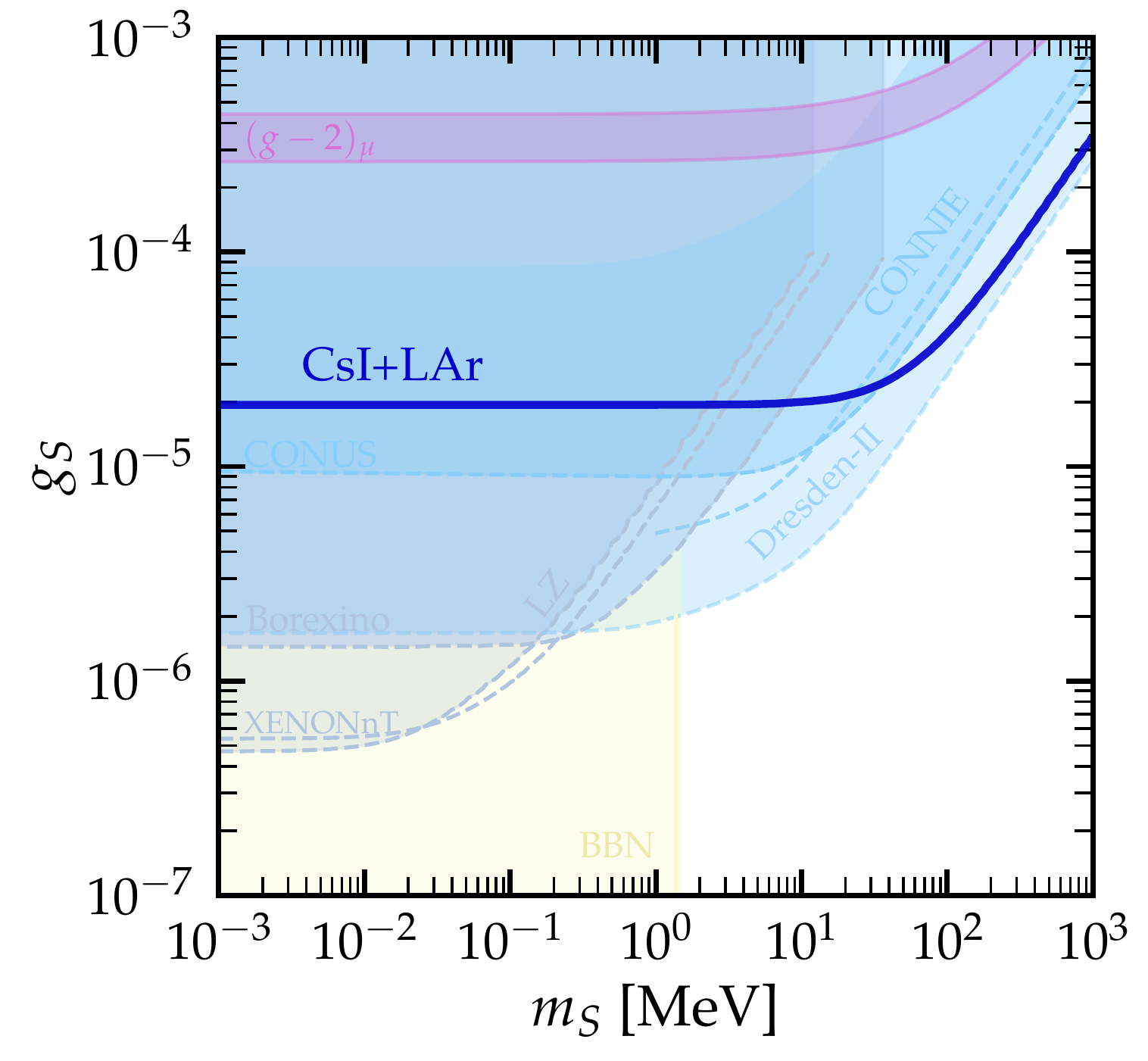}
\caption{
    90\% C.L. exclusion regions from our combined CsI+LAr analysis of vector (left) and scalar (right) light mediators coupled universally to neutrinos and quarks,
    compared with other available experimental constraints. See text for more details.}
\label{fig:Light_med_compare}
\end{figure}

In order to make a comparison with the available constraints from other experimental probes, we reproduce in Fig.~\ref{fig:Light_med_compare} the exclusion curves from our combined analysis (dark blue line) for the cases of the universal vector mediator (see left panel of Fig.~\ref{fig:lightV}) and the universal scalar mediator (see left panel of Fig.~\ref{fig:lightST}).
  We choose  the latter as benchmark scenarios  since they can be easily recast into more dedicated theoretical models, where neutrinos and quarks couple differently to the new light mediator.
  We compare our results with the existing limits from other \cevns~experiments, in particular CONNIE~\cite{CONNIE:2019xid}, CONUS~\cite{CONUS:2021dwh} and Dresden-II~\cite{AristizabalSierra:2022axl},
  from the analysis given in Ref.~\cite{A:2022acy} of two multi-ton DM experiments (namely XENONnT~\cite{XENONCollaboration:2022kmb} and LZ~\cite{LZ:2022ufs}),
  from the analysis of solar neutrino data collected by Borexino~\cite{Coloma:2022umy}, from collider experiments~\cite{AtzoriCorona:2022moj}, including BaBar~\cite{BaBar:2014zli} and
LHCb~\cite{LHCb:2017trq}, and from rare meson decays at NA48~\cite{NA482:2015wmo}.
In the $\sim 1$ GeV mass region, we also show bounds from beam-dump experiments~\cite{Bauer:2018onh,Cadeddu:2020nbr}, including
E141~\cite{Riordan:1987aw,Bjorken:2009mm},
E137~\cite{Bjorken:1988as},
E774~\cite{Bross:1989mp},
KEK~\cite{Konaka:1986cb},
Orsay~\cite{Andreas:2012mt},
U70/$\nu$-CAL~I~\cite{Blumlein:2011mv,Blumlein:2013cua},
CHARM~\cite{CHARM:1985anb,Gninenko:2012eq},
NOMAD~\cite{NOMAD:2001eyx}, NA64~\cite{NA64:2016oww,NA64:2019auh} and the 
fixed target experiments
A1~\cite{Merkel:2014avp} and APEX~\cite{APEX:2011dww}.
In the low mass regime, we show the bound from Big Bang Nucleosynthesis~\cite{Blinov:2019gcj,Suliga:2020jfa} (BBN) obtained by requiring that the nonstandard mediator couples to neutrinos only.
Finally, we also indicate the $2\sigma$ preferred region to account for the muon anomalous magnetic moment $(g-2)_\mu$~ \cite{AtzoriCorona:2022moj}.
While the COHERENT constraints  obtained here slightly improve  the existing bounds only in a small mass region around $m_X \sim 0.1$ GeV, they nonetheless constitute a complementary test of these
new physics scenarios.

\subsection{Neutrino EM properties}
\label{sec:magnetic}

\subsubsection{Effective neutrino magnetic moment}

\begin{figure}[t]
\includegraphics[width=  \textwidth]{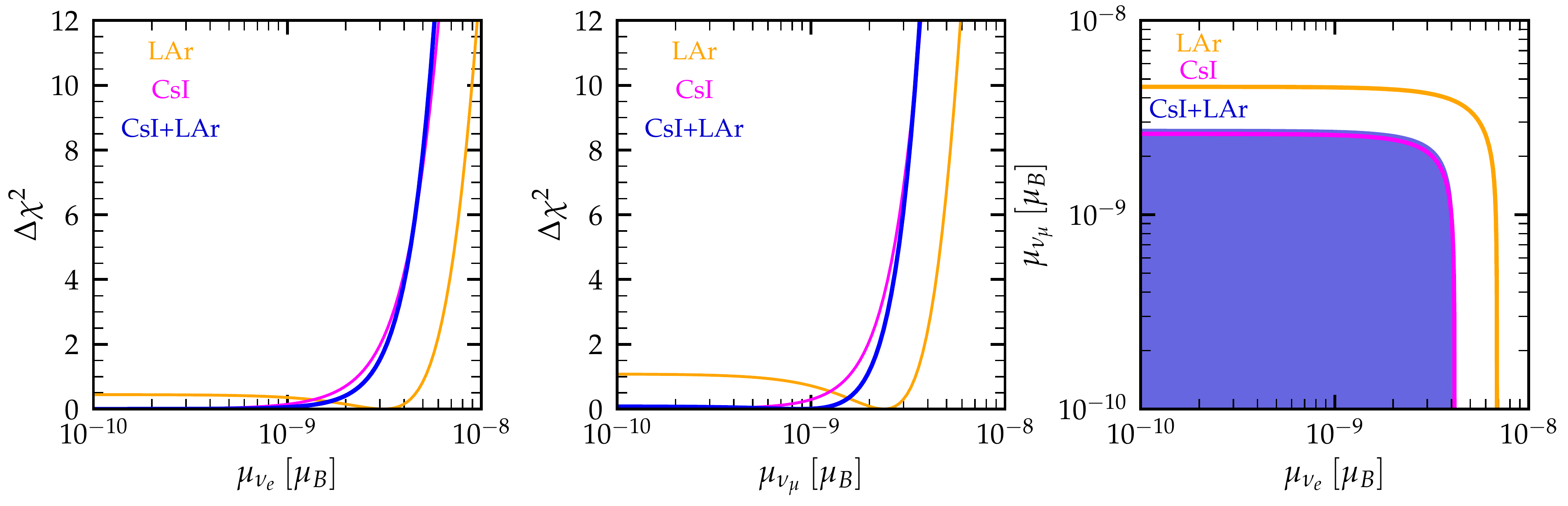}
\caption{Left and middle panels:
  $\Delta\chi^2$ profile for the effective neutrino magnetic moments $\mu_{\nu_{e}}$ and $\mu_{\nu_{\mu}}$ obtained from the analysis of CsI (magenta), LAr (orange) and CsI + LAr (blue) data.
Right panel: $90 \%$ C.L. allowed regions (2 d.o.f.) when two effective MM are taken simultaneously. The analysis of CsI data includes CE$\nu$NS + ES interactions.}
\label{fig:magnetic-CsI}
\end{figure}

The constraints on neutrino effective MM $\mu_{\nu_{e}}$ and $\mu_{\nu_{\mu}}$ are summarised in Fig.~\ref{fig:magnetic-CsI}.
In the left (middle) panel, we present the $\Delta \chi^2$ profiles as a function of the effective electron neutrino (muon neutrino) MM.
We assume that only one MM is nonzero at a time, and we include ES events in the CsI analysis.
One finds an improvement of a factor $\sim 2$ with the recent CsI data in comparison to the LAr data set.
From the combined analysis of CsI+LAr data, at 90\% C.L. we find the upper limits
\begin{equation}
\begin{aligned}
     \mu_{\nu_{e}}  <~ & 3.6~(3.8) \times 10^{-9}~\mu_B \, , \\ 
   \mu_{\nu_{\mu }} <~ & 2.4~(2.6) \times 10^{-9}~\mu_B \, ,
    \end{aligned}
\end{equation}
where the limits in parenthesis indicate the results from the CE$\nu$NS-only analysis.  
We note that the inclusion of ES events improves the constraints only slightly, {in agreement with Ref.~\cite{AtzoriCorona:2022qrf} where a similar analysis was presented.
Likewise, only a small improvement is found with respect to the results obtained in previous analyses of COHERENT-CsI (2017) data in Refs.~\cite{Papoulias:2017qdn, Cadeddu:2020lky}.
This comes from the fact that, even though the 2021 dataset  increased the statistics considerably, the  threshold of the COHERENT-CsI detector remained unchanged for the two measurements.}
Indeed, due to their larger threshold, \cevns~ experiments cannot compete with the very low-threshold dark matter direct detection experiments,
for which the sensitivity reach on the neutrino magnetic moment is in the few $\times 10^{-12}~\mathrm{\mu_B}$ ballpark~\cite{A:2022acy}.
We have also performed a combined analysis, allowing both effective MM to vary simultaneously.
The corresponding result is presented in the right panel of Fig.~\ref{fig:magnetic-CsI}.
Similarly to the NGI case discussed previously, there is a tiny part of the region allowed in the combined analysis which falls outside the CsI-driven contour.
Again, this is due to the fact that the analysis of LAr data leads to a nonzero best fit coupling, in contrast to the CsI data (see upper panel).

\subsubsection{Neutrino charge radius} 
\label{sec:chargerad}

\begin{figure}[t]
\includegraphics[width= \textwidth]{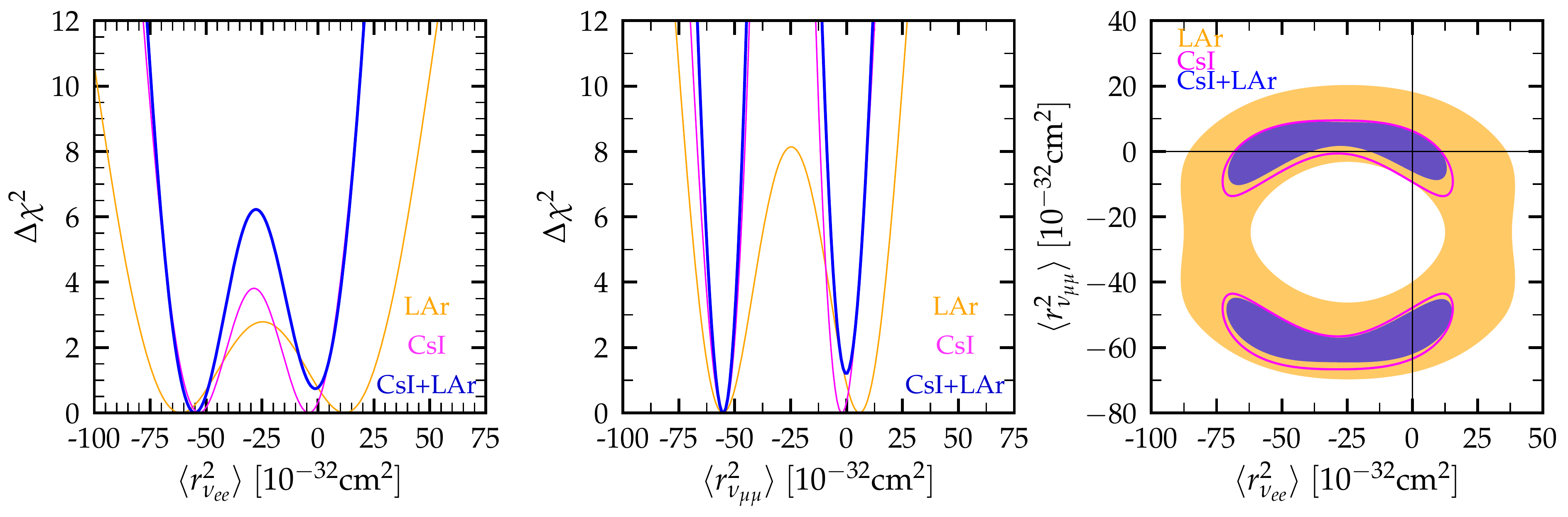}
\caption{Left and middle panels: $\Delta\chi^2$ profile
for the neutrino charge radii obtained from the analysis of CsI (magenta), LAr (orange) and CsI + LAr (blue) data.
Right panel: $90\%$ C.L. (2 d.o.f.) allowed regions for the neutrino charge radii, in units of $10^{-32}~\text{cm}^2$.   The analysis of CsI data includes only CE$\nu$NS interactions.} 
\label{fig:charge_radius-CsI}
\end{figure}

Bounds on flavor-diagonal neutrino CR are shown in Fig.~\ref{fig:charge_radius-CsI}. The left and middle panels show the $\Delta\chi^2$ profile for
$\langle r^2_{\nu_{ee}}\rangle$ and $\langle r^2_{\nu_{\mu \mu}}\rangle$ from the analysis of CsI (magenta), LAr (orange) and CsI + LAr (blue) data,
in units of $10^{-32}~\text{cm}^2$. We find the following $1\sigma$ allowed regions: 
\begin{equation}
\begin{aligned}
    \langle r^2_{\nu_{ee}} \rangle \in & \, [-61.2, -48.2] \cup [-4.7, 2.2] \times 10^{-32}~\mathrm{cm^2} \, , \\
   \langle r^2_{\nu_{\mu \mu }} \rangle \in & \, [-58.2, -52.1] \times 10^{-32}~\mathrm{cm^2}\,.
    \end{aligned}
\end{equation}
While our results are less stringent, they are still in reasonable agreement with those reported in Ref.~\cite{AtzoriCorona:2022qrf}. 
The agreement is not perfect due to our different analysis strategy, in which the efficiency and time uncertainties are taken into account in a more systematic manner.
The right panel of Fig.~\ref{fig:charge_radius-CsI} summarizes the $90 \%$ C.L. (2 d.o,f.) allowed contours in the  $\langle r^2_{\nu_{ee}}\rangle$-$\langle r^2_{\nu_{\mu \mu}}\rangle$ plane.
Note that the analysis of CsI data includes only CE$\nu$NS interactions, since the ES event contribution on CsI is negligible. 
From a closer inspection of this figure, one sees a substantial improvement due to the CsI data:
the LAr data alone lead to a single region, while the CsI data result in a more constrained allowed area, with two separate regions.  
We should also stress the fact that our CsI result differs from the one obtained in Ref.~\cite{AtzoriCorona:2022qrf},
where the authors obtained four distinct contour islands at 90\%~C.L. 
 Let us finally note that the current CE$\nu$NS limits on the neutrino charge radius are weaker than the existing limits on $\langle r^2_{\nu_e} \rangle$ from TEXONO~\cite{TEXONO:2009knm} and
  $\langle r^2_{\nu_\mu} \rangle$ from BNL-E734~\cite{Ahrens:1990fp} by about one order of magnitude.  

\subsubsection{Neutrino millicharge}
\label{sec:milich}

\begin{figure}[t]
\includegraphics[width= \textwidth]{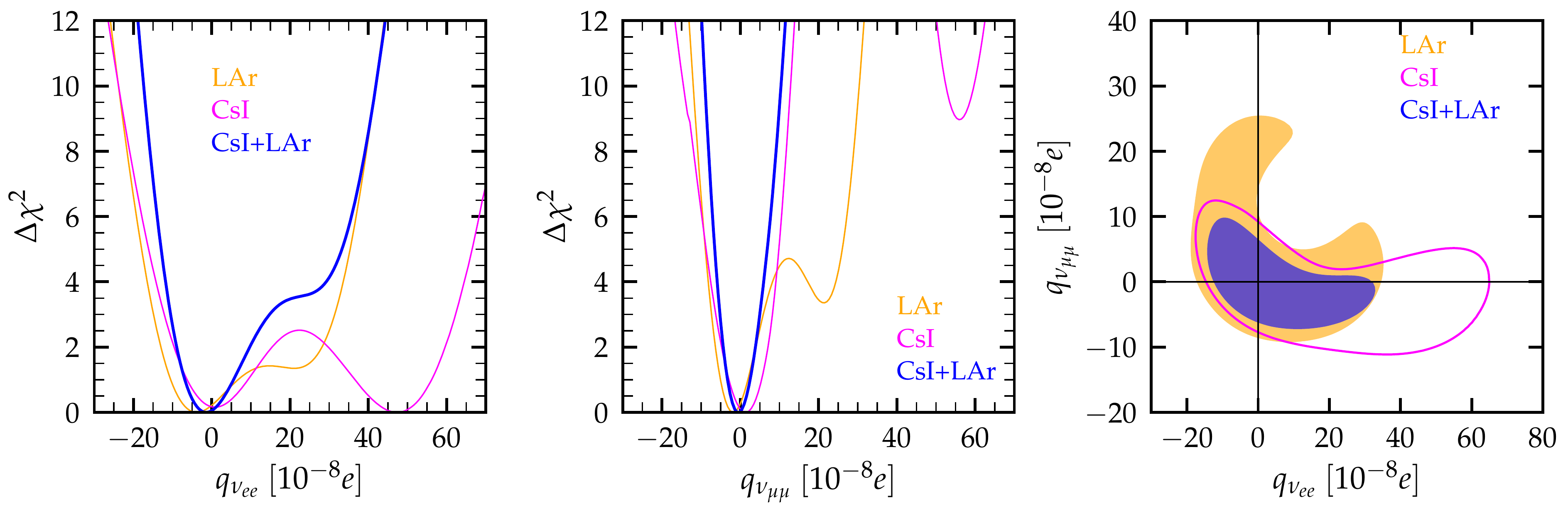}
\includegraphics[width= \textwidth]{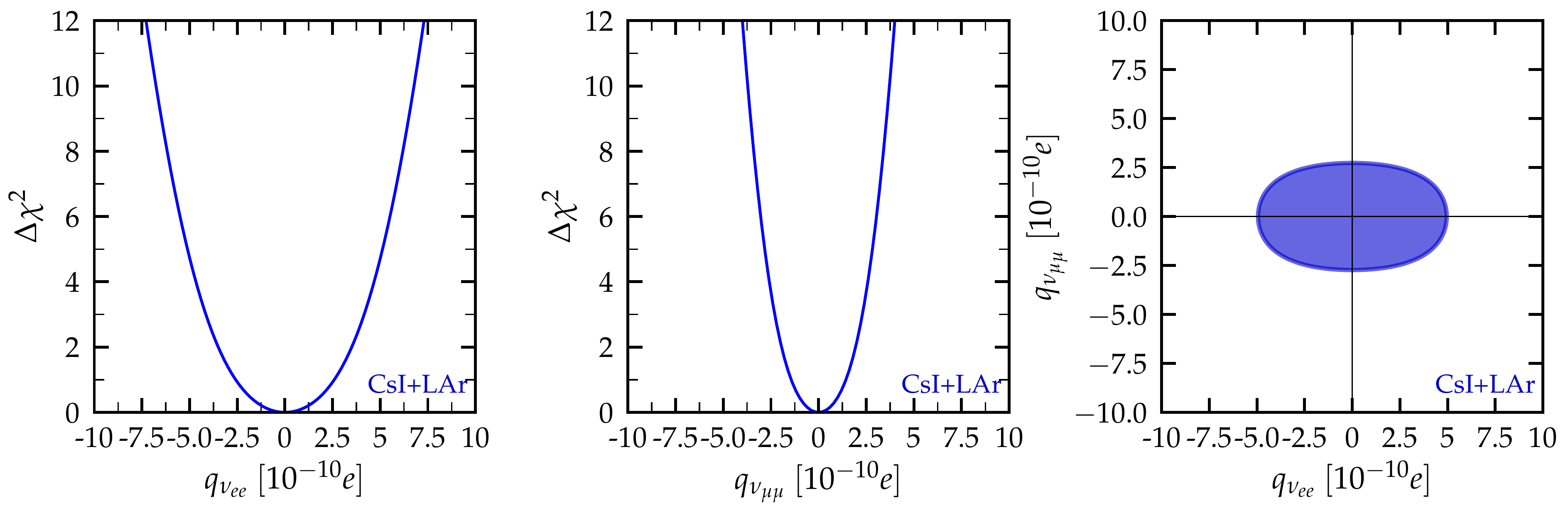}
\caption{Left and middle panels: $\Delta\chi^2$ profile
for the neutrino electric charges obtained from the analysis of CsI (magenta), LAr (orange) and CsI + LAr (blue) data.
Right panels: $90 \%$ C.L. (2 dofs) allowed regions for the neutrino electric charges, in units of the elementary charge $e$. In the upper (lower) panels the analysis of CsI data includes only CE$\nu$NS (CE$\nu$NS+ES) interactions. In the lower panels, we do not show LAr contours as the combined analysis is driven by CsI data.}
\label{fig:milicharge-CsI}
\end{figure}

The $\Delta\chi^2$ profile for the neutrino EC parameters $q_{\nu_{e e}}$ and $q_{\nu_{\mu \mu}}$ is given in the upper-left and upper-middle plots of Fig.~\ref{fig:milicharge-CsI}. 
The upper-right plot shows the 90\% C.L. (2 d.o.f.) allowed regions in the plane ($q_{\nu_{e e}}$, $q_{\nu_{\mu \mu}}$), when both parameters are allowed to vary simultaneously.
From the combined analysis of CsI+LAr data, we find the following $1\sigma$ allowed intervals
\begin{equation}
\begin{aligned}
     q_{\nu_{ee}} \in &  \, [-6.9,~5.6]\times 10^{-8}~e  \, , \\
    q_{\nu_{\mu \mu }} \in & \, [-3.3,~2.5]\times 10^{-8}~e \, .
    \end{aligned}
\end{equation}
Note that this result is obtained considering only \cevns ~interactions in CsI data, neglecting the ES contribution.
We find appreciable differences when comparing our results with those obtained in Ref.~\cite{Khan:2022not}, 
mainly due to the fact that the latter did not take into account the timing information in COHERENT data.
Indeed, in our analysis we obtain two minima in the $\Delta \chi^2$ profile of $q_{\nu_{ee}}$. 
In general, our present results are in good agreement with those of Ref.~\cite{AtzoriCorona:2022moj},
the slight differences being due to the different analysis methods, as discussed above.  

Note that, since in ES the momentum transferred is much smaller than in \cevns ~interactions,
the inclusion of ES events in the analysis strongly enhances the sensitivity of COHERENT data to neutrino EC.
This is shown in the lower panels of Fig.~\ref{fig:milicharge-CsI}, where a dramatic improvement of two orders or magnitude is gained when including ES events. 
In the latter case, the constraints are completely dominated by ES-induced events in CsI and as a consequence, the combined analysis is totally driven by CsI data.
We find the following $1\sigma$ allowed ranges:
\begin{equation}
\begin{aligned}
     q_{\nu_{ee}} \in \, & [-2.6,~2.6]\times 10^{-10}~e  \, ,\\
    q_{\nu_{\mu \mu }} \in \, & [-1.4,~1.4]\times 10^{-10}~e \, .
    \end{aligned}
\end{equation}
 As in the case of the effective neutrino magnetic moment, the present limits are less severe compared to the currently most stringent constraints derived from the analysis of
  LZ and XENONnT data in Ref.~\cite{A:2022acy} by two orders of magnitude (for reactor-based experiments see Ref.~\cite{Gninenko:2006fi}).

\subsection{Conversion to sterile neutrinos} 
\label{sec:vsterile_results}

\subsubsection{Active-to-sterile neutrino oscillations}

\begin{figure}[t]
\includegraphics[width= 0.48 \textwidth]{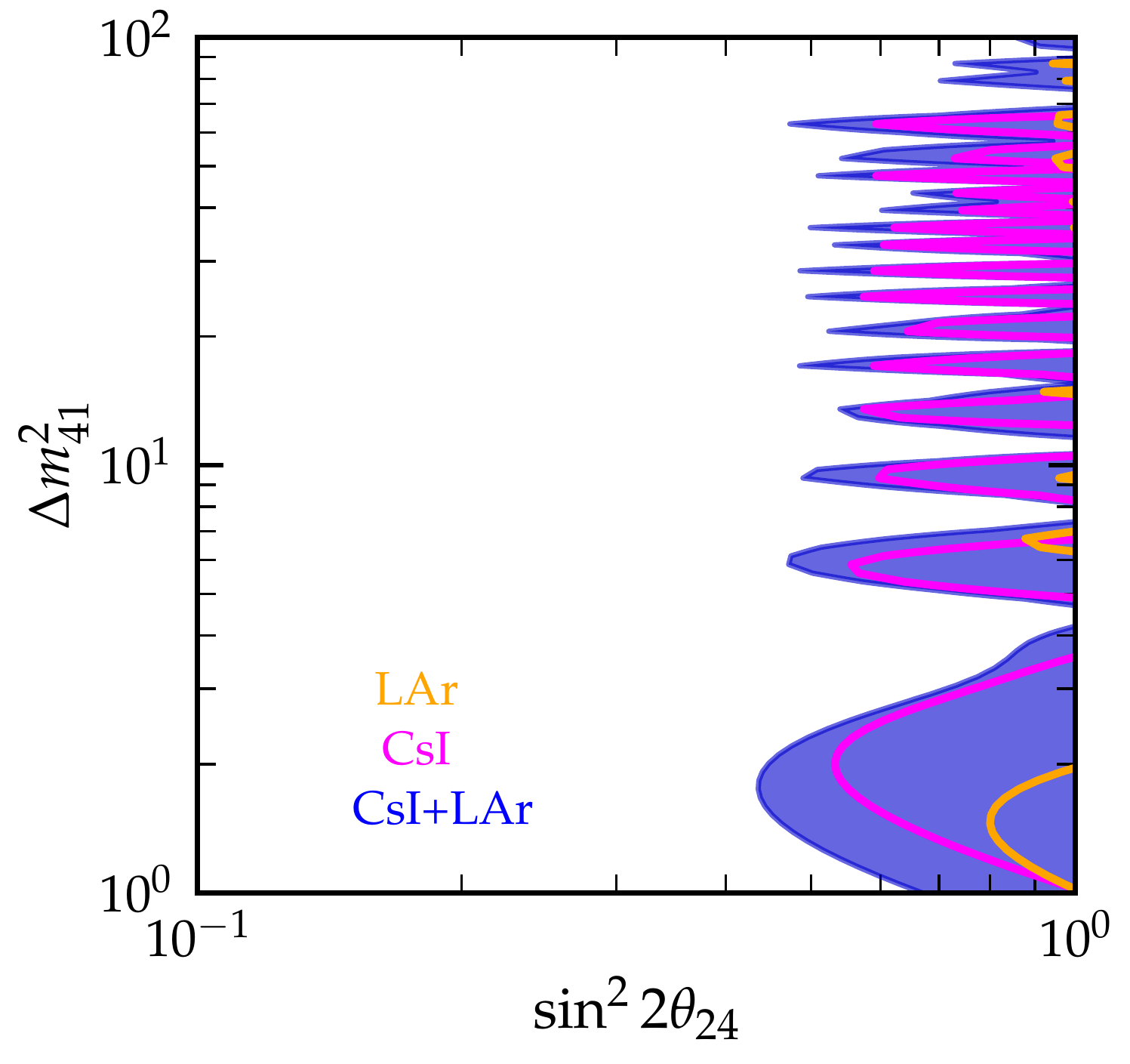}
\caption{ $90 \%$ C.L. (2 d.o.f.) exclusion regions from the analysis of CsI (magenta), LAr (orange) and CsI + LAr (blue) data, in the ($\Delta m^2_{41} $, $\sin^2 2\theta_{24}$) plane.
}
\label{fig:sterilev}
\end{figure}

Neutrino oscillations involving a sterile state may be studied with neutrinos produced at a Spallation Neutron Source.
Given the small electron neutrino component in the SNS flux, one does not expect any meaningful sensitivity in the ($\Delta m^2_{41}$, $\sin^2 2\theta_{14}$) plane. 
However, muon neutrino oscillations to a sterile state can be constrained with COHERENT data. 
In Fig.~\ref{fig:sterilev} we show the $90 \%$ C.L. (2 d.o.f.) exclusion regions from the analysis of CsI (magenta), LAr (orange) and CsI + LAr (blue) data.
Our analysis of CsI data includes only CE$\nu$NS interactions.  Concerning the sterile neutrino hypothesis we find a slightly improved fit for the CsI data when compared to the SM, while for the case of LAr it leads to a poorer result,  very far from the promising prospects for this scenario expected from the SBN program or the upcoming reactor SBL experiments~\cite{Machado:2019oxb}.

\subsubsection{Active-to-sterile EM interactions}  

\begin{figure}[!htb]
\includegraphics[width=  0.95\textwidth]{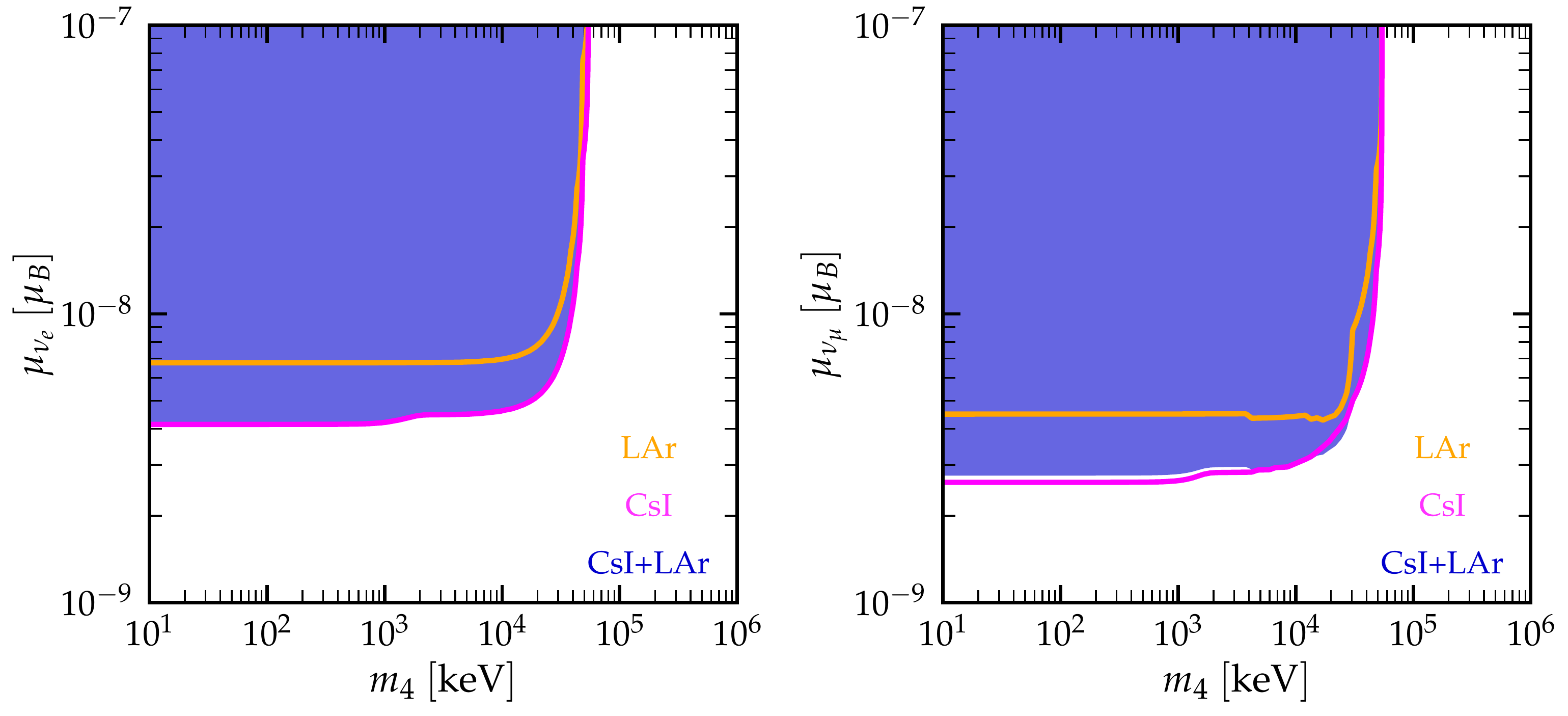}
\caption{ $90 \%$ C.L. (2 d.o.f.) exclusion regions for the neutrino magnetic moments concerning active-to-sterile transitions, from the analysis of CsI (magenta), LAr (orange) and CsI + LAr (blue) data.
The analysis of CsI data includes CE$\nu$NS + ES interactions.}
\label{fig:dipole-CsI}
\end{figure}

We now consider active-sterile transitions in the presence of a nonzero neutrino transition magnetic moment.
The corresponding results are shown in Fig.~\ref{fig:dipole-CsI}.  
Given the nuclear recoil and neutrino energies typical of the COHERENT experiment, the maximum sensitivity to the sterile state mass is about $50$ MeV due to kinematics. 
Similarly to the case of active-active MM, ES events contribute sizeably to the \cevns ~ event rate, and their inclusion in the analysis leads to a slight improvement in the resulting constraints,  reflected in a small kink around $\sim 2~\mathrm{MeV}$. 
From the combined analysis of the full COHERENT data, the 90\% C.L. contours shown in Fig.~\ref{fig:dipole-CsI} exclude values of $\mu_{\nu_{e (\mu)}}$  as low as $4 (3)\times 10^{-8} \mu_B$ for $m_4 \lesssim 10$ MeV. 
At this point, we would like to highlight the complementarity of the COHERENT measurements with existing \cevns~reactor experiments.
Although the present results for the case of electron neutrinos are slightly less stringent than those obtained from the analysis of Dresden-II data~\cite{AristizabalSierra:2022axl}, the COHERENT data analyzed here can probe larger values of $m_4$. One should also stress that COHERENT data can be used to probe TMMs to sterile neutrinos in the muon sector, unlike reactor-based experiments.

\section{Summary and conclusions} 
\label{sec:conclusions}

In this paper we have analyzed the updated CsI data release from the COHERENT experiment and combined this result with the previous LAr dataset. 
By performing a thorough statistical analysis including all systematic errors plus the relevant nuisance parameters associated to signal shape uncertainties,
we have probed Standard Model parameters such as the weak mixing angle, nuclear physics as well as several new physics scenarios. We have shown that the inclusion
of the recent CsI data significantly  improves the \cevns~sensitivity in most of these physics cases.\\[-.4cm]

In order to compare the various scenarios examined above with respect to SM expectations we now give a summary of our results.
In this paper we have analyzed the impact of recent CsI+LAr COHERENT data on:
\begin{itemize}
[leftmargin=*,topsep=0pt]
  \setlength{\itemsep}{1pt}
  \setlength{\parskip}{0pt}
  \setlength{\parsep}{0pt}
\item
     Standard Model physics (weak mixing angle in Fig.~\ref{fig:thweak-CsI} and neutron RMS radii in Fig.~\ref{fig:Rn})
\item flavor-preserving neutrino NSI (Fig.~\ref{fig:NSI-CsI_ee} and Fig.~\ref{fig:NSI-CsI_ee_mumu})
\item co-existence of nonuniversal and flavor-changing neutrino NSI (Fig.~\ref{fig:NSI-CsI_FC-a} and Fig.~\ref{fig:NSI-CsI_FC-b})
 \item  neutrino generalized interactions  (Fig.~\ref{fig:NGI-CsI})
  \item light mediators (\Cref{fig:lightV,fig:lightST,fig:Light_med_compare})
  \item effective neutrino magnetic moments (Fig.~\ref{fig:magnetic-CsI})
   \item  neutrino charge radii (Fig.~\ref{fig:charge_radius-CsI})
    \item neutrino millicharges (Fig.~\ref{fig:milicharge-CsI})
    \item oscillations into sterile states (Fig.~\ref{fig:sterilev}) 
   \item TMMs to sterile neutrinos (Fig.~\ref{fig:dipole-CsI}).
   \end{itemize}
     
In Table~\ref{tab:summary} we present the $\chi^2_\text{min}$ values obtained in each physics case, normalized to the total number of degrees of freedom~\footnote{These are calculated as $\#_{\rm{dof}}^{\rm{CsI+LAr}} = \#_{\rm{CsI}} + \#_{\rm{LAr}} - \#_{\rm{free~pars}}$, where the number of degrees of freedom for CsI data is $\#_{\rm{CsI}} = 99$ (data bins) $-1$ (time) $=98$, while $\#_{\rm{LAr}} =120$. }.
At this point, we would like to remark that both CsI and LAr datasets are in good agreement with the SM expectation, with $ \chi^2/\#_{\rm{dof}}^{\rm{CsI}} = 0.849$ and $ \chi^2/\#_{\rm{dof}}^{\rm{LAr}} = 0.887$, which implies a preference for the presence of \cevns\ as given by the SM over the background hypothesis of $11.5 \sigma$ and $3.4 \sigma$, respectively (for the background-only hypothesis one finds $ \chi^2/\#_{\rm{dof}}^{\rm{CsI}} = 2.19$ and $ \chi^2/\#_{\rm{dof}}^{\rm{LAr}} = 0.984$).
We also find that the combined CsI+LAr analysis leads to $\chi^2/\#_{\rm{dof}}^{\rm{CsI+LAr}} = 0.870$, favoring the SM over the background explanation at almost $12 \sigma$.

  Regarding the new physics scenarios considered in this work, for the case of CsI, as well as for the combined CsI+LAr analysis, only the addition of a light vector mediator tends to improve the fit,
  while this is not the case for the LAr data alone.
  In contrast, the presence of a NGI with nonzero $C_S^q$ couplings leads to the best fit for the LAr dataset. 
  NGI with nonzero $C_V^q$ as well as a light vector mediator with universal couplings also fit well the combined datasets CsI+LAr,
  slightly improving the quality of the fit (by $\sim 1 \sigma$) in comparison with the SM.
  On the other hand, active-active TMM, sterile conversion through TMM, NSI, as well as the NGI analysis with nonzero $C_S^q$ and $C_T^q$ couplings, all lead to poorer fits than the SM.
  Finally, a scenario in which muon neutrinos oscillate into a light sterile state leads to the worst fit, both analyzing the two datasets separately and combined.
  Indeed, the addition of the new CsI data disfavor the sterile neutrino oscillations scenario. 
  All in all, in this paper we have provided a survey of the main physics implications of a combined analysis of COHERENT CsI and LAr data.

\begin{table}[!ht]
\resizebox{ 0.75 \textwidth}{!}{%
{\renewcommand{\arraystretch}{1.2}
\setlength{\tabcolsep}{0.3mm}
\begin{tabular}{|l|c|c|c|c|}
\specialrule{.2em}{.1em}{.1em}
 \rowcolor{gainsboro!60} \textbf{Scenario}   & \textbf{SM}     & \textbf{\begin{tabular}[c]{@{}c@{}}weak mixing angle\\ ($\sin^2 \theta_{W}$)\end{tabular}}                                                 &  \textbf{\begin{tabular}[c]{@{}c@{}}nuclear neutron \\   radius ($R_n$)\end{tabular}}    & \textbf{\begin{tabular}[c]{@{}c@{}}$\boldsymbol{\mathrm{MM_{active}}}$\\ ($\mu_{\nu_e}, \mu_{\nu_\mu}$)\end{tabular}}                                                                                                                                  \\ \specialrule{.1em}{.05em}{.05em} 
{\color{magenta} \textbf{CsI } }
& 83.2  (0.849)                                                                                                   & 82.8  (0.854)                                                                                                          & 81.9  (0.845)                                                                                                       & 83.2  (0.867)                                                                                                           \\ \hline
{\color{darktangerine}\textbf{LAr}} 
& 106.5 (0.887)                                                                                                  & 105.5 (0.887)                                                                                                         & 105.5 (0.887)                                                                                                     & 105.4 (0.893)                                                                                                          \\\hline
{\color{ceruleanblue}\textbf{CsI+LAr }} 
& 189.7 (0.870)                                                                                                    & 189.7 (0.874)                                                                                                      & ---                                                                                                 & 189.6 (0.877)                                                                                                    \\  \specialrule{.1em}{.05em}{.05em} 
 \rowcolor{gainsboro!60}\textbf{Scenario}                             & \textbf{\begin{tabular}[c]{@{}c@{}}NSI NU\\ $(\epsilon_{ee}^{dV}, \epsilon_{ee}^{uV})$\end{tabular}}    & \textbf{\begin{tabular}[c]{@{}c@{}}NSI NU\\ $(\epsilon_{\mu \mu}^{dV}, \epsilon_{\mu \mu}^{uV})$\end{tabular}} & \textbf{\begin{tabular}[c]{@{}c@{}}NSI NU\\ $(\epsilon_{ee}^{dV}, \epsilon_{\mu \mu}^{d V})$\end{tabular}} & \textbf{\begin{tabular}[c]{@{}c@{}}NSI NU\\ $(\epsilon_{ee}^{uV}, \epsilon_{\mu \mu}^{u V})$\end{tabular}}      \\ \specialrule{.1em}{.05em}{.05em} 
{\color{magenta} \textbf{CsI } }
& 82.9 (0.863)                                                                                                   & 82.9 (0.863)                                                                                                   & 82.8 (0.863)                                                                                                      & 82.8 (0.863)                                                                                                      \\ \hline
{\color{darktangerine}\textbf{LAr }} 
& 105.7 (0.896)                                                                                                  & 105.6 (0.895)                                                                                                         & 105.5 (0.894)                                                                                                     & 105.5 (0.894)                                                                                                     \\\hline
{\color{ceruleanblue}\textbf{CsI+LAr }} 
& 188.9 (0.874)                                                                                                   & 188.5 (0.873)                                                                                                      & 188.9 (0.875)                                                                                                 & 188.5 (0.872)                                                                                                    \\  \specialrule{.1em}{.05em}{.05em} 
 \rowcolor{gainsboro!60} \textbf{Scenario}                             & \textbf{\begin{tabular}[c]{@{}c@{}}NSI FC\\ $(\epsilon_{ee}^{dV}, \epsilon_{e \mu}^{dV})$\end{tabular}} & \textbf{\begin{tabular}[c]{@{}c@{}}NSI FC\\ $(\epsilon_{\mu \mu}^{uV}, \epsilon_{e \mu}^{uV})$\end{tabular}}   & \textbf{\begin{tabular}[c]{@{}c@{}}NSI FC\\ $(\epsilon_{ee}^{dV}, \epsilon_{e \tau}^{dV})$\end{tabular}}   & \textbf{\begin{tabular}[c]{@{}c@{}}NSI FC\\ $(\epsilon_{\mu \mu}^{dV}, \epsilon_{\mu \tau}^{dV})$\end{tabular}} \\ \specialrule{.1em}{.05em}{.05em} 
{\color{magenta} \textbf{CsI } }
& 82.9   (0.863)                                                                                             & 82.9   (0.863)                                                                                                  & 82.9   (0.863)                                                                                                & 82.9   (0.863)                                                                                                      \\ \hline
{\color{darktangerine}\textbf{LAr }} 
& 105.5  (0.894)                                                                                                 & 105.5  (0.894)                                                                                                     & 105.7  (0.896)                                                                                               & 105.6  (0.895)                                                                                                     \\\hline
{\color{ceruleanblue}\textbf{CsI+LAr }} 
& 189.4  (0.877)                                                                                                & 189.1  (0.876)                                                                                                    & 189.4  (0.877)                                                                                               & 189.1  (0.876)                                                                                                   \\  \specialrule{.1em}{.05em}{.05em} 
\rowcolor{gainsboro!60} \textbf{Scenario}                             & \textbf{\begin{tabular}[c]{@{}c@{}}NGI (V)\\ ($C_V^q$)\end{tabular}}                                                                                    & \textbf{\begin{tabular}[c]{@{}c@{}}NGI (S)\\ ($C_S^q$)\end{tabular}}                                                                                            & \textbf{\begin{tabular}[c]{@{}c@{}}NGI (T)\\ ($C_T^q$)\end{tabular}}                                                                                        &                                                                                                           \\ \specialrule{.1em}{.05em}{.05em} 
{\color{magenta} \textbf{CsI } }
&  82.8 (0.854)                                                                                                 &  83.2 (0.858)                                                                                                 &  83.2 (0.858)                                                                                                 & ---                                                                                                                \\ \hline
{\color{darktangerine}\textbf{LAr }}
&   105.5  (0.887)                                                                                           &   103.2  (0.867)                                                                                                  &   104.6 (0.879)                                                                                              & ---                                                                                                          \\\hline
{\color{ceruleanblue}\textbf{CsI+LAr }} 
&  188.6 (0.869)                                                                                                &  189.7 (0.874)                                                                                                &  189.7 (0.874)                                                                                          & ---                                                                                                    \\  \specialrule{.1em}{.05em}{.05em} 
 \rowcolor{gainsboro!60} \textbf{Scenario}                             & \textbf{\begin{tabular}[c]{@{}c@{}}NGI (V-T)\\ ($C_V^q,C_T^q$)\end{tabular}}                                                                                      & \textbf{\begin{tabular}[c]{@{}c@{}}NGI (V-S)\\ ($C_V^q,C_S^q$)\end{tabular}}                                                                                             & \textbf{\begin{tabular}[c]{@{}c@{}}NGI (S-T)\\ ($C_S^q,C_T^q$)\end{tabular}}                                                                                         &                                                                                                           \\ \specialrule{.1em}{.05em}{.05em} 
{\color{magenta} \textbf{CsI } }
& 82.8  (0.863)                                                                                                 & 82.9  (0.863)                                                                                                      & 83.2  (0.867)                                                                                                     & ---                                                                                                                \\ \hline
{\color{darktangerine}\textbf{LAr }}
& 103.3  (0.875)                                                                                                & 102.6 (0.870)                                                                                                         & 103.2  (0.874)                                                                                                   & ---                                                                                                          \\\hline
{\color{ceruleanblue}\textbf{CsI+LAr }} 
& 188.6 (0.873)                                                                                                  & 188.6  (0.873)                                                                                                    & 189.7 (0.878)                                                                                                & ---                                                                                                    \\  \specialrule{.1em}{.05em}{.05em} 
 \rowcolor{gainsboro!60}\textbf{Scenario}                             & \textbf{\begin{tabular}[c]{@{}c@{}}LV universal\\ ($m_V, g_V$)\end{tabular}}                          & \textbf{\begin{tabular}[c]{@{}c@{}}LV B-L \\ ($m_V, g_V$)\end{tabular}}                       & \textbf{\begin{tabular}[c]{@{}c@{}}LS \\ ($m_S, g_S$)\end{tabular}}                            & \textbf{\begin{tabular}[c]{@{}c@{}}LT \\ ($m_T, g_T$)\end{tabular}}                                                                                                      \\ \specialrule{.1em}{.05em}{.05em} 
{\color{magenta} \textbf{CsI } } 
& 81.4   (0.848)                                                                                                 & 83.2   (0.867)                                                                                                       & 83.2   (0.867)                                                                                                   & 83.2   (0.867)                                                                                                  \\ \hline
{\color{darktangerine}\textbf{LAr }}
& 105.6   (0.895)                                                                                               & 105.5   (0.894)                                                                                                     & 102.9   (0.872)                                                                                                  & 104.6   (0.887)                                                                                                       \\\hline
{\color{ceruleanblue}\textbf{CsI+LAr }} 
& 187.8  (0.869)                                                                                                 & 189.6  (0.878)                                                                                                   & 189.4  (0.877)                                                                                               & 189.5  (0.877)                                                                                                 \\  \specialrule{.1em}{.05em}{.05em} 
 \rowcolor{gainsboro!60} \textbf{Scenario}                             & \textbf{\begin{tabular}[c]{@{}c@{}}millicharge\\ ($ q_{\nu_{ee}}, q_{\nu_{\mu \mu}}$)\end{tabular}}                         & \textbf{\begin{tabular}[c]{@{}c@{}}charge radius\\ ($\langle r^2_{\nu_{ee}} \rangle, \langle r^2_{\nu_{\mu \mu}} \rangle$)\end{tabular}}                     & \textbf{\begin{tabular}[c]{@{}c@{}}$\boldsymbol{\mathrm{TMM_{sterile}}}$\\ ($m_4, \mu_{\nu_\mu}$)\footnote{For the active-sterile TMM case, we show only the limits coming out from the $(m_4, \mu_{\nu_{\mu}})$ analysis since we have verified that the case of  $(m_4, \mu_{\nu_{e}})$ leads to approximately the same result.}\end{tabular}}                                                                                    &                 \textbf{\begin{tabular}[c]{@{}c@{}}Sterile Osc.\\ ($\sin^2 2 \theta_{24}, \Delta m^2_{42}$)\end{tabular}}                                                                              \\ \specialrule{.1em}{.05em}{.05em} 
{\color{magenta} \textbf{CsI } }
& 83.2   (0.867)                                                                                                & 82.8   (0.863)                                                                                                      & 83.2   (0.867)                                                                                 & 82.1  (0.855)                                                                                             \\ \hline
{\color{darktangerine}\textbf{LAr }} 
& 106.4  (0.902)                                                                                                 & 105.5  (0.894)                                                                                                     & 105.1  (0.891)                                                                                          & 106.5  (0.902)                                                                                              \\
\hline
{\color{ceruleanblue}\textbf{CsI+LAr }} 
                                             & 189.7  (0.878)                                                                                                  & 188.4  (0.872)                                                                                                   & 189.5  (0.877)                                                                                                   & 188.6  (0.881)                                                                                                \\ \specialrule{.2em}{.1em}{.1em}
\end{tabular}}%
}
\caption{$\chi^2_\text{min}$ values for the various physics scenarios analyzed in the present work, in parenthesis normalized to $\#_{\rm{dof}}$ in each case. }
\label{tab:summary}
\end{table}

\section*{Acknowledgements}

 We thank Oscar Sanders for collaborating in the early stages of this project.
We are grateful to Dan Pershey for sharing insightful details about the COHERENT analysis performed in Ref.~\cite{COHERENT:2021xmm}.
We also thank Luis Flores and Anirban Majumdar for fruitful discussions. 
This work has been supported by the Spanish grants PID2020-113775GB-I00 (AEI/10.13039/501100011033) and CIPROM/2021/054 (Generalitat Valenciana).
VDR acknowledges financial support by the SEJI/2020/016 grant (Generalitat Valenciana).
DKP was supported by the Hellenic Foundation for Research and Innovation (H.F.R.I.) under the “3rd Call for H.F.R.I. Research Projects to support Post-Doctoral Researchers” (Project Number: 7036). The work of O. G. M. and G. S. G. has been supported in part by CONACYT-Mexico under grant A1-S-23238. O. G. M. has been
supported by SNI (Sistema Nacional de Investigadores, Mexico).

\noindent

\appendix
\section{Details of the CsI signal simulation}
\label{sec:appendix}

\begin{figure}[!htb]
\includegraphics[width=  0.5\textwidth]{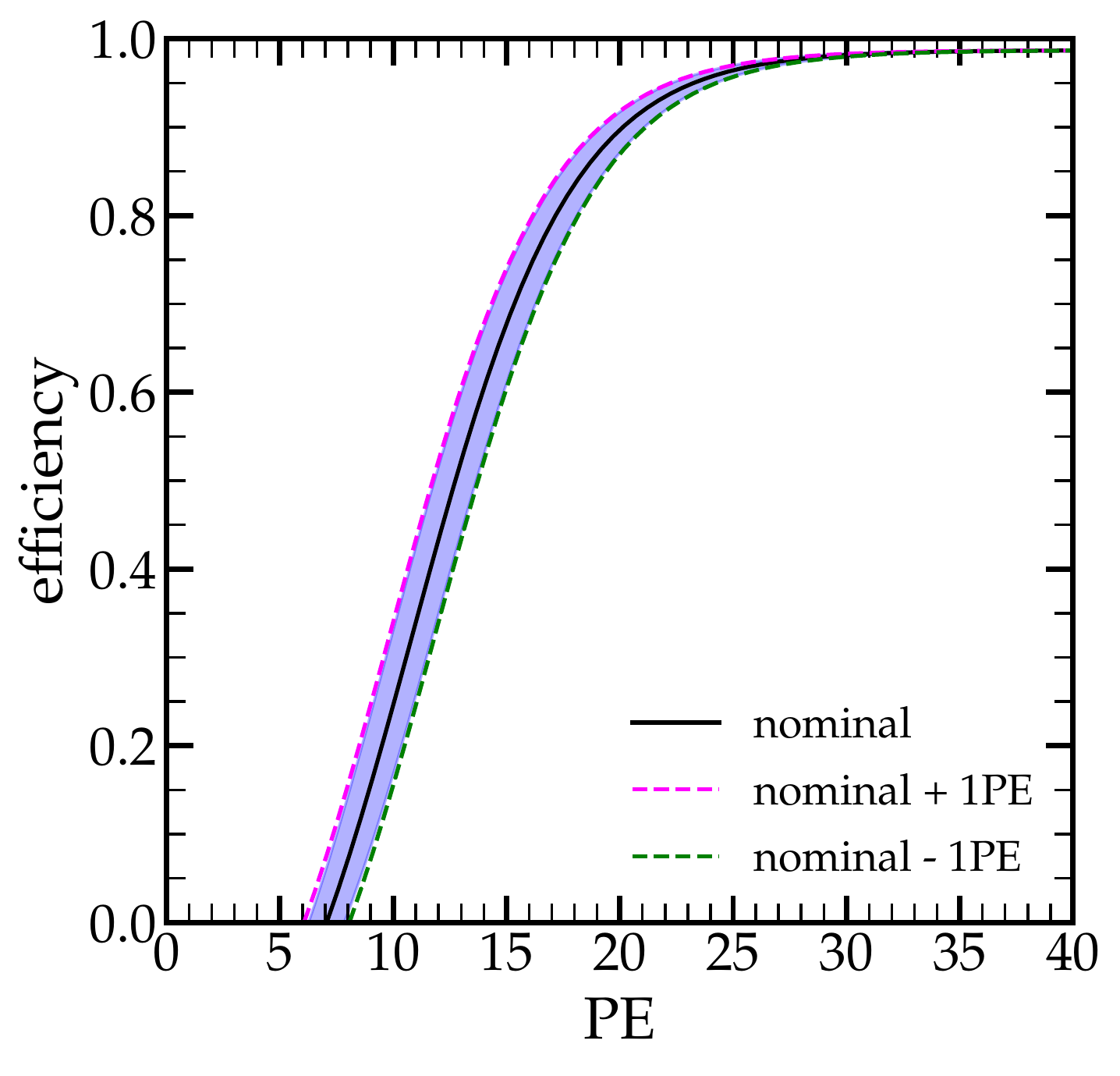}
\caption{COHERENT-CsI efficiency as a function of PE along with its $1\sigma$ uncertainty.}
\label{fig:eff}
\end{figure}

In this Appendix we provide some further details regarding the analysis of COHERENT-CsI data. First, in Fig.~\ref{fig:eff} we present the efficiency as a function of the reconstructed PE,
  where the width of the curve illustrates the $\pm 1 \sigma$ uncertainty given in Ref.~\cite{COHERENT:2021xmm}.
  We also show that $\pm 1$PE variations of the reconstructed photoelectron have an effect equivalent with varying the parameters $a,b,c,d$ entering
  Eq.~(\ref{eq:CsI_E_efficiency}) within $\pm 1 \sigma$. 

\begin{figure}[!htb]
\includegraphics[width=  \textwidth]{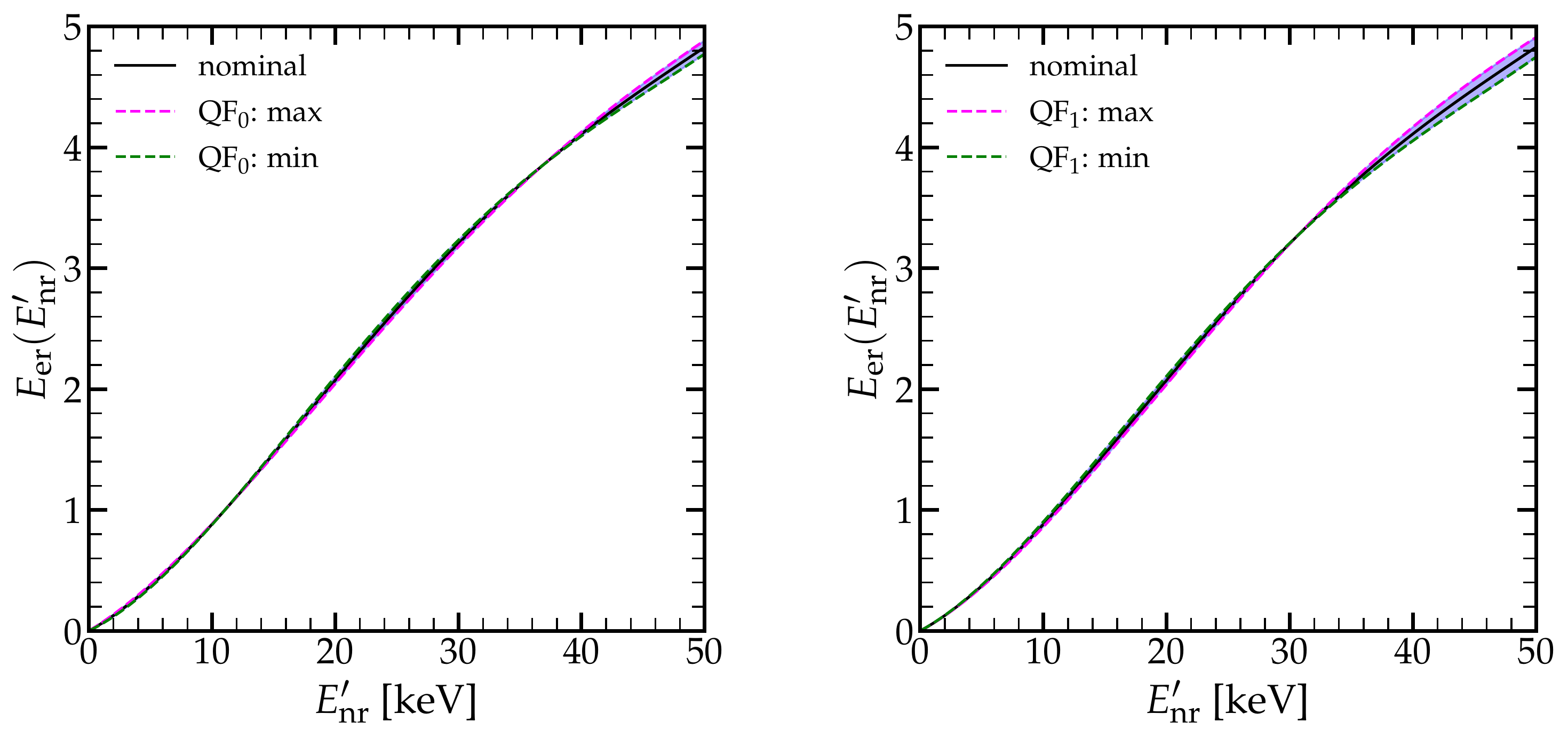}
\caption{COHERENT-CsI scintillation curve as function of the true nuclear recoil energy for the two QF models reported by COHERENT~\cite{COHERENT:2021xmm}.}
\label{fig:QF}
\end{figure}

\begin{figure}[!htb]
\includegraphics[width=  \textwidth]{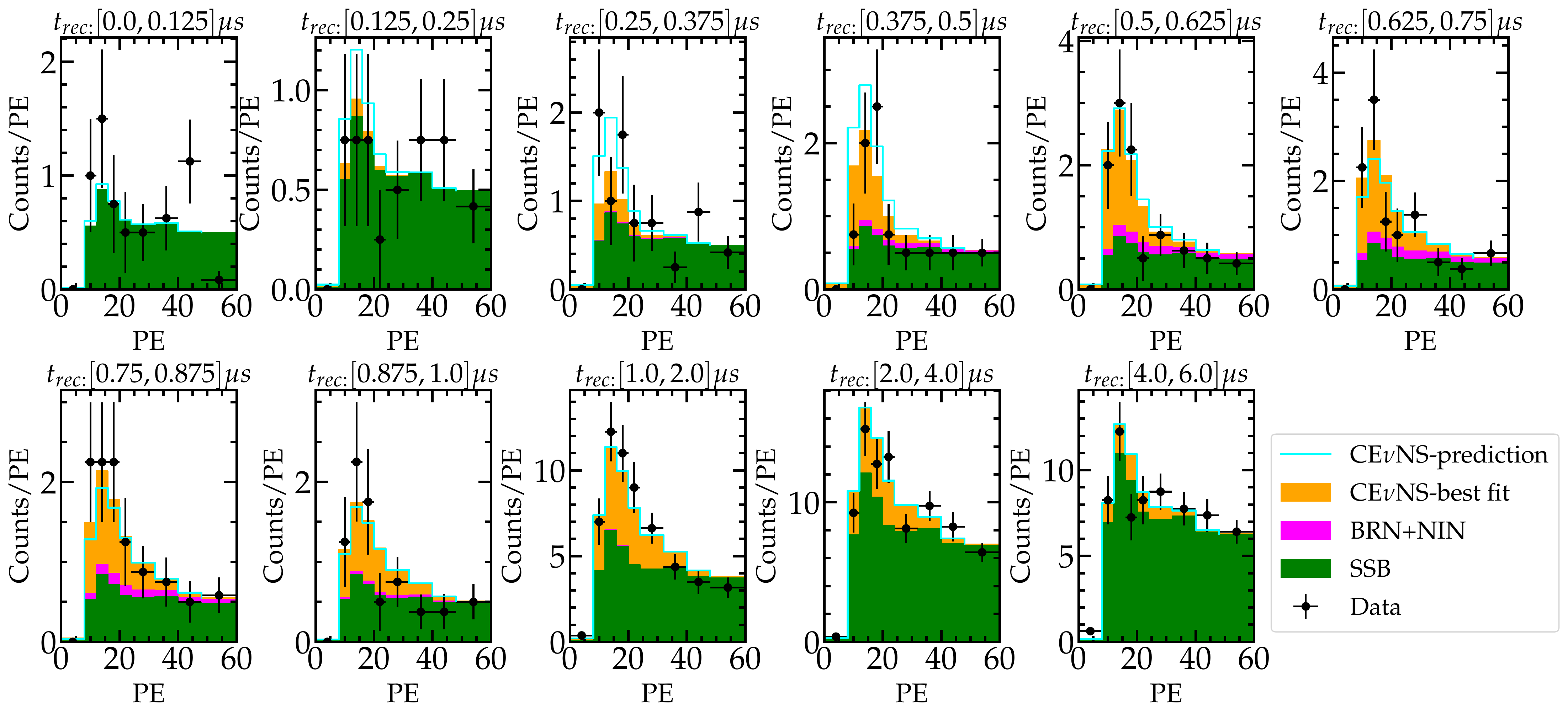}
\caption{Simulation of the event spectra at  COHERENT-CsI along with the experimental data of COHERENT-CsI measurement~\cite{COHERENT:2021xmm} (in agreement with~\cite{Pershey:M7s}).}
\label{fig:event_spectra}
\end{figure}

In Fig.~\ref{fig:QF} we present the scintillation response curve $E_\text{er}$ as a function of the true nuclear recoil energy $E^\prime_\text{nr}$ (see Sec.~\ref{sec:csi-number-events})
  using the best fit coefficients $x_i$ provided in the supplemental material of Ref.~\cite{COHERENT:2021xmm} for the two QF models reported by COHERENT Collaboration.
  The width of the curve indicates the $\pm 1 \sigma$ uncertainty which is obtained by varying the coefficients $x_i$ within their $\pm 1 \sigma$ uncertainties. As can be seen, the variation of the individual $x_i$ coefficients is expected to have a rather small effect on the shape uncertainty of the  CE$\nu$NS event rate. Thus, in order to reduce computational time in our analysis we assigned a
  flat 3.8\% uncertainty to the \cevns~ normalization (through the nuisance $\alpha_5$)~\cite{COHERENT:2021xmm}.
We finally compare our theoretical calculation of the event spectra with the experimental data from the COHERENT-CsI measurement in Fig.~\ref{fig:event_spectra}. Also superimposed is the best fit of the predicted event spectra from where one can notice the effect of timing and threshold uncertainties taken into account in the present work (compare cyan and orange histograms).

\bibliographystyle{utphys}
\bibliography{bibliography}  

\end{document}